%% file: main.tex
\IfFileExists{patch-revtex4-2.tex}{\input{patch-revtex4-2.tex}}{}

\documentclass[letterpaper,11pt,superscriptaddress,nofootinbib]{quantumarticle}

\pdfoutput=1

\usepackage{xcolor}
\usepackage{amsmath,amsthm,amsfonts,amssymb,braket}
\usepackage[pdftex,colorlinks=true,linkcolor=blue,citecolor=darkred,urlcolor=darkblue]{hyperref}
\usepackage[all]{xy}
\usepackage{enumitem}	
\usepackage{fullpage}
\usepackage{mathtools}

\usepackage{makecell}

\usepackage{tikz}   
\usetikzlibrary{shapes}

\usepackage{soul}

\newtheorem{proposition-definition}[lemma]{Proposition-Definition}
\theoremstyle{definition}

\usetikzlibrary{decorations.text}

\definecolor{darkblue}{rgb}{0.,0.,0.4}
\definecolor{darkred}{rgb}{0.5,0.,0.}
\definecolor{darkpurple}{rgb}{0.5,0.,0.5}
\definecolor{ltgreen}{rgb}{0.1,.59,.43}
\definecolor{orange}{rgb}{1.0, 0.5, 0.0}
\definecolor{meh}{rgb}{1.0,0.95,0.7}

\newcommand{\replace}[1]{\textcolor{red}{\st{#1}}}
\renewcommand{\replace}[1]{}

\newcommand*\leadauthor{\thanks{Lead authors}}

\renewcommand{\leadauthor}{}
\newcommand{\leadauthorquantum}{*}
\AtBeginDocument{%
  \lfoot{*Lead authors}%
  \AtBeginShipoutNext{%
    \lfoot{}%
  }%
}

\newcommand{\onlinecite}[1]{\cite{#1}}

\makeatletter
\def\l@subsubsection#1#2{}
\makeatother

\newcommand*{\vcenteredhbox}[1]{\begingroup
\setbox0=\hbox{#1}\parbox{\wd0}{\box0}\endgroup}

\input{figurecommands}

\begin{document}
\title{Improved Pairwise Measurement-Based Surface Code}

\author{Linnea Grans-Samuelsson\leadauthorquantum}\leadauthor
\affiliation{Microsoft Station Q, Santa Barbara, California 93106-6105 USA}
\author{Ryan V.~Mishmash\leadauthorquantum}\leadauthor
\affiliation{Microsoft Station Q, Santa Barbara, California 93106-6105 USA}
\author{David Aasen\leadauthorquantum}\leadauthor
\affiliation{Microsoft Station Q, Santa Barbara, California 93106-6105 USA}
\author{Christina Knapp}
\affiliation{Microsoft Station Q, Santa Barbara, California 93106-6105 USA}
\author{Bela Bauer}
\affiliation{Microsoft Station Q, Santa Barbara, California 93106-6105 USA}
\author{Brad Lackey}
\affiliation{Microsoft Quantum, Redmond, Washington 98052, USA}
\author{Marcus P. da Silva}
\affiliation{Microsoft Quantum, Redmond, Washington 98052, USA}
\author{Parsa Bonderson\leadauthorquantum}\leadauthor
\affiliation{Microsoft Station Q, Santa Barbara, California 93106-6105 USA}

\begin{abstract}

We devise a new realization of the surface code on a rectangular lattice of qubits utilizing single-qubit and nearest-neighbor two-qubit Pauli measurements and three auxiliary qubits per plaquette.
This realization gains substantial advantages over prior pairwise measurement-based realizations of the surface code.
It has a short operation period of 4 steps and our performance analysis for a standard circuit noise model yields a high fault-tolerance threshold of approximately $0.66\% $.
The syndrome extraction circuits avoid bidirectional hook errors, so we can achieve full code distance by choosing appropriate boundary conditions.
We also construct variants of the syndrome extraction circuits that entirely prevent hook errors, at the cost of larger circuit depth.
This achieves full distance regardless of boundary conditions, with only a modest decrease in the threshold.
Furthermore, we propose an efficient strategy for dealing with dead components (qubits and measurements) in our surface code realization, which can be adopted more generally for other surface code realizations.
This new surface code realization is highly optimized for Majorana-based hardware, accounting for constraints imposed by layouts and the implementation of measurements, making it competitive with the recently proposed Floquet codes. 

\end{abstract}

\maketitle


\section{Introduction}

The inherent fragility of quantum states has presented a formidable challenge in the pursuit of a scalable quantum computer. 
Quantum error correction will undoubtedly be essential in any practical realization. 
Due to its high threshold and local connectivity, the surface code~\cite{Kitaev2003,Bravyi1998,Dennis2002} is a leading candidate for a scalable quantum error correcting code. 
Realizing a surface code subject to hardware constraints is a challenge, and different realizations will have varying performances. 
In particular, designing a syndrome extraction circuit composed of native operations, while maintaining code performance, is essential. 
In the broader context of the implementation of useful quantum algorithms, the resources required can be greatly impacted by the choice of code and how well matched it is to the hardware constraints~\cite{Beverland2022,Paetznick2023}.

Most of the proposed implementations of the surface code in hardware have followed the CNOT gate-based realization of stabilizer measurement circuits of Ref.~\onlinecite{Dennis2002}, or variants thereof.
More recent proposals motivated by measurement-based Majorana quantum computing hardware~\cite{Karzig2017} have considered pairwise measurement-based realizations of the surface code~\cite{Tran2019,Chao2020,Gidney2022a}, which all utilized two auxiliary qubits for each bulk plaquette stabilizer measurement circuit.
These pairwise measurement-based proposals each exhibited various significant drawbacks, such as complicated layouts and difficult measurements in Majorana hardware\footnote{In Majorana hardware, the double ancilla realizations~\cite{Tran2019,Chao2020} require mixed tetron-hexon layouts and measurement loops involving coherent links; the windmill realization~\cite{Chao2020} requires measurement loops involving coherent links; and the pentagonal tiling realization~\cite{Gidney2022a} require measurement loops involving coherent links or long semiconductor segments.}, relatively long circuits~\cite{Chao2020}, or bidirectional hook errors~\cite{Gidney2022a}.
The Floquet codes developed in Refs.~\onlinecite{Hastings2021,Haah2022,Gidney2022c,Paetznick2023} provided a major advancement for the realization of pairwise measurement-based codes, largely eliminating such drawbacks and achieving better performance with a high fault-tolerance threshold.

In this paper, we devise and analyze a new implementation of the surface code using single- and two-qubit Pauli measurements on a rectangular array of qubits.
This implementation utilizes three auxiliary qubits per stabilizer in the bulk.
This property motivates us to refer to this surface code implementation as the ``3aux'' code when a short descriptive name is useful.
Moreover, the stabilizer measurement circuits only utilize $X$, $Z$, $XX$, and $ZZ$ Pauli measurements, with the pairwise measurements being directionally correlated between nearest-neighbor qubits, e.g. $XX$ measurements are always between horizontal neighbors and $ZZ$ are always between vertical neighbors.
It is worth observing that our stabilizer measurement circuits and those of Refs.~\onlinecite{Tran2019,Chao2020,Gidney2022a} are all closely related through circuit equivalences, for example using $ZX$-calculus relations (see e.g. Ref.~\onlinecite{Wetering2020} for a review of $ZX$-calculus).
Despite these close relations, there are important physical differences that translate into implementation and performance advantages for our surface code realization.
For example, our stabilizer measurement circuits can be pipelined in a manner that yields a minimal operation period of 4 steps for running the code, in which case no qubit is idle in any step.

The stabilizer measurement circuits in our surface code implementation have hook errors, similar to the original CNOT-based realizations discussed in Ref.~\onlinecite{Dennis2002}. 
Hook errors stem from single physical faults that, for a given circuit, are equivalent to higher-weight (e.g. two-qubit) errors on the data qubits (and not equivalent to single data qubit errors).
These can reduce the {\em fault distance}~\cite{bombinlogical2023}, i.e., the minimum number of gate faults that causes an undetectable logical error; in particular, in the toric and surface codes, where logical operators are associated with a direction, hook errors reduce the fault distance when aligned with a logical operator of the same type. Unidirectionality or bidirectionality of hook errors indicate whether each type occurs in one or two directions with respect to the surface.
In our case, the hook errors are unidirectional, with $X$ and $Z$ type hook errors in perpendicular directions.
As such, our implementation can be utilized for the ``rotated'' surface code~\cite{Bombin2007} without halving the fault distance by choosing boundary conditions and operation schedules accordingly, similar to the strategy used in Ref.~\onlinecite{Tomita2014}.
This is in contrast to realizations with hook errors that cannot be favorably oriented, such as the bidirectional hook errors of Ref.~\onlinecite{Gidney2022a}.  
Additionally, for situations where it is desirable, we can modify the circuit to prevent hook errors from occurring, at the expense of increasing the operation period to 7 steps.
Given these properties, our code provides an interesting test bed for analyzing the effects of hook errors on code performance.

We can modify the bulk 4-gon (four data qubit) stabilizer measurement circuits to provide circuits for measuring 1-gon, 2-gon, and 3-gon operators that interlock with the bulk 4-gon circuits.
These $n$-gon measurement circuits utilize measurements from the same operation set, so constitute a minimal modification of the bulk.
We can use the $n$-gons to implement code patches with any desired boundary conditions and alter the shape of code patches during operation, without disrupting the operation cycle.
Moreover, we can utilize the $n$-gon measurement circuits to implement a protocol for dealing with dead components (i.e. dead qubits and pairwise measurements).
In particular, we present an improved and maximally efficient variant of the protocol of Ref.~\onlinecite{Auger2017} for dealing with dead components by measuring stabilizers for reduced $n$-gons that exclude dead components.
Our protocol can be incorporated in a natural manner that requires minimal modification to the code operation, i.e. no addition to the set of measurements and no change to the bulk operation cycle.
Furthermore, our circuit pipelining effectively alternates between $Z$-type and $X$-type $n$-gon measurements within each round, which allows all the ``superplaquette'' operators to be measured by measuring the reduced $n$-gons (``damaged plaquettes'') within each round.
We expect this to provide better performance than interleaved circuits would for such dead component protocols.

Since a motivation of this work was to develop code implementations for measurement-based Majorana quantum computing hardware, in particular in arrays of ``tetron'' qubits~\cite{Karzig2017}, our surface code realization is highly optimized for such hardware.
In this regard, the rectangular lattice used for our code represents the simplest possible layout for such Majorana hardware.
Moreover, for this hardware and layout, the measurements utilized in this code are extremely simple and likely to yield the lowest error rates of any set of physical Majorana parity measurements capable of generating a quantum error correcting code.
Another consideration for this hardware is whether to use single or double columns of semiconductor ``rails,'' which run between Majorana qubits to enable the measurements.
Double-rail semiconductor layouts avoid physical conflicts between simultaneous measurements of adjacent qubits, generally allowing circuits to be implemented more efficiently in time.
On the other hand, single-rail semiconductor layouts significantly simplify the requirements on fabrication and control for Majorana-based hardware, as compared to double-rail layouts.
Using single-rail semiconductor layouts instead of double-rail generally increases the operation period by up to a factor of two; for example, using single-rail layouts instead of double-rail will double the operation period for the 4.8.8 Floquet code implementation in Majorana hardware described in Ref.~\onlinecite{Paetznick2023}.
We find that the use of single-rail layouts for our surface code realization can be implemented with a mild one step increase in the period.

Finally, we analyze the performance of our code for various scenarios.
Our results indicate significant improvement compared to the previous pairwise measurement-based surface code realizations.
In particular, simulating the logical failure rate for the standard circuit noise model, we find the fault-tolerance threshold to be approximately $0.66\%$, and achieve full distance sub-threshold scaling (when hook errors are appropriately addressed).
Examining the effects of hook errors when using different boundary conditions and hook-preventing circuit modifications, we generally see the code performing better and achieving full distance in the deep sub-threshold regime for the scenarios where hook errors are benign or absent.
Interestingly, the threshold and near threshold behavior are only modestly reduced by the hook-preventing modifications, in contrast to hook-flagging modifications of CNOT-based circuits~\cite{Chamberland2018}.
Moreover, we compare our code to the state-of-the-art pairwise measurement-based 4.8.8 Floquet code~\cite{Hastings2021,Haah2022,Paetznick2023}, and find it to be reasonably competitive, especially at lower physical error rates.
For variants of these two codes that are compatible with implementation in Majorana hardware with single-rail layouts, their performance becomes even more competitive and their thresholds nearly identical.
We provide a summary comparison of key properties of the various pairwise measurement-based codes in Table~\ref{tab:comparison}.
(For the detailed comparison of performance and resource requirements between our surface code realizations and the 4.8.8 Floquet code, see Sec.~\ref{sec:performance_results}.)

The structure of this paper is as follows.
In Sec.~\ref{sec:Circuits}, we introduce our code, presenting the pairwise measurement-based stabilizer measurement circuits and pipelining to realize the surface code on a rectangular lattice.
In Sec.~\ref{sec:Hook}, we describe the occurrence of hook errors in our circuits and present modifications of our measurement circuits that prevent the occurrence of hook errors.
In Sec.~\ref{sec:boundaries}, we detail the measurement circuits for all $n$-gons, with $n=1,2,3$, and utilize these for surface code patches with boundaries.
In Sec.~\ref{sec:DeadComponents}, we describe our proposed strategy for dealing with dead components and its application to our surface code realization. 
In Sec.~\ref{sec:Majorana}, we discuss the implementation in Majorana hardware and provide a modification of the pipelining to make the code compatible with single-rail semiconductor layouts.
In Sec.~\ref{sec:Performance}, we describe our numerical simulations of code performance and resource estimation and present the results.

\begin{table}[]
    \centering
\begin{tabular}{c c c c c}
protocol &  qubit count & depth & threshold & Majorana hardware \\
\hline
3aux  &  $O(4 d_{\rm f}^2)$  &   4  & 0.66\%  &  simple \\
3aux, single-rail  &  $O(4 d_{\rm f}^2)$  &   5  & 0.51\%  &  simple \\
\hline
double ancilla &  $O(3 d_{\rm f}^2)$  &   10 & 0.24\%  &  complicated \\
windmill &  $O(2 d_{\rm f}^2)$  &   20  &  0.15\%  &  complicated  \\
pentagonal &  $O(12 d_{\rm f}^2)$  &   6  & 0.4\% $^\ast$  &  complicated \\
\hline
4.8.8 &  $O(4 d_{\rm f}^2)$  &   3  & 1.3\%  &  simple \\
honeycomb &  $O(6 d_{\rm f}^2)$  &   3  &  1.3\%  &  simple \\
4.8.8, single-rail &  $O(4 d_{\rm f}^2)$  &   6  & 0.52\%  &  simple \\
\hline
\end{tabular}
    \caption{
    Comparison between key properties of pairwise measurement-based codes, including our realization of the surface code (denoted ``3aux''), the double ancilla and windmill realizations of Ref.~\protect\onlinecite{Chao2020}, the pentagonal tiling realization of Ref.~\protect\onlinecite{Gidney2022a}, and the Floquet code on the 4.8.8 and honeycomb lattices~\cite{Hastings2021,Haah2022,Gidney2022c,Paetznick2023}.
    The term ``single-rail'' indicates variants of the corresponding codes that are needed to make them compatible with Majorana hardware using single-rail semiconductor layouts.
    Total qubit count for a logical patch is given to leading order as function of fault distance $d_{\rm{f}}$.
    Circuit depth is given per round of syndrome extraction.
    Fault tolerance thresholds are computed with respect to the noise model of  Ref.~\protect\onlinecite{Chao2020}, except for that of the pentagonal tiling realization from Ref.~\protect\onlinecite{Gidney2022a} (indicated by $^*$), which uses a slightly different noise model.
    We indicate whether the code can be implemented in Majorana hardware using simple layouts and measurements, or if it requires complicated layouts and measurements that are likely prohibitive.
    }
    \label{tab:comparison}
\end{table}


\section{The Code Circuits}
\label{sec:Circuits}

\begin{figure}[t!]
\centering

\begin{tikzpicture}
\newcommand{\halfscaling}{0.6}
\newcommand{\nodesize}{1.5}
\draw [fill=white,opacity=0] (-0.5*\halfscaling,-0.5*\halfscaling) rectangle (2.5*\halfscaling,2.5*\halfscaling); 
\draw [fill=red,opacity=0.1] (0,0) rectangle (2*\halfscaling,2*\halfscaling);
\draw [fill=blue,opacity=0.15] (2*\halfscaling,0) rectangle (4*\halfscaling,2*\halfscaling);
\draw [fill=red,opacity=0.1] (2*\halfscaling,2*\halfscaling) rectangle (4*\halfscaling,4*\halfscaling);
\draw [fill=blue,opacity=0.15] (0*\halfscaling,2*\halfscaling) rectangle (2*\halfscaling,4*\halfscaling);
\foreach \i in {0,...,4}{\foreach \j in {0,...,4}{
\draw (\i*\halfscaling, \j*\halfscaling) node[circle, draw, fill=black, minimum size = \nodesize mm, inner sep = 0]{}; }; };
\foreach \i in {0,2,4}{\foreach \j in {0,2,4}{
\draw (\i*\halfscaling, \j*\halfscaling) node[circle, draw, fill=white, minimum size = \nodesize mm, inner sep = 0]{};};};
\end{tikzpicture}
\hspace{1cm}
\begin{tikzpicture}
\newcommand{\halfscaling}{0.6}
\newcommand{\nodesize}{1.5}
\draw [fill=white,opacity=0] (-0.5*\halfscaling,-0.5*\halfscaling) rectangle (2.5*\halfscaling,2.5*\halfscaling); 
\draw [fill=blue,opacity=0.15] (0,0) rectangle (2*\halfscaling,2*\halfscaling);
\foreach \i in {0,...,2}{\foreach \j in {1}{
\draw (\i*\halfscaling, \j*\halfscaling) node[circle, draw, fill=black, minimum size = \nodesize mm, inner sep = 0]{}; }; };
\foreach \i in {0,2}{\foreach \j in {0,2}{
\draw (\i*\halfscaling, \j*\halfscaling) node[circle, draw, fill=white, minimum size = \nodesize mm, inner sep = 0]{};};};
\node at (0*\halfscaling, 1*\halfscaling)[below] {A};
\node at (1*\halfscaling, 1*\halfscaling)[below] {B};
\node at (2*\halfscaling, 1*\halfscaling)[below] {C};
\end{tikzpicture}
\hspace{1cm}
\begin{tikzpicture}
\newcommand{\halfscaling}{0.6}
\newcommand{\nodesize}{1.5}
\draw [fill=white,opacity=0] (-0.5*\halfscaling,-0.5*\halfscaling) rectangle (2.5*\halfscaling,2.5*\halfscaling); 
\draw [fill=red,opacity=0.1] (0,0) rectangle (2*\halfscaling,2*\halfscaling);
\foreach \i in {1}{\foreach \j in {0,...,2}{
\draw (\i*\halfscaling, \j*\halfscaling) node[circle, draw, fill=black, minimum size = \nodesize mm, inner sep = 0]{}; }; };
\foreach \i in {0,2}{\foreach \j in {0,2}{
\draw (\i*\halfscaling, \j*\halfscaling) node[circle, draw, fill=white, minimum size = \nodesize mm, inner sep = 0]{};};};
\node at ( 1*\halfscaling,0*\halfscaling)[left] {C};
\node at ( 1*\halfscaling,1*\halfscaling)[left] {B};
\node at ( 1*\halfscaling,2*\halfscaling)[left] {A};
\end{tikzpicture}
\hspace{1cm}
\begin{tikzpicture}
\newcommand{\halfscaling}{0.6}
\newcommand{\nodesize}{1.5}
\draw [fill=white,opacity=0] (-0.5*\halfscaling,-0.5*\halfscaling) rectangle (2.5*\halfscaling,2.5*\halfscaling); 
\draw (0,0.5) node[circle, draw, fill=white, minimum size = \nodesize mm, inner sep = 0]{};
\node at (0,0.5)[right] {$\;$ data qubit};
\draw (0,0) node[circle, draw, fill=black, minimum size = \nodesize mm, inner sep = 0]{};
\node at (0,0)[right] {$\;$ auxiliary qubit};
\end{tikzpicture}

\caption{
A rectangular lattice of qubits used for a pairwise measurement-based realization of the surface code.
Data qubits are shown as open dots and auxiliary qubits are shown as solid dots.
The blue and red squares will correspond to $Z$-type and $X$-type plaquettes ($4$-gons), respectively.
Each plaquette exclusively utilizes three auxiliary qubits (labeled $A$-$C$) to execute its stabilizer measurement circuit.
}
\label{fig:rectangular_lattice}
\end{figure}
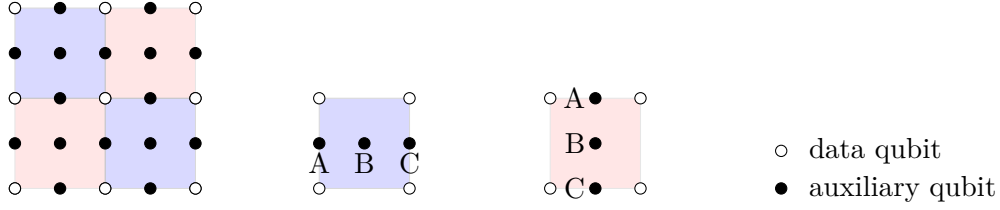

We use a rectangular lattice of qubits to implement a pairwise measurement-based realization of the surface code, as shown in Fig.~\ref{fig:rectangular_lattice}.
The plaquettes correspond to $4$-gon stabilizer measurements, which are arranged in a checkerboard pattern of $Z$-type ($ZZZZ$ stabilizers) and $X$-type ($XXXX$ stabilizers).
Each plaquette exclusively utilizes three auxiliary qubits to perform the stabilizer measurement circuit.

\begin{figure}[t!]
\centering
\input{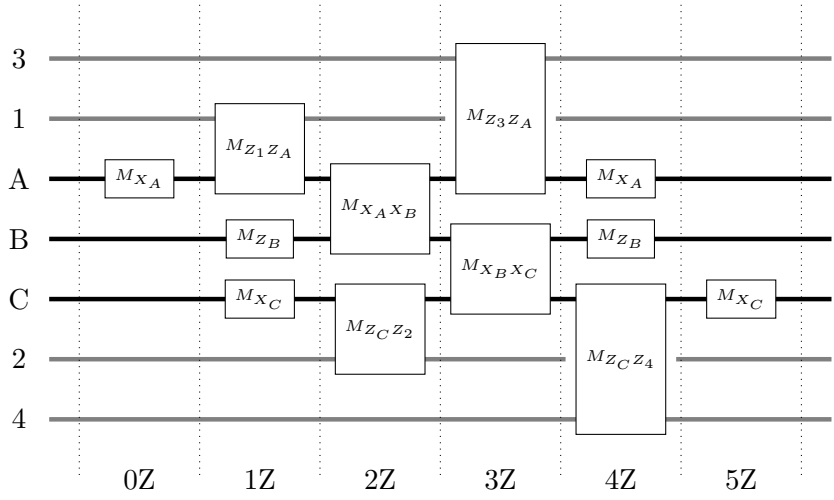}
\caption{
The $M_{ZZZZ}$ circuit for measuring $ZZZZ$ on four data qubits using three auxiliary qubits.
The data qubits are labeled $1$-$4$ in the order that they are addressed in this circuit. 
The auxiliary qubits are labeled $A$-$C$.
The measurement schedule shown here is useful for repeatedly applying the measurement circuit with a four step period, which can be done by repeating steps $1$-$4$.
}
\label{fig:MZ4circuit}
\end{figure}

In Fig.~\ref{fig:MZ4circuit}, we present a circuit diagram for $M_{ZZZZ}$, the measurement of $ZZZZ$ on four data qubits, using three auxiliary qubits.
The $ZZZZ$ stabilizer measurement outcome is given by the product of the measurement outcomes of the six $Z$-type measurements, i.e. $M_{Z_B}$ (initial), $M_{Z_A Z_1}$, $M_{Z_C Z_2}$, $M_{Z_A Z_3}$, $M_{Z_C Z_4}$, and $M_{Z_B}$ (final).
There are various methods for verify the functioning of this circuit.
One straightforward method is to compute the instantaneous stabilizer group (ISG)~\cite{Hastings2021} following each of the six steps of the circuit:
\begin{align}
\label{eq:MZ4circuit_ISG}
& 0Z: \quad \left\langle X_{A} \right\rangle \notag \\
& 1Z: \quad \left\langle Z_{1}Z_{A} , Z_{B}, X_{C} \right\rangle \notag \\
& 2Z: \quad \left\langle X_{A}X_{B}, Z_{C}Z_{2} ,  Z_{1}Z_{A}Z_{B} \right\rangle \notag \\
& 3Z: \quad \left\langle Z_{3}Z_{A}, X_{B}X_{C}, Z_{1}Z_{A}Z_{B}Z_{C}Z_{2} \right\rangle \notag  \\
& 4Z: \quad \left\langle X_{A}, Z_{B}, Z_{C}Z_{4}, Z_{1}Z_{B}Z_{C}Z_{2}Z_{3} \right\rangle \notag \\
& 5Z: \quad \left\langle X_{C}, X_{A}, Z_{B}, Z_{1}Z_{2}Z_{3}Z_{4} \right\rangle 
.
\end{align}
Alternatively, one may verify the functioning of the circuit through a straightforward application of $ZX$-calculus~\cite{Wetering2020}.

\begin{figure}[t!]
\centering
\input{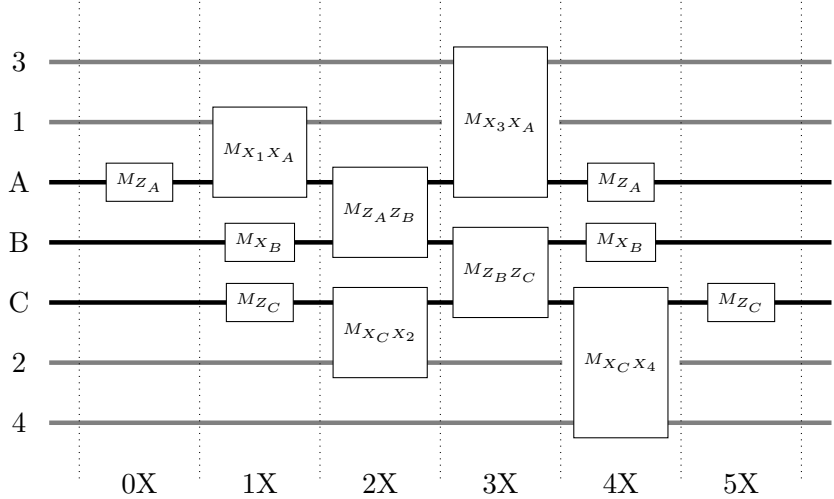}
\caption{
The $M_{XXXX}$ circuit for measuring $XXXX$ on four data qubits using three auxiliary qubits may be obtained from $M_{ZZZZ}$ by interchanging $X \leftrightarrow Z$ for all qubits in the circuit.
}
\label{fig:MX4circuit}
\end{figure}

By replacing $X \leftrightarrow Z$ in this circuit, we obtain the circuit for measuring $XXXX$ on four data qubits shown in Fig.~\ref{fig:MX4circuit}.
We note that there are numerous equivalences that can be applied to produce equivalent stabilizer circuits.
For example, one could apply basis changes, permute data or auxiliary qubit labels, or modify the operation schedule, e.g. by adding/removing time steps and sliding measurements into different time steps (without sliding past other measurements on the same qubit lines).
We will use this flexibility to incorporate certain appealing features in the implementation.

One desirable feature is to minimize the operation time.
In order to compress the circuit into the shortest possible operation period, we can use the measurement schedules shown for $M_{ZZZZ}$ and $M_{XXXX}$ in Figs.~\ref{fig:MZ4circuit} and \ref{fig:MX4circuit}.
While these circuits are shown with six steps, if we are repeating a stabilizer measurement circuit, the single-qubit measurements on auxiliary qubits can serve as both the final measurement of one round of running the stabilizer measurement circuit and the initial measurement of the subsequent round.
Thus, we can avoid repeating the single-qubit measurements on auxiliary qubits $A$ and $C$ between every round, removing steps 0 and 5 from all but the very first and last rounds of applying the circuit, respectively.
In this way, the number of steps necessary for applying $r$ rounds of the stabilizer measurement circuit is $4r+2$, i.e. the circuit can be implemented with a $4$ step period.
Here, the initial and final rounds correspond to the initial measurement of qubit $A$ (step 0) and final measurement of qubit $C$ (step 5), while steps 1-4 are repeated $r$ times in between.
We will find it useful to repeat the single-qubit measurement of auxiliary $B$, as readout errors of that measurement would otherwise affect the 4-gon stabilizer readout of the two successive stabilizer measurements. 
This single-qubit measurement can be repeated without slowing down the circuit by incorporating it in steps 1 and 4.

Next, we consider applying these $M_{ZZZZ}$ and $M_{XXXX}$ circuits on the rectangular lattice with a checkerboard pattern in order to implement the surface code.
(For now, we focus on the bulk plaquettes and will return to the matter of boundaries and smaller $n$-gons in Sec.~\ref{sec:boundaries}.)
When designing this implementation, we must ensure that operations in neighboring plaquettes do not conflict with each other.
Specifically, the measurement sequences should avoid multiply-addressing any given qubit at any step.
(One should also avoid conflicting uses of measurement components in hardware layouts where this may occur; we will consider this for Majorana hardware in Sec.~\ref{sec:Majorana}.) 
Furthermore, the operation schedule must coordinate between adjacent plaquettes in a manner that correctly builds up the instantaneous stabilizer group to yield the desired $M_{ZZZZ}$ and $M_{XXXX}$ measurements needed for a surface code.

\begin{figure}[t!]
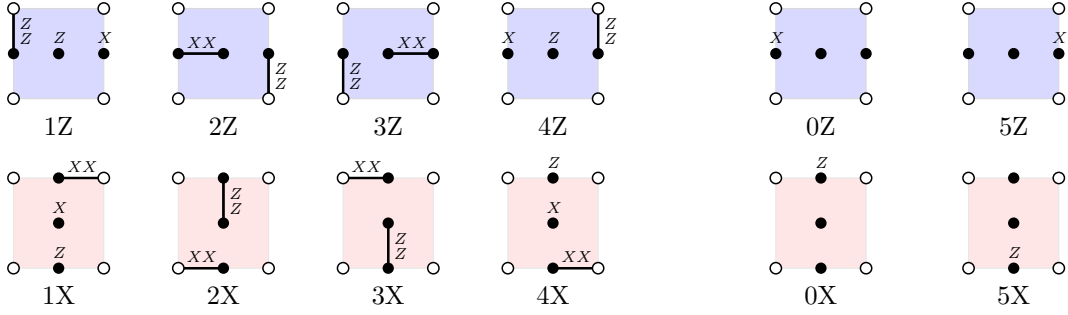

\centering

\input{ZZZZ}\hspace{1.5cm}\input{ZZZZ_0}\hspace{0.5cm}\input{ZZZZ_5}

\input{XXXX}\hspace{1.5cm}\input{XXXX_0}\hspace{0.5cm}\input{XXXX_5}

\caption{
The $M_{ZZZZ}$ and $M_{XXXX}$ measurement circuit steps for $Z$-type and $X$-type plaquettes.
Labels indicate the operators measured on the respective qubits in the given step, with connecting lines indicating pairwise measurement.
These circuits can be applied in parallel with a relative shift in their operation schedules.
A 4 step period for repeated application of these surface code stabilizer measurement circuits is obtained using the pipelining: $\cdots, (1Z,3X), (2Z,4X), (3Z,1X), (4Z,2X), \cdots$.
Steps 1-4 of a given circuit are applied repeatedly, whereas steps 0 and 5 are only used in the ramp up and down of the repetition of circuits, as indicated in Eqs.~\eqref{eq:compressed_pipeline}-\eqref{eq:pipeline_rampdown}.
}
\label{fig:Code_compressed}
\end{figure}

We can achieve the necessary properties by carefully pipelining the circuits with the data qubits addressed in the manner shown in Fig.~\ref{fig:Code_compressed}, with the steps pipelined as
\begin{equation}
\label{eq:compressed_pipeline}
\cdots, (1Z,3X), (2Z,4X), (3Z,1X), (4Z,2X), \cdots
.
\end{equation}
In this way, applying $r$ rounds of the surface code stabilizer measurements takes $4r+4$ steps, i.e. has period $4$, where we ramp up with the steps
\begin{equation}
(0Z, - ), (1Z, - ), (2Z,0X), (3Z, 1X), \cdots
\end{equation}
and ramp down with the steps
\begin{equation}
\label{eq:pipeline_rampdown}
\cdots, (4Z,2X), (5Z, 3X), (-,4X), (-,5X)
.
\end{equation}
Alternatively, we could interchange $X$ and $Z$ in the ramp up and down.
With the pipelining in Eq.~\eqref{eq:compressed_pipeline}, the value of the $ZZZZ$ stabilizer between time steps $(2Z,4X)$ and $(3Z,1X)$ is measured by the $M_{ZZZZ}$ circuit, and similarly the value of the $XXXX$ stabilizer between time steps $(4Z,2X)$ and $(1Z,3X)$ is measured by the $M_{XXXX}$ circuit.
The reason for specifying the time steps for the data qubit stabilizers is because, when considering the full code, the data qubit operators at different times will generally not be equivalent due to other plaquettes' stabilizer circuits addressing those data qubits.

We note that it is also possible to correctly implement the $M_{ZZZZ}$ and $M_{XXXX}$ circuits by interleaving them on a synchronous schedule by appropriately choosing the order in which the circuit addresses the data qubits, the details of which are given in Appendix~\ref{app:interleave}.
This turns out to have significant disadvantages compared to the pipelined measurement schedule presented above, so we do not focus on it in this paper.

\section{Hook Errors and their Prevention}
\label{sec:Hook}

The stabilizer measurement circuits of Figs.~\ref{fig:MZ4circuit} and \ref{fig:MX4circuit} exhibit hook errors, which are errors that stem from a single fault, but that are equivalent to two-qubit errors on the data qubits.\replace{circuit noise errors that are equivalent to two-qubit errors on the data qubits.}
In this paper, we distinguish between faults and errors as in Ref.~\onlinecite{Paetznick2023}: a fault is a failure of a circuit component (in our case, a measurement or an idling qubit) which results in a set of errors, and an error is represented by a single-qubit Pauli operator applied after the circuit component or a flipped measurement outcome, i.e. a readout error.
For example, a fault on a two-qubit measurement can result in a readout error and a Pauli error on either or both qubits involved.
Hook errors are concerning because they can reduce the fault distance and harm performance.
We use ``code distance'' $d$ to mean the minimal number of single data qubit errors that combine to produce a logical error, and ``fault distance'' $d_{\rm f}$ to mean the minimal number of faults such that the resulting errors combine to produce a logical error.\replace{circuit noise errors that combine to produce a logical error.}
The fault distance~\cite{bombinlogical2023} is the effective distance achieved by a particular realization of the code, which depends on the specific details of the circuits.
In some situations, the directionality of the hook errors allows them to be oriented perpendicular to the corresponding logical operators through a judicious choice of boundary conditions, thereby making their effect on the encoded logical state benign~\cite{Tomita2014}.
We now discuss how hook errors occur in the stabilizer measurement circuits presented in Sec.~\ref{sec:Circuits}.
We find that they are unidirectional and can be made benign, e.g. for a rotated surface code patch, using an appropriate choice of boundary conditions.  
Moreover, we discuss how to modify these circuits to prevent hook errors altogether.

\begin{figure}[t!]
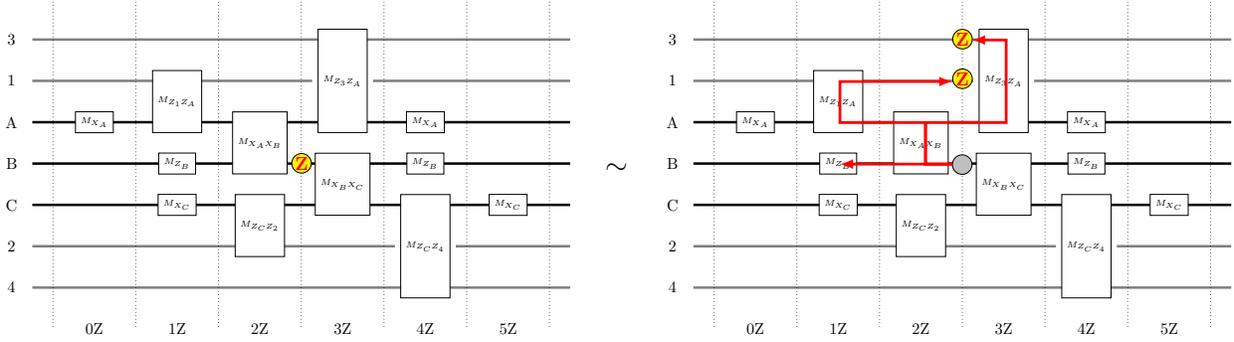

\centering
\vcenteredhbox{\scalebox{0.55}{
\input{ZZZZ_sparser}}}\vcenteredhbox{ \hspace{0.25cm}$\sim $\hspace{0.25cm}}\vcenteredhbox{
\scalebox{0.55}{
\input{ZZZZ_sparser}}}
\put(-359,0){\scalebox{0.65}{\begin{tikzpicture}
\draw[fill = yellow] (0,0) circle(0.2);
\node at (0,0) {\color{red} \bf Z};
\end{tikzpicture}}}
\put(-109,32){\scalebox{0.65}{\begin{tikzpicture}
\draw[fill = yellow] (0,0) circle(0.2);
\node at (0,0) {\color{red} \bf Z};
\end{tikzpicture}}}
\put(-109,47){\scalebox{0.65}{\begin{tikzpicture}
\draw[fill = yellow] (0,0) circle(0.2);
\node at (0,0) {\color{red} \bf Z};
\end{tikzpicture}}}
\put(-151,0.5){\scalebox{0.65}{\begin{tikzpicture}
\draw[fill = gray!50!white] (0,0) circle(0.2);
\draw[-latex, red, ultra thick] plot [smooth, tension=0] coordinates {(-0.2,0.01) (-0.75,0.01) (-0.75,0.86)  (0.9,0.86)  (0.9,2.55)  (0.2,2.55)};
\draw[-latex, red, ultra thick] plot [smooth, tension=0] coordinates {(-0.2,0.01) (-0.75,0.01) (-0.75,0.86) (-2.5,0.86)  (-2.5,1.7)  (-0.2,1.7)};
\draw[-latex, red, ultra thick] plot [smooth, tension=0] coordinates {(-0.2,0.01) (-2.5,0.01)};
\end{tikzpicture}}}
\caption{
An example of a hook error in the $M_{ZZZZ}$ measurement circuit is given by a $Z$ error on auxiliary qubit $B$ that occurs between steps 2 and 3, as shown in the circuit on the left.
This error is equivalent to a $ZZ$ error on data qubits 1 and 3, as shown in the circuit on the right. (Errors are marked in yellow. Red lines indicate the path through which the Pauli operator is ``pushed'' through the circuit elements using equivalences.) 
}
\label{fig:MZ4_hooks}
\end{figure}

For the stabilizer measurement circuits of Figs.~\ref{fig:MZ4circuit} and \ref{fig:MX4circuit}, readout errors and two-qubit errors stemming from faults on the pairwise measurements\replace{at the pairwise measurements} of auxiliary qubits in our circuits are equivalent to two-qubit errors on data qubits.
In more detail for the $M_{ZZZZ}$ circuit, a Pauli error $Z_B$ between steps 2 and 3 is equivalent to a $Z_1 Z_3$ or $Z_2 Z_4$ error on the data qubits, as shown in Fig.~\ref{fig:MZ4_hooks}.\footnote{
When the stabilizer circuits are pipelined as in Eq.~\eqref{eq:compressed_pipeline}, this hook error is equivalent to a data $ZZ$ error occurring at that same time interval (between steps 2 and 3).
This $ZZ$ error on data qubits is not necessarily equivalent to a $ZZ$ error at a different time step, since the neighboring plaquettes' $M_{XXXX}$ circuits also address these data qubits.
}
A readout error stemming from a fault on the\replace{A readout error at the} $M_{X_A X_B}$ or $M_{X_B X_C}$ measurement is also equivalent to a $Z_1 Z_3$ or $Z_2 Z_4$ error on the data qubits.
Similarly, for the $M_{XXXX}$ circuit, a Pauli error $X_B$ between steps 2 and 3 or a readout error at the $M_{Z_A Z_B}$ or $M_{Z_B Z_C}$ measurement is equivalent to a $X_1 X_3$ or $X_2 X_4$ error on the data qubits.
If we do not repeat the single-qubit measurements on auxiliary qubits $A$ and $C$, then readout errors stemming from faults on\replace{readout errors at} these non-repeated measurements would be equivalent to these same two-qubit errors on data qubits listed above for the respective circuit types.

We observe that the hook errors in our circuits for a given plaquette type are unidirectional, with the direction in our code realization correlated with the error type: the $Z$-plaquettes' hook errors correspond to $ZZ$ data qubit errors in the vertical direction and the $X$-plaquettes' hook errors correspond to $XX$ data qubit errors in the horizontal direction.
This is a useful property that, for example, allows us to choose boundary conditions for the ``rotated'' surface code on a planar patch such that the logical operators are aligned perpendicular to the corresponding type of hook errors; that is, the logical qubit's $Z$-logical string operators are horizontal and the $X$-logical string operators are vertical.
This choice prevents the hook errors from reducing the fault distance of the code (at least during logical idle).
We will return to this matter in Sec.~\ref{sec:boundaries}, after describing the circuits for boundary stabilizer (3-gon, 2-gon, and 1-gon) measurements.

\begin{figure}[t!]
\centering

\input{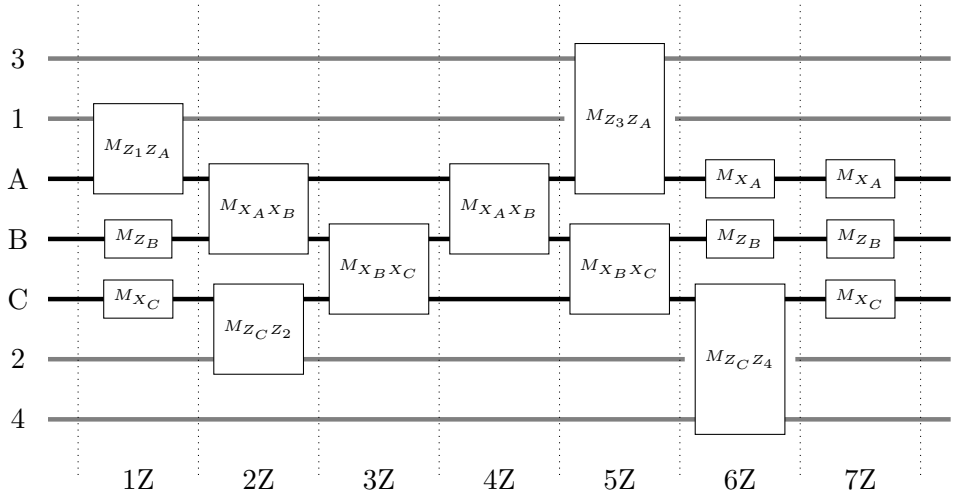}

\caption{
The modified circuit for $M_{ZZZZ}$ in which there are no hook errors. 
Steps 3, 4, and 7 shown here are the additional steps that have been inserted into the original circuit from Fig.~\ref{fig:MZ4circuit}.
(The $M_{Z_B}$ auxiliary qubit measurement only needs to occur twice per cycle, but we have shown it occurring three times.) 
Here, we only show the steps for one round of repeated application of the measurement circuit.
In order to ramp up the circuit, one needs to begin with a step 0 that applies a $M_{X_A}$ measurement before step 1; this could be achieved (with additional redundant measurements) by applying step 7.
}
\label{fig:hook_detect_circuit}
\end{figure}

\begin{figure}[t!]
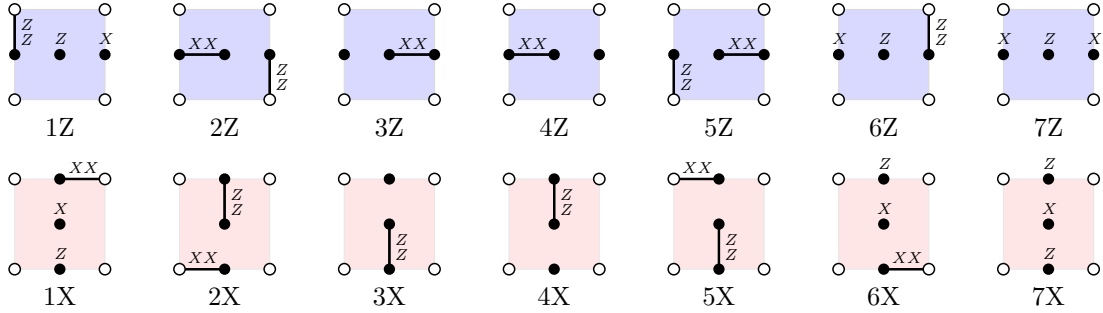

\centering

\input{ZZZZ_hook}

\input{XXXX_hook}

\caption{
The hook-preventing $M_{ZZZZ}$ and $M_{XXXX}$ measurement circuit steps for $Z$-type and $X$-type plaquettes.
These can be pipelined as in Eq.~\eqref{eq:hook_preventing_pipelines} to achieve a 7 step period.
}
\label{fig:hook_detect_pipeline}
\end{figure}

Another way we can address the hook errors is to modify our stabilizer measurement circuits so that they detect and distinguish the occurrence of these problematic faults\replace{physical errors}, and prevent the resulting errors\replace{them} from being equivalent to two data qubit errors. 
This strategy could be useful in scenarios where the logical qubit operators are not (or cannot be) aligned strictly perpendicular to the direction of their corresponding hook errors, such as when performing certain logical gate operations (see, e.g., Ref.~\onlinecite{bombinlogical2023}).
One way of modifying our circuits to prevent the hook errors is to repeat the measurements with which they are associated.
For the pairwise auxiliary qubit measurements, immediately repeating each measurement does not fix the problem, but repeating the pair of measurements (alternating between the $AB$ and $BC$ measurements) does. 
With these modifications, the resulting hook-preventing circuit for $M_{ZZZZ}$ is shown in Fig.~\ref{fig:hook_detect_circuit}. 
The repeated measurements add new ``detectors'' to the circuit, which allow us to distinguish the errors stemming from the problematic faults\replace{problematic physical errors} from the previously equivalent two data qubit errors.
(A detector is a set of measurements for which the product (or parity) of their outcomes is fixed in the absence of errors~\cite{Gidney2021b}; see Sec.~\ref{sec:Performance} for further discussion.)
The $M_{XXXX}$ circuit can be obtained from this circuit by replacing $X \leftrightarrow Z$.
Using these hook-preventing stabilizer measurement circuits, the surface code can be implemented with a 7-step period.
Using the same order of addressing data qubits as used for the unmodified circuits, as shown in Fig.~\ref{fig:hook_detect_pipeline}, we find several options for pipelining the circuits to achieve this periodicity, namely:
\begin{equation}
\label{eq:hook_preventing_pipelines}
\begin{array}{lll}
1: & & \cdots, (1Z,6X), (2Z,7X), (3Z,1X), (4Z,2X), (5Z,3X), (6Z,4X), (7Z,5X), \cdots \\
2: & & \cdots, (1Z,5X), (2Z,6X), (3Z,7X), (4Z,1X), (5Z,2X), (6Z,3X), (7Z,4X), \cdots \\
3: & & \cdots, (1Z,4X), (2Z,5X), (3Z,6X), (4Z,7X), (5Z,1X), (6Z,2X), (7Z,3X), \cdots \\
4: & & \cdots, (1Z,3X), (2Z,4X), (3Z,5X), (4Z,6X), (5Z,7X), (6Z,1X), (7Z,2X), \cdots 
\end{array}
\end{equation}

It it worth mentioning that this hook-preventing method of repeating the problematic measurements can be applied to other measurement-based code implementations, such as the pentagonal tiling realization of the surface code devised in Ref.~\onlinecite{Gidney2022a}.
We present details for this example in Appendix~\ref{app:pentagonal_hook_prevention}.

\section{Smaller \texorpdfstring{$n$}{n}-gons and Boundaries}
\label{sec:boundaries}

\begin{figure}[t!]
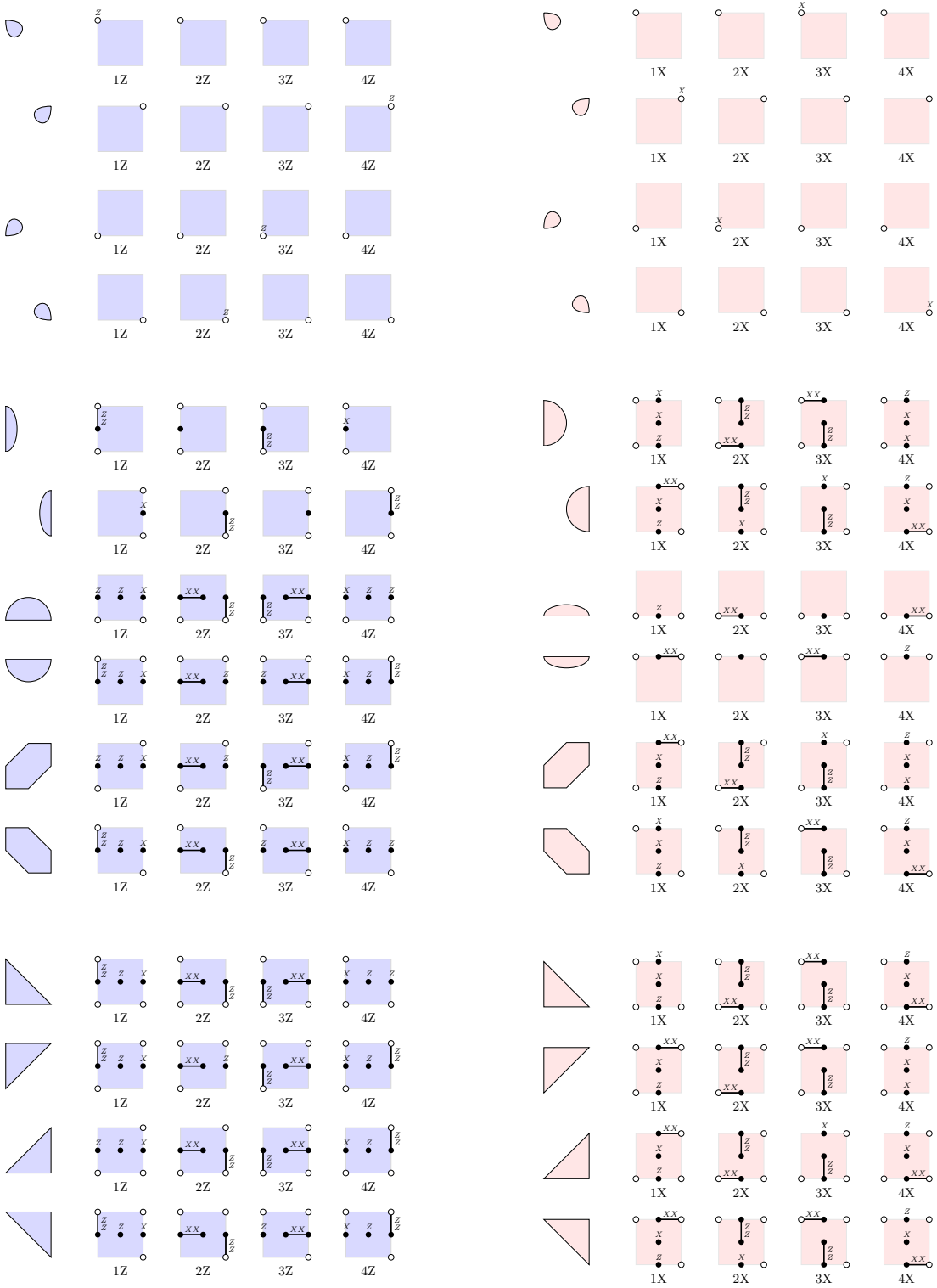

\centering 
\scalebox{0.985}{

\begin{minipage}{0.45\linewidth}
    
\scalebox{0.6}{
\begin{tikzpicture}
\newcommand{\scaling}{1.2}
   \draw [fill = white, opacity = 0] (-0.3*\scaling,0.3*\scaling) rectangle (1.3*\scaling,-1.3*\scaling);
\draw (0.5*\scaling, -1.3*\scaling) node[]{\phantom{\small X}};
\draw [fill = white, opacity = 0] (0,0) rectangle (1*\scaling,-1*\scaling);
\draw[scale=2,fill = blue, opacity = 0.15] (0,0)  to[in=0,out=-90,loop] (0,0);
\draw[scale=2] (0,0)  to[in=0,out=-90,loop] (0,0);
\end{tikzpicture}
\hspace{0.25cm}
\input{Z_NW}
}

\scalebox{0.6}{
\begin{tikzpicture}
\newcommand{\scaling}{1.2}
   \draw [fill = white, opacity = 0] (-0.3*\scaling,0.3*\scaling) rectangle (1.3*\scaling,-1.3*\scaling);
\draw (0.5*\scaling, -1.3*\scaling) node[]{\phantom{\small X}};
\draw [fill = white, opacity = 0] (0,0) rectangle (1*\scaling,-1*\scaling);
\draw[scale=2,fill = blue, opacity = 0.15] (0.5*\scaling,0)  to[in=-90,out=180,loop] (0.5*\scaling,0);
\draw[scale=2] (0.5*\scaling,0)  to[in=-90,out=180,loop] (0.5*\scaling,0);
\end{tikzpicture}
\hspace{0.25cm}
\input{Z_NE}
}

\scalebox{0.6}{
\begin{tikzpicture}
\newcommand{\scaling}{1.2}
   \draw [fill = white, opacity = 0] (-0.3*\scaling,0.3*\scaling) rectangle (1.3*\scaling,-1.3*\scaling);
\draw (0.5*\scaling, -1.3*\scaling) node[]{\phantom{\small X}};
\draw [fill = white, opacity = 0] (0,0) rectangle (1*\scaling,-1*\scaling);
\draw[scale=2,fill = blue, opacity = 0.15] (0,-0.5*\scaling)  to[in=90,out=0,loop] (0,-0.5*\scaling);
\draw[scale=2] ((0,-0.5*\scaling) to[in=90,out=0,loop] (0,-0.5*\scaling);
\end{tikzpicture}
\hspace{0.25cm}
\input{Z_SW}
}

\scalebox{0.6}{
\begin{tikzpicture}
\newcommand{\scaling}{1.2}
   \draw [fill = white, opacity = 0] (-0.3*\scaling,0.3*\scaling) rectangle (1.3*\scaling,-1.3*\scaling);
\draw (0.5*\scaling, -1.3*\scaling) node[]{\phantom{\small X}};
\draw [fill = white, opacity = 0] (0,0) rectangle (1*\scaling,-1*\scaling);
\draw[scale=2,fill = blue, opacity = 0.15] (0.5*\scaling,-0.5*\scaling)  to[in=90,out=180,loop] (0.5*\scaling,-0.5*\scaling);
\draw[scale=2] ((0.5*\scaling,-0.5*\scaling) to[in=90,out=180,loop] (0.5*\scaling,-0.5*\scaling);
\end{tikzpicture}
\hspace{0.25cm}
\input{Z_SE}
}

\vspace{0.75cm}

\scalebox{0.6}{
\begin{tikzpicture}
   \newcommand{\scaling}{1.2}
   \newcommand{\circlesize}{2}
   \draw [fill = white, opacity = 0] (-0.3*\scaling,0.3*\scaling) rectangle (1.3*\scaling,-1.3*\scaling);
\draw (0.5*\scaling, -1.3*\scaling) node[]{\phantom{\small X}};
\begin{scope}
    \clip (0,0) rectangle (1*\scaling,-1*\scaling);
    \draw[fill = blue, opacity = 0.15] (0,-0.5*\scaling) ellipse (0.25*\scaling cm and 0.5*\scaling cm);
    \draw[] (0,-0.5*\scaling) ellipse (0.25*\scaling cm and 0.5*\scaling cm);
    \draw (0,0) -- (0,-18\scaling);
\end{scope}
\end{tikzpicture}
\hspace{0.25cm}
\input{ZZ_right}
}

\scalebox{0.6}{
\begin{tikzpicture}
   \newcommand{\scaling}{1.2}
   \newcommand{\circlesize}{2}
   \draw [fill = white, opacity = 0] (-0.3*\scaling,0.3*\scaling) rectangle (1.3*\scaling,-1.3*\scaling);
\draw (0.5*\scaling, -1.3*\scaling) node[]{\phantom{\small X}};
\begin{scope}
    \clip (0,0) rectangle (1*\scaling,-1*\scaling);
    \draw[fill = blue, opacity = 0.15] (1*\scaling,-0.5*\scaling) ellipse (0.25*\scaling cm and 0.5*\scaling cm);
    \draw[] (1*\scaling,-0.5*\scaling) ellipse (0.25*\scaling cm and 0.5*\scaling cm);
    \draw (1*\scaling,0) -- (1*\scaling,-1*\scaling);
\end{scope}
\end{tikzpicture}
\hspace{0.25cm}
\input{ZZ_left}
}

\scalebox{0.6}{
\begin{tikzpicture}
   \newcommand{\scaling}{1.2}
   \newcommand{\circlesize}{2}
   \draw [fill = white, opacity = 0] (-0.3*\scaling,0.3*\scaling) rectangle (1.3*\scaling,-1.3*\scaling);
\draw (0.5*\scaling, -1.3*\scaling) node[]{\phantom{\small X}};
\begin{scope}
    \clip (0,0) rectangle (1*\scaling,-1*\scaling);
    \draw[fill = blue, opacity = 0.15] (0.5*\scaling,-1*\scaling) circle(0.5*\scaling);
    \draw[]  (0.5*\scaling,-1*\scaling) circle(0.5*\scaling);
    \draw (0,-1*\scaling) -- (1*\scaling,-1*\scaling);
\end{scope}
\end{tikzpicture}
\hspace{0.25cm}
\input{ZZ_top}
}

\scalebox{0.6}{
\begin{tikzpicture}
   \newcommand{\scaling}{1.2}
   \newcommand{\circlesize}{2}
   \draw [fill = white, opacity = 0] (-0.3*\scaling,0.3*\scaling) rectangle (1.3*\scaling,-1.3*\scaling);
\draw (0.5*\scaling, -1.3*\scaling) node[]{\phantom{\small X}};
\begin{scope}
    \clip (0,0) rectangle (1*\scaling,-1*\scaling);
    \draw[fill = blue, opacity = 0.15] (0.5*\scaling,0) circle(0.5*\scaling);
    \draw[] (0.5*\scaling,0) circle(0.5*\scaling);
    \draw (0,0) -- (1*\scaling,0);
\end{scope}
\end{tikzpicture}
\hspace{0.25cm}
\input{ZZ_bottom}
}

\scalebox{0.6}{
\begin{tikzpicture}
\newcommand{\scaling}{1.2}
\draw [fill = white, opacity = 0] (-0.3*\scaling,0.3*\scaling) rectangle (1.3*\scaling,-1.3*\scaling);
\draw (0.5*\scaling, -1.3*\scaling) node[]{\phantom{\small X}};
\draw [fill = blue, opacity = 0.15]  
(0.5*\scaling,0) -- (1*\scaling,0) -- (1*\scaling, -0.5*\scaling) 
  -- (0.5*\scaling,-1*\scaling) -- (0,-1*\scaling) -- (0, -0.5*\scaling) 
  -- cycle;
\draw []  
(0.5*\scaling,0) -- (1*\scaling,0) -- (1*\scaling, -0.5*\scaling) 
  -- (0.5*\scaling,-1*\scaling) -- (0,-1*\scaling) -- (0, -0.5*\scaling) 
  -- cycle;
\end{tikzpicture}
\hspace{0.25cm}
\input{ZZ_slash}
}

\scalebox{0.6}{
\begin{tikzpicture}
\newcommand{\scaling}{1.2}
\draw [fill = white, opacity = 0] (-0.3*\scaling,0.3*\scaling) rectangle (1.3*\scaling,-1.3*\scaling);
\draw (0.5*\scaling, -1.3*\scaling) node[]{\phantom{\small X}};
\draw [fill = blue, opacity = 0.15]  
(0.5*\scaling,0) -- (0*\scaling,0) -- (0*\scaling, -0.5*\scaling) 
  -- (0.5*\scaling,-1*\scaling) -- (1*\scaling,-1*\scaling) -- (1*\scaling, -0.5*\scaling) 
  -- cycle;
\draw []  
(0.5*\scaling,0) -- (0*\scaling,0) -- (0*\scaling, -0.5*\scaling) 
  -- (0.5*\scaling,-1*\scaling) -- (1*\scaling,-1*\scaling) -- (1*\scaling, -0.5*\scaling) 
  -- cycle;
\end{tikzpicture}
\hspace{0.25cm}
\input{ZZ_backslash}
}

\vspace{0.75cm}

\scalebox{0.6}{
\begin{tikzpicture}
\newcommand{\scaling}{1.2}
\draw [fill = white, opacity = 0] (-0.3*\scaling,0.3*\scaling) rectangle (1.3*\scaling,-1.3*\scaling);
\draw (0.5*\scaling, -1.3*\scaling) node[]{\phantom{\small X}};
\draw [fill = blue, opacity = 0.15]  (0,0) -- (0, -1*\scaling) 
  -- (1*\scaling,-1*\scaling) 
  -- cycle;
\draw []  (0,0) -- (0, -1*\scaling) 
  -- (1*\scaling,-1*\scaling) 
  -- cycle;
\end{tikzpicture}
\hspace{0.25cm}
\input{ZZZ_SW}
}

\scalebox{0.6}{
\begin{tikzpicture}
\newcommand{\scaling}{1.2}
\draw [fill = white, opacity = 0] (-0.3*\scaling,0.3*\scaling) rectangle (1.3*\scaling,-1.3*\scaling);
\draw (0.5*\scaling, -1.3*\scaling) node[]{\phantom{\small X}};
\draw [fill = blue, opacity = 0.15]  (0,0*\scaling) -- (1*\scaling, 0*\scaling) 
  -- (0*\scaling,-1*\scaling) 
  -- cycle;
\draw []   (0,0*\scaling) -- (1*\scaling, 0*\scaling) 
  -- (0*\scaling,-1*\scaling) 
  -- cycle;
\end{tikzpicture}
\hspace{0.25cm}
\input{ZZZ_NW}
}

\scalebox{0.6}{
\begin{tikzpicture}
\newcommand{\scaling}{1.2}
\draw [fill = white, opacity = 0] (-0.3*\scaling,0.3*\scaling) rectangle (1.3*\scaling,-1.3*\scaling);
\draw (0.5*\scaling, -1.3*\scaling) node[]{\phantom{\small X}};
\draw [fill = blue, opacity = 0.15]  (1*\scaling,0) -- (1*\scaling, -1*\scaling) 
  -- (0*\scaling,-1*\scaling) 
  -- cycle;
\draw []  (1*\scaling,0) -- (1*\scaling, -1*\scaling) 
  -- (0*\scaling,-1*\scaling) 
  -- cycle;
\end{tikzpicture}
\hspace{0.25cm}
\input{ZZZ_SE}
}

\scalebox{0.6}{
\begin{tikzpicture}
\newcommand{\scaling}{1.2}
\draw [fill = white, opacity = 0] (-0.3*\scaling,0.3*\scaling) rectangle (1.3*\scaling,-1.3*\scaling);
\draw (0.5*\scaling, -1.3*\scaling) node[]{\phantom{\small X}};
\draw [fill = blue, opacity = 0.15]  (0,0) -- (1*\scaling,0) 
  -- (1*\scaling,-1*\scaling) 
  -- cycle;
\draw []  (0,0) -- (1*\scaling,0) 
  -- (1*\scaling,-1*\scaling) 
  -- cycle;
\end{tikzpicture}
\hspace{0.25cm}
\input{ZZZ_NE}
}

\end{minipage}\hspace{1cm}
\begin{minipage}{0.45\linewidth}

\scalebox{0.6}{
\begin{tikzpicture}
\newcommand{\scaling}{1.2}
   \draw [fill = white, opacity = 0] (-0.3*\scaling,0.3*\scaling) rectangle (1.3*\scaling,-1.3*\scaling);
\draw (0.5*\scaling, -1.3*\scaling) node[]{\phantom{\small X}};
\draw [fill = white, opacity = 0] (0,0) rectangle (1*\scaling,-1*\scaling);
\draw[scale=2,fill = red, opacity = 0.1] (0,0)  to[in=0,out=-90,loop] (0,0);
\draw[scale=2] (0,0)  to[in=0,out=-90,loop] (0,0);
\end{tikzpicture}
\hspace{0.25cm}
\input{X_NW}
}

\scalebox{0.6}{
\begin{tikzpicture}
\newcommand{\scaling}{1.2}
   \draw [fill = white, opacity = 0] (-0.3*\scaling,0.3*\scaling) rectangle (1.3*\scaling,-1.3*\scaling);
\draw (0.5*\scaling, -1.3*\scaling) node[]{\phantom{\small X}};
\draw [fill = white, opacity = 0] (0,0) rectangle (1*\scaling,-1*\scaling);
\draw[scale=2,fill = red, opacity = 0.1] (0.5*\scaling,0)  to[in=-90,out=180,loop] (0.5*\scaling,0);
\draw[scale=2] (0.5*\scaling,0)  to[in=-90,out=180,loop] (0.5*\scaling,0);
\end{tikzpicture}
\hspace{0.25cm}
\input{X_NE}
}

\scalebox{0.6}{
\begin{tikzpicture}
\newcommand{\scaling}{1.2}
   \draw [fill = white, opacity = 0] (-0.3*\scaling,0.3*\scaling) rectangle (1.3*\scaling,-1.3*\scaling);
\draw (0.5*\scaling, -1.3*\scaling) node[]{\phantom{\small X}};
\draw [fill = white, opacity = 0] (0,0) rectangle (1*\scaling,-1*\scaling);
\draw[scale=2,fill = red, opacity = 0.1] (0,-0.5*\scaling)  to[in=90,out=0,loop] (0,-0.5*\scaling);
\draw[scale=2] ((0,-0.5*\scaling) to[in=90,out=0,loop] (0,-0.5*\scaling);
\end{tikzpicture}
\hspace{0.25cm}
\input{X_SW}
}

\scalebox{0.6}{
\begin{tikzpicture}
\newcommand{\scaling}{1.2}
   \draw [fill = white, opacity = 0] (-0.3*\scaling,0.3*\scaling) rectangle (1.3*\scaling,-1.3*\scaling);
\draw (0.5*\scaling, -1.3*\scaling) node[]{\phantom{\small X}};
\draw [fill = white, opacity = 0] (0,0) rectangle (1*\scaling,-1*\scaling);
\draw[scale=2,fill = red, opacity = 0.1] (0.5*\scaling,-0.5*\scaling)  to[in=90,out=180,loop] (0.5*\scaling,-0.5*\scaling);
\draw[scale=2] ((0.5*\scaling,-0.5*\scaling) to[in=90,out=180,loop] (0.5*\scaling,-0.5*\scaling);
\end{tikzpicture}
\hspace{0.25cm}
\input{X_SE}
}

\vspace{0.75cm}

\scalebox{0.6}{
\begin{tikzpicture}
   \newcommand{\scaling}{1.2}
   \newcommand{\circlesize}{2}
   \draw [fill = white, opacity = 0] (-0.3*\scaling,0.3*\scaling) rectangle (1.3*\scaling,-1.3*\scaling);
\draw (0.5*\scaling, -1.3*\scaling) node[]{\phantom{\small X}};
\begin{scope}
    \clip (0,0) rectangle (1*\scaling,-1*\scaling);
    \draw[fill = red, opacity = 0.1] (0,-0.5*\scaling) circle(0.5*\scaling);
    \draw[] (0,-0.5*\scaling) circle(0.5*\scaling);
    \draw (0,0) -- (0,-1*\scaling);
\end{scope}
\end{tikzpicture}
\hspace{0.25cm}
\input{XX_right}
}

\scalebox{0.6}{
\begin{tikzpicture}
   \newcommand{\scaling}{1.2}
   \newcommand{\circlesize}{2}
   \draw [fill = white, opacity = 0] (-0.3*\scaling,0.3*\scaling) rectangle (1.3*\scaling,-1.3*\scaling);
    \draw (0.5*\scaling, -1.3*\scaling) node[]{\phantom{\small X}};
\begin{scope}
    \clip (0,0) rectangle (1*\scaling,-1*\scaling);
    \draw[fill = red, opacity = 0.1] (1*\scaling,-0.5*\scaling) circle(0.5*\scaling);
    \draw[] (1*\scaling,-0.5*\scaling) circle(0.5*\scaling);
    \draw (1*\scaling,0) -- (1*\scaling,-1*\scaling);
\end{scope}
\end{tikzpicture}
\hspace{0.25cm}
\input{XX_left}
}

\scalebox{0.6}{
\begin{tikzpicture}
   \newcommand{\scaling}{1.2}
   \newcommand{\circlesize}{2}
   \draw [fill = white, opacity = 0] (-0.3*\scaling,0.3*\scaling) rectangle (1.3*\scaling,-1.3*\scaling);
   \draw (0.5*\scaling, -1.3*\scaling) node[]{\phantom{\small X}};
\begin{scope}
    \clip (0,0) rectangle (1*\scaling,-1*\scaling);
    \draw[fill = red, opacity = 0.1]  (0.5*\scaling, -1*\scaling)  ellipse (0.5*\scaling cm and 0.25*\scaling cm);
    \draw[] (0.5*\scaling, -1*\scaling) ellipse (0.5*\scaling cm and 0.25*\scaling cm);
    \draw (0,-1*\scaling) -- (1*\scaling,-1*\scaling);
\end{scope}
\end{tikzpicture}
\hspace{0.25cm}
\input{XX_top}
}

\scalebox{0.6}{
\begin{tikzpicture}
   \newcommand{\scaling}{1.2}
   \newcommand{\circlesize}{2}
   \draw [fill = white, opacity = 0] (-0.3*\scaling,0.3*\scaling) rectangle (1.3*\scaling,-1.3*\scaling);
    \draw (0.5*\scaling, -1.3*\scaling) node[]{\phantom{\small X}};
\begin{scope}
    \clip (0,0) rectangle (1*\scaling,-1*\scaling);
    \draw[fill = red, opacity = 0.1] (0.5*\scaling,0) ellipse (0.5*\scaling cm and 0.25*\scaling cm);
    \draw[] (0.5*\scaling,0) ellipse (0.5*\scaling cm and 0.25*\scaling cm);
    \draw (0,0) -- (1*\scaling,0);
\end{scope}
\end{tikzpicture}
\hspace{0.25cm}
\input{XX_bottom}
}

\scalebox{0.6}{
\begin{tikzpicture}
\newcommand{\scaling}{1.2}
\draw [fill = white, opacity = 0] (-0.3*\scaling,0.3*\scaling) rectangle (1.3*\scaling,-1.3*\scaling);
\draw (0.5*\scaling, -1.3*\scaling) node[]{\phantom{\small X}};
\draw [fill = red, opacity = 0.1]  
(0.5*\scaling,0) -- (1*\scaling,0) -- (1*\scaling, -0.5*\scaling) 
  -- (0.5*\scaling,-1*\scaling) -- (0,-1*\scaling) -- (0, -0.5*\scaling) 
  -- cycle;
\draw []  
(0.5*\scaling,0) -- (1*\scaling,0) -- (1*\scaling, -0.5*\scaling) 
  -- (0.5*\scaling,-1*\scaling) -- (0,-1*\scaling) -- (0, -0.5*\scaling) 
  -- cycle;
\end{tikzpicture}
\hspace{0.25cm}
\input{XX_slash}
}

\scalebox{0.6}{
\begin{tikzpicture}
\newcommand{\scaling}{1.2}
\draw [fill = white, opacity = 0] (-0.3*\scaling,0.3*\scaling) rectangle (1.3*\scaling,-1.3*\scaling);
\draw (0.5*\scaling, -1.3*\scaling) node[]{\phantom{\small X}};
\draw [fill = red, opacity = 0.1]  
(0.5*\scaling,0) -- (0*\scaling,0) -- (0*\scaling, -0.5*\scaling) 
  -- (0.5*\scaling,-1*\scaling) -- (1*\scaling,-1*\scaling) -- (1*\scaling, -0.5*\scaling) 
  -- cycle;
\draw []  
(0.5*\scaling,0) -- (0*\scaling,0) -- (0*\scaling, -0.5*\scaling) 
  -- (0.5*\scaling,-1*\scaling) -- (1*\scaling,-1*\scaling) -- (1*\scaling, -0.5*\scaling) 
  -- cycle;
\end{tikzpicture}
\hspace{0.25cm}
\input{XX_backslash}
}

\vspace{0.75cm}

\scalebox{0.6}{
\begin{tikzpicture}
\newcommand{\scaling}{1.2}
\draw [fill = white, opacity = 0] (-0.3*\scaling,0.3*\scaling) rectangle (1.3*\scaling,-1.3*\scaling);
\draw (0.5*\scaling, -1.3*\scaling) node[]{\phantom{\small X}};
\draw [fill = red, opacity = 0.1]  (0,0) -- (0, -1*\scaling) 
  -- (1*\scaling,-1*\scaling) 
  -- cycle;
\draw []  (0,0) -- (0, -1*\scaling) 
  -- (1*\scaling,-1*\scaling) 
  -- cycle;
\end{tikzpicture}
\hspace{0.25cm}
\input{XXX_SW}
}

\scalebox{0.6}{
\begin{tikzpicture}
\newcommand{\scaling}{1.2}
\draw [fill = white, opacity = 0] (-0.3*\scaling,0.3*\scaling) rectangle (1.3*\scaling,-1.3*\scaling);
\draw (0.5*\scaling, -1.3*\scaling) node[]{\phantom{\small X}};
\draw [fill = red, opacity = 0.1]  (0,0*\scaling) -- (1*\scaling, 0*\scaling) 
  -- (0*\scaling,-1*\scaling) 
  -- cycle;
\draw []   (0,0*\scaling) -- (1*\scaling, 0*\scaling) 
  -- (0*\scaling,-1*\scaling) 
  -- cycle;
\end{tikzpicture}
\hspace{0.25cm}
\input{XXX_NW}
}

\scalebox{0.6}{
\begin{tikzpicture}
\newcommand{\scaling}{1.2}
\draw [fill = white, opacity = 0] (-0.3*\scaling,0.3*\scaling) rectangle (1.3*\scaling,-1.3*\scaling);
\draw (0.5*\scaling, -1.3*\scaling) node[]{\phantom{\small X}};
\draw [fill = red, opacity = 0.1]  (1*\scaling,0) -- (1*\scaling, -1*\scaling) 
  -- (0*\scaling,-1*\scaling) 
  -- cycle;
\draw []  (1*\scaling,0) -- (1*\scaling, -1*\scaling) 
  -- (0*\scaling,-1*\scaling) 
  -- cycle;
\end{tikzpicture}
\hspace{0.25cm}
\input{XXX_SE}
}

\scalebox{0.6}{
\begin{tikzpicture}
\newcommand{\scaling}{1.2}
\draw [fill = white, opacity = 0] (-0.3*\scaling,0.3*\scaling) rectangle (1.3*\scaling,-1.3*\scaling);
\draw (0.5*\scaling, -1.3*\scaling) node[]{\phantom{\small X}};
\draw [fill = red, opacity = 0.1]  (0,0) -- (1*\scaling,0) 
  -- (1*\scaling,-1*\scaling) 
  -- cycle;
\draw []  (0,0) -- (1*\scaling,0) 
  -- (1*\scaling,-1*\scaling) 
  -- cycle;
\end{tikzpicture}
\hspace{0.25cm}
\input{XXX_NE}
}

\end{minipage}

}

\caption{
$Z$-type (left) and $X$-type (right) measurement circuits for 1-gons, 2-gons, and 3-gons.
}
\label{fig:ngons}
\end{figure}

In order to operate the code on surfaces with a boundary, we need circuits for measuring the multi-qubit Pauli operators of 1, 2, or 3 data qubits, which we call 1-gons, 2-gons, or 3-gons, respectively.
A convenient way to produce stabilizer measurement circuits for these boundary stabilizers is to start with the corresponding 4-gon measurement circuits and remove the appropriate data qubits.
In doing so, we may reduce or remove certain measurements from the circuit, where reduction changes a two-qubit measurement involving a removed data qubit into a single-qubit measurement of the same type on the corresponding auxiliary qubit, e.g. $M_{Z_A Z_1} \rightarrow M_{Z_A}$.
We may also remove auxiliary qubits from the $n$-gon when removing data qubits in this way, depending on which type of $n$-gon we are producing.
In particular, 2-gons with data qubits addressed by a common auxiliary qubit only require that auxiliary qubit (the other two can be removed), and 1-gons require no auxiliary qubits.
The resulting $n$-gon measurement circuits are shown in Fig.~\ref{fig:ngons}.
(The ramp up and down steps 0 and 5 are not shown; we can simply use those of the 4-gon circuits when the step is required.)
A convenient property of producing $n$-gon measurement circuits in this manner is that they interlock with the bulk 4-gon circuits.
In other words, for a system that uses these $n$-gon measurement circuits, we can operate all $Z$-type $n$-gons on the same schedule and all $X$-type $n$-gons on the same schedule, without further modification, e.g. we can use the pipelining in Eq.~\eqref{eq:compressed_pipeline} of $Z$- and $X$-plaquette circuits for all the $n$-gons.

\begin{figure}[t!]
\centering
\scalebox{0.65}{
\begin{tikzpicture}
\newcommand{\halfscaling}{0.6}
\newcommand{\nodesize}{1.5}
\draw [fill=white,opacity=0] (-1.5*\halfscaling,-1.5*\halfscaling) rectangle (5.5*\halfscaling,5.5*\halfscaling); 
\draw[fill=blue, opacity = 0.15] (4*\halfscaling,3*\halfscaling) circle(1*\halfscaling);
\draw[fill=red, opacity=0.1] (1*\halfscaling,4*\halfscaling) circle(1*\halfscaling) ;
\draw[fill=red,opacity=0.1] (3*\halfscaling,0) circle(1*\halfscaling) ;
\draw[fill=blue,opacity=0.15] (0,1*\halfscaling) circle(1*\halfscaling) ;
\draw [fill=white,white] (0,0) rectangle (4*\halfscaling,4*\halfscaling);
\draw [opacity=0.1] (0,0) rectangle (4*\halfscaling,4*\halfscaling);
\draw [fill=red,opacity=0.1] (0,0) rectangle (2*\halfscaling,2*\halfscaling);
\draw [fill=blue, opacity = 0.15] (2*\halfscaling,0) rectangle (4*\halfscaling,2*\halfscaling);
\draw [fill=red,opacity=0.1] (2*\halfscaling,2*\halfscaling) rectangle (4*\halfscaling,4*\halfscaling);
\draw [fill=blue, opacity = 0.15] (0*\halfscaling,2*\halfscaling) rectangle (2*\halfscaling,4*\halfscaling);
\foreach \i in {0,...,4}{\foreach \j in {0,...,4}{
\draw (\i*\halfscaling, \j*\halfscaling) node[circle, draw, fill=black, minimum size = \nodesize mm, inner sep = 0]{};};};
\foreach \i in {0,2,4}{\foreach \j in {0,2,4}{
\draw (\i*\halfscaling, \j*\halfscaling) node[circle, draw, fill=white, minimum size = \nodesize mm, inner sep = 0]{};};};
\node at (2*\halfscaling,-2*\halfscaling) {\large (a)};
\end{tikzpicture}}
\hspace{1cm}
%
%
%
\scalebox{0.65}{
\begin{tikzpicture}
\newcommand{\halfscaling}{0.6}
\newcommand{\nodesize}{1.5}
\draw [fill=white,opacity=0] (-1.5*\halfscaling,-1.5*\halfscaling) rectangle (5.5*\halfscaling,5.5*\halfscaling); 
\draw[fill=red, opacity = 0.1] (8*\halfscaling,1*\halfscaling) circle(1*\halfscaling);
\draw[fill=red, opacity = 0.1] (8*\halfscaling,5*\halfscaling) circle(1*\halfscaling);
\draw[fill=blue, opacity=0.15] (3*\halfscaling,8*\halfscaling) circle(1*\halfscaling) ;
\draw[fill=blue, opacity=0.15] (7*\halfscaling,8*\halfscaling) circle(1*\halfscaling) ;
\draw[fill=blue,opacity=0.15] (1*\halfscaling,0) circle(1*\halfscaling) ;
\draw[fill=blue,opacity=0.15] (5*\halfscaling,0) circle(1*\halfscaling) ;
\draw[fill=red,opacity=0.1] (0,3*\halfscaling) circle(1*\halfscaling) ;
\draw[fill=red,opacity=0.1] (0,7*\halfscaling) circle(1*\halfscaling) ;
\draw [fill=white, white] (0,0) rectangle (8*\halfscaling,8*\halfscaling);
\foreach \i in {0,4}{\foreach \j in {0,4}{
\begin{scope}[shift={(\i*\halfscaling,\j*\halfscaling)}]
\draw [opacity=0.1] (0,0) rectangle (4*\halfscaling,4*\halfscaling);
\draw [fill=red,opacity=0.1] (0,0) rectangle (2*\halfscaling,2*\halfscaling);
\draw [fill=blue, opacity = 0.15] (2*\halfscaling,0) rectangle (4*\halfscaling,2*\halfscaling);
\draw [fill=red,opacity=0.1] (2*\halfscaling,2*\halfscaling) rectangle (4*\halfscaling,4*\halfscaling);
\draw [fill=blue, opacity = 0.15] (0*\halfscaling,2*\halfscaling) rectangle (2*\halfscaling,4*\halfscaling);
\end{scope}
};};
\foreach \i in {1,...,7}{\foreach \j in {1,...,7}{
\draw (\i*\halfscaling, \j*\halfscaling) node[circle, draw, fill=black, minimum size = \nodesize mm, inner sep = 0]{};};};
\foreach \i in {3,7}{
\draw (0*\halfscaling, \i*\halfscaling) node[circle, draw, fill=black, minimum size = \nodesize mm, inner sep = 0]{};
\draw (\i*\halfscaling, 8*\halfscaling) node[circle, draw, fill=black, minimum size = \nodesize mm, inner sep = 0]{};
};
\foreach \i in {1,5}{
\draw (\i*\halfscaling, 0*\halfscaling) node[circle, draw, fill=black, minimum size = \nodesize mm, inner sep = 0]{};
\draw (8*\halfscaling, \i*\halfscaling) node[circle, draw, fill=black, minimum size = \nodesize mm, inner sep = 0]{};
};
\foreach \i in {0,2,4,6,8}{\foreach \j in {0,2,4,6,8}{
\draw (\i*\halfscaling, \j*\halfscaling) node[circle, draw, fill=white, minimum size = \nodesize mm, inner sep = 0]{};};};
\foreach \j in {2,3,4,  6,7,8}{
\draw ( -1*\halfscaling, \j*\halfscaling) node[circle, draw, fill=black, minimum size = \nodesize mm, inner sep = 0]{};
\draw ( \j*\halfscaling, 9*\halfscaling) node[circle, draw, fill=black, minimum size = \nodesize mm, inner sep = 0]{};
};
\foreach \j in {0,1,2,  4,5,6}{
\draw ( 9*\halfscaling, \j*\halfscaling) node[circle, draw, fill=black, minimum size = \nodesize mm, inner sep = 0]{};
\draw ( \j*\halfscaling, -1*\halfscaling) node[circle, draw, fill=black, minimum size = \nodesize mm, inner sep = 0]{};
};
\node at (4*\halfscaling,-2*\halfscaling) {\large (b)};
\end{tikzpicture}}
\hspace{1cm}
\scalebox{0.65}{
\begin{tikzpicture}
\newcommand{\halfscaling}{0.6}
\newcommand{\nodesize}{1.5}
\draw [fill=white,opacity=0] (-2.5*\halfscaling,-2.5*\halfscaling) rectangle (6.5*\halfscaling,6.5*\halfscaling); 
\draw [fill=red,opacity=0.1] (0,0) rectangle (2*\halfscaling,2*\halfscaling);
\draw [fill = blue, opacity = 0.15]  (2*\halfscaling,6*\halfscaling) -- (2*\halfscaling, 2*\halfscaling) 
  -- (6*\halfscaling,2*\halfscaling) 
  -- cycle;
\draw [fill = blue, opacity = 0.15]  (2*\halfscaling,-2*\halfscaling) -- (2*\halfscaling, 2*\halfscaling) 
  -- (-2*\halfscaling,2*\halfscaling) 
  -- cycle;
\draw [fill = red, opacity = 0.1]  (2*\halfscaling,6*\halfscaling) -- (2*\halfscaling, 2*\halfscaling) 
  -- (-2*\halfscaling,2*\halfscaling) 
  -- cycle;
\draw [fill = red, opacity = 0.1]  (2*\halfscaling,-2*\halfscaling) -- (2*\halfscaling, 2*\halfscaling) 
  -- (6*\halfscaling,2*\halfscaling) 
  -- cycle;
\draw [fill=white,white] (0,0) rectangle (4*\halfscaling,4*\halfscaling);
\draw [opacity=0.1] (0,0) rectangle (4*\halfscaling,4*\halfscaling);
\draw [fill=red,opacity=0.1] (0,0) rectangle (2*\halfscaling,2*\halfscaling);
\draw [fill=blue,opacity=0.15] (2*\halfscaling,0) rectangle (4*\halfscaling,2*\halfscaling);
\draw [fill=red,opacity=0.1] (2*\halfscaling,2*\halfscaling) rectangle (4*\halfscaling,4*\halfscaling);
\draw [fill=blue,opacity=0.15] (0*\halfscaling,2*\halfscaling) rectangle (2*\halfscaling,4*\halfscaling);
\foreach \i in {0,...,4}{\foreach \j in {0,...,4}{
\draw (\i*\halfscaling, \j*\halfscaling) node[circle, draw, fill=black, minimum size = \nodesize mm, inner sep = 0]{};};};
\foreach \i in {0,2,4}{\foreach \j in {0,2,4}{
\draw (\i*\halfscaling, \j*\halfscaling) node[circle, draw, fill=white, minimum size = \nodesize mm, inner sep = 0]{};};};
\foreach \j in {2,3,4}{
\draw ( -1*\halfscaling, \j*\halfscaling) node[circle, draw, fill=black, minimum size = \nodesize mm, inner sep = 0]{};
\draw ( \j*\halfscaling, 5*\halfscaling) node[circle, draw, fill=black, minimum size = \nodesize mm, inner sep = 0]{};
};
\foreach \j in {0,1,2}{
\draw ( 5*\halfscaling, \j*\halfscaling) node[circle, draw, fill=black, minimum size = \nodesize mm, inner sep = 0]{};
\draw ( \j*\halfscaling, -1*\halfscaling) node[circle, draw, fill=black, minimum size = \nodesize mm, inner sep = 0]{};
};
\foreach \j in {4,5,6}{
\draw ( 1*\halfscaling, \j*\halfscaling) node[circle, draw, fill=black, minimum size = \nodesize mm, inner sep = 0]{};
\draw ( \j*\halfscaling, 3*\halfscaling) node[circle, draw, fill=black, minimum size = \nodesize mm, inner sep = 0]{};
};
 \foreach \j in {-2,-1,0}{
 \draw ( 3*\halfscaling, \j*\halfscaling) node[circle, draw, fill=black, minimum size = \nodesize mm, inner sep = 0]{};
 \draw ( \j*\halfscaling, 1*\halfscaling) node[circle, draw, fill=black, minimum size = \nodesize mm, inner sep = 0]{};
 };
 \draw (-2*\halfscaling, 2*\halfscaling) node[circle, draw, fill=white, minimum size = \nodesize mm, inner sep = 0]{};
 \draw (6*\halfscaling, 2*\halfscaling) node[circle, draw, fill=white, minimum size = \nodesize mm, inner sep = 0]{};
 \draw (2*\halfscaling, 6*\halfscaling) node[circle, draw, fill=white, minimum size = \nodesize mm, inner sep = 0]{};
  \draw (2*\halfscaling, -2*\halfscaling) node[circle, draw, fill=white, minimum size = \nodesize mm, inner sep = 0]{};
\node at (2*\halfscaling,-3*\halfscaling) {\large (c)};
\end{tikzpicture}}
\caption{
Different boundary conditions for a square patch of surface code, all shown for fault distance $d_{\rm f}=3$ when using the period 4 circuits of Figs.~\ref{fig:Code_compressed} and \ref{fig:ngons}.
(a) The rotated surface code with a good choice of boundary conditions aligns the logical strings perpendicular to the corresponding hook errors of the same type, so $d_{\rm f}=d$.
The relation between number of physical qubits and  distance in this case is $N = 4d^2 -4d +1$.
(b) The rotated surface code with a bad choice of boundary conditions aligns the logical strings parallel to the corresponding hook errors of the same type, which halves the fault distance, so $d_{\rm f}= \lceil d/2 \rceil$.
The resulting relation between number of physical qubits and the distance in this case is $N = 4 d^2 - 3$, giving the scaling with fault distance of $N=O(16 d_{\rm f}^2$).
(Using the period 7 hook-preventing circuits of Fig.~\ref{fig:hook_detect_pipeline} would avoid halving the distance.)
(c) The original (unrotated) surface code has the logical strings aligned diagonal to the direction of the hook errors, so $d_{\rm f}=d$.
The relation between number of physical qubits and distance in this case is $N = 8d^2 -8d +1$.}
\label{fig:patches}
\end{figure}
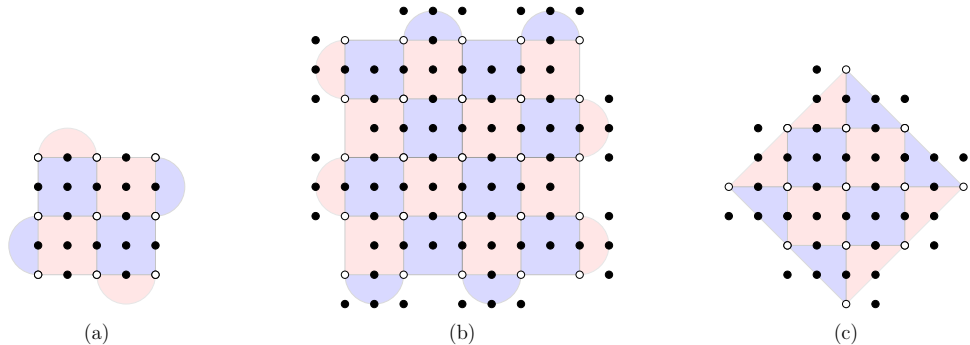

With these $n$-gon measurement circuits, it is straightforward to define the code on surfaces with boundaries.
However, an issue that arises is that certain choices of boundary conditions may be more advantageous than others.
For example, when putting the surface code on a square patch of data qubits with the ``rotated surface code'' boundary conditions~\cite{Bombin2007}, there are two options for how to assign boundary types to the edges of the square patch.
The good choice for our code realization, shown in Fig.~\ref{fig:patches}(a), corresponding to using $Z$-type 2-gons along vertical edges and $X$-type 2-gons along horizontal edges.
This choice aligns the $Z$ and $X$ logical string operators perpendicular to the direction of the $Z$- and $X$-plaquettes' hook errors, respectively.
The bad choice for our surface code realization, shown in Fig.~\ref{fig:patches}(b), corresponding to using $Z$-type 2-gons along horizontal edges and $X$-type 2-gons along vertical edges.
This choice utilizes more auxiliary qubits at the boundary and, more importantly, aligns the logical string operators with the corresponding hook errors, which halves the fault distance of the code.
In contrast, the original (unrotated) surface code boundary conditions~\cite{Bravyi1998} is implemented using 3-gons along the boundary edges, with $Z$-type and $X$-type 3-gons corresponding to ``rough'' and ``smooth'' boundaries.
In this case, the logical string operators are aligned diagonally across the plaquettes, so the hook errors do not affect the code distance.
However, the relation between distance and number of qubits for the original surface code is worse than that of the (good) rotated surface code due to this diagonal alignment of logical string operators with respect to the plaquettes.

\section{Fault Tolerance in the Presence of Dead Components}
\label{sec:DeadComponents}

An important problem to address when implementing error-correcting codes in physical hardware is maintaining fault tolerance in the presence of physical components that are nonfunctional or exhibit substantially higher error rates than most of the components.
We can map the effects of faulty physical components to an effective computational model, where they are specified in terms of qubits, computational gates, and measurements.
For our purposes, we will refer to qubits, computational gates, and measurements that are nonfunctional or exhibiting atypically higher error rates as ``dead components.''
The dead components can potentially be identified during the bring-up and calibration phase of operating the hardware (if dead at the time) or also during the operation of the error-correcting code when error syndromes indicate a qubit or operation is exhibiting a high error rate.

Ref.~\onlinecite{Stace2009} introduced a strategy for dealing with dead data qubits by removing them from the code operation and forming ``superplaquette'' operators, which are products of the original plaquette operators (of the same type) that exclude the dead qubits.\footnote{We refer to all the stabilizers of the surface code as plaquette operators, distinguishing them as $Z$-type and $X$-type, rather than ``plaquette'' and ``star'' operators, respectively.}
Building on this idea, Ref.~\onlinecite{Auger2017} proposed to generate the superplaquette measurements by measuring the ``damaged stabilizers,'' i.e. the original plaquette operators reduced by removal of the dead data qubits, in a manner such that they combine to yield all the superplaquette stabilizers.
In general, a damaged stabilizer will not commute with all other stabilizers, so they cannot all be simultaneously measured.
In light of this, Ref.~\onlinecite{Auger2017} proposed to deal with dead components by successively measuring the damaged stabilizers of $Z$-type and $X$-type in alternating rounds of stabilizer measurements, while continuing to measure both types of undamaged stabilizers every round.
In this way, damaged stabilizers would be measured half as often as the undamaged stabilizers.
Using this alternation between measuring $Z$-type and $X$-type damaged stabilizers, the instantaneous stabilizer group after a given round includes the damaged stabilizers of the type from that round, together with the superplaquette stabilizers of the other type formed from the previous round's damaged stabilizers (but not the previous round's individual damaged stabilizers).

As a concrete realization, Ref.~\onlinecite{Auger2017} considered the CNOT-based implementation of plaquette stabilizer measurements, using an auxiliary qubit for each plaquette and CNOT gates connecting auxiliary to data qubits, with a suitable interleaving of the circuits of neighboring plaquettes.
For a dead data qubit, the CNOT gates involving that qubit are simply removed from the respective stabilizer measurement circuits.
For these circuits on damaged plaquettes, Ref.~\onlinecite{Auger2017} stated that the normal circuit schedule could not be used because it would randomize the superplaquette values due to the anti-commutation of damaged stabilizers, and they consequently required measurement circuits for $Z$-type and $X$-type damaged stabilizers to be applied in separate alternating rounds.
This claim is incorrect; the normal schedule for the interleaved CNOT-based implementation can, in fact, be utilized for the damaged stabilizers.~\footnote{We thank Reviewer 1 for bringing this to our attention.}
One can understand this, at least qualitatively, by using circuit equivalences, e.g. via $ZX$-calculus, and deforming the circuits so that damaged plaquette measurements appear as layered (in fictitious time), rather than interleaved.
In this manner, we see that the CNOT-based stabilizer measurements are not actually simultaneous and that the order in which data qubits are addressed by the interleaved circuits determines an effective order (the layering order in fictitious time) in which the damaged stabilizers are measured (Ref.~\onlinecite{Auger2017} observed this property in their Fig.~11).
Moreover, each layer of the effective ordering contains stabilizers of either only $Z$-type or only $X$-type, but not both.
Tracking the corresponding effective instantaneous stabilizer group with respect to this effective measurement order reveals that the superplaquette stabilizers are formed over multiple rounds of applying the measurement circuits.
Attempting to form the superplaquette stabilizers from the damaged plaquette measurements of a single round would indeed yield randomized values, but appropriate compositions of pieces from multiple rounds produce values for superplaquette operators that are deterministic in the absence of errors.
The number of rounds across which these superplaquette stabilizers are composed can grow with the size of a connected region of dead components.
In particular, a region with $l$ effective measurement layers for the damaged stabilizers in each round will require $\lceil (l-1)/2 \rceil$ and $\lceil l/2 \rceil$ to build up the two respective types of superplaquette stabilizers.
For example, a region with a single dead data qubit has $l=3$, a region with two adjacent dead data qubits has $l=4$, and a region with a smallest square of dead data qubits (a dead plaquette) has $l=5$. (An undamaged region can be thought of as having $l=2$.)
Pulling this understanding back to the interleaved circuits running with the normal circuit schedule, we see that damaged plaquette stabilizers would be measured at the same rate as undamaged stabilizers, but this would yield a reduced number of superplaquette measurements due to finite time.
In particular, $r$ rounds of measurement circuits would respectively yield $r+1-\lceil (l-1)/2 \rceil$ and $r+1-\lceil l/2 \rceil$ measurements of the two types of superplaquette stabilizers.
We typically expect to be operating in a regime where $l < d_{\rm f}$ and $r \approx d_{\rm f}$, in which case this circuit schedule provides a greater number of superplaquette stabilizer measurements and detectors than alternating rounds between measuring $Z$-type and $X$-type damaged stabilizers.
As such, we expect operating the circuits this way to provide better performance in the presence of dead components.
Interleaved versions of our pairwise measurement-based stabilizer circuits, as described in Appendix~\ref{app:interleave}, behavior similarly with respect to effective time ordering of damaged plaquettes and forming superplaquette stabilizers over multiple rounds.
Again, for $r$ rounds of measurement circuits, this will yield $r+1-\lceil (l-1)/2 \rceil$ and $r+1-\lceil l/2 \rceil$ measurements of the two types of superplaquette stabilizers.

In contrast, the pipelining we use for our measurement-based circuits, e.g. in Eq.~\eqref{eq:compressed_pipeline}, effectively operates as alternation between all $Z$-type and all $X$-type $n$-gon measurements.
Again, this can be seen by tracking when the data qubits of neighboring plaquettes are addressed by the circuits for the different plaquette types, which allows one to isolate each plaquette's circuit in fictitious time.
Regardless of how many components are dead, there will be exactly two layers in the effective ordering in each round.
Thus, $r$ rounds of pipelined measurement circuits yields $r$ measurements for each of the two types of superplaquette stabilizers.
In light of this, we expect circuits that are pipelined in this manner to provide better performance in the presence of dead components than interleaved circuits.
We note that a similar pipelining can be applied to the CNOT-based implementation of stabilizer measurements, which would yield the same advantage, though with the potential drawback that the auxiliary qubit initialization and measurement steps for the $Z$-type and $X$-type stabilizers would not be concurrent, potentially complicating timing of physical operations.

Returning to the situation where all undamaged stabilizers are measured in each measurement round, but only one type of the damaged stabilizers are measured, Ref.~\onlinecite{Strikis2023} found improved performance for larger regions of dead components by modifying the protocol of Ref.~\onlinecite{Auger2017} to alternate between $l$ repeated rounds of damaged stabilizer measurements of each type, where $l$ is the linear size of the dead region.
For our pipelined circuits, we could follow a similar protocol of repeating damaged stabilizer measurements of each type $l$ times before alternating, but this would require halving the rate at which damaged stabilizers are measured.
It is not obvious whether the trade-off for employing this strategy would improve performance in our case; in fact, we expect it to decrease performance, since it would require halving the rate at which damaged plaquettes are measured.

Ref.~\onlinecite{Auger2017} additionally described a strategy for dealing with dead auxiliary qubits and CNOT gates (which they called syndrome qubit and link fabrication errors, respectively).
For this, they identify all the data qubits directly interacting with a dead CNOT gate or auxiliary qubit, i.e. the data qubit acted upon by a given CNOT gate or all data qubits in the plaquette associated with the given auxiliary qubit, respectively.
Then, all such data qubits (even though not dead) are removed from the code, together with all the dead components.

We propose a different strategy for dealing with the dead auxiliary qubits and connections that avoids unnecessarily removing data qubits that are not dead, and which we expect should improve performance.
Here, ``connection'' refers to a multi-qubit operation, which may include computational gates (e.g. CNOTs) or measurements (e.g. pairwise Pauli measurements) acting on multiple physical qubits.
Our strategy for determining the modification of the code operation to remove dead components can be divided into three steps:
\begin{enumerate}
\item 
For any $n$-gon involving $m$ dead data qubits, reduce it to a $(n-m)$-gon by removing the dead data qubits.
\item 
For any $n$-gon (possibly the result of a reduction in step 1) involving dead auxiliary qubits, split it up into a $n_1$-gon, ... , and $n_k$-gon, where $n_1 + \cdots +n_k = n$, such that $k$ is minimized and none of the resulting measurement circuits utilize the dead auxiliary qubits.
\item
For any $n$-gon (possibly the result of a reduction and/or splitting in steps 1 and 2) involving dead connections, split it up into a $n_1$-gon, ... , and $n_k$-gon, where $n_1 + \cdots +n_k = n$, such that $k$ is minimized and none of the resulting measurement circuits utilize the dead connections.
\end{enumerate}
The result of applying each step is unique, i.e. no choices need to be made (at least for simple enough realizations, including all those of interest examined in this paper).
Within each step, the reduction or splitting of a $n$-gon can be determined iteratively by assessing one component at a time, until all dead components of that step's type have been removed.
Splitting $n$-gons (steps 2 and 3) always preserves the number of data qubits.
On the other hand, the reductions and splittings (all steps) may result in auxiliary qubits or connections that are not dead, but are no longer utilized in the stabilizer measurement circuits; such auxiliary qubits and connections are collaterally removed from the code operation, as they no longer participate.
For example, if an auxiliary qubit is dead, then all connections to it are removed; similarly, if all connections to a qubit are dead, that qubit is removed.
We will see more examples of this when we consider specific realizations.
In examples (not considered explicitly in this paper) where certain auxiliary qubits or connections are shared by multiple plaquettes' stabilizer measurement circuits, they may, in principle, be dead or removed with respect to one plaquette, but not another.
For these situations, the functionality of the components may be assessed for each plaquette, which can then be independently modified accordingly.
Though it would not change the functioning of the code, if there are regions that become completely disconnected from the main region of the code, e.g. a single data qubit with no connections to other qubits, these can safely be removed as well.

This protocol provides the most efficient salvaging of functioning components (without adding components or elements to the set of operations) when removing the dead components -- no functioning data qubits are removed from the code operation (unless they are completely disconnected from the rest of the code) and functioning auxiliary qubits and connections are only removed if they are not needed to implement the (minimally split) $n$-gons formed from the reducing and splitting process.
Moreover, it minimizes the damage to the code in terms of the gaps created in the code, i.e. the number and size of superplaquette operators which effectively reduce the code and fault distances.
Since the above strategy is more efficient with respect to removing components and minimizes the damage to the code, we expect it to improve performance with respect to using the code reduction strategy of Ref.~\onlinecite{Auger2017}.

The strategy presented here applies rather generally to different realizations of the surface code (and possibly other codes), assuming they have certain reasonable properties.
One such assumption is that there are natural circuits for measuring all the possible reduced and split plaquette stabilizers that only use previously existing components.
The application of step 1 only depends on the code, not the detailed realization.
In contrast, the application of steps 2 and 3 will depend on the details of the particular realization.

\begin{figure}[t!]
\centering

\begin{minipage}{0.8\linewidth}
{\flushleft
\input{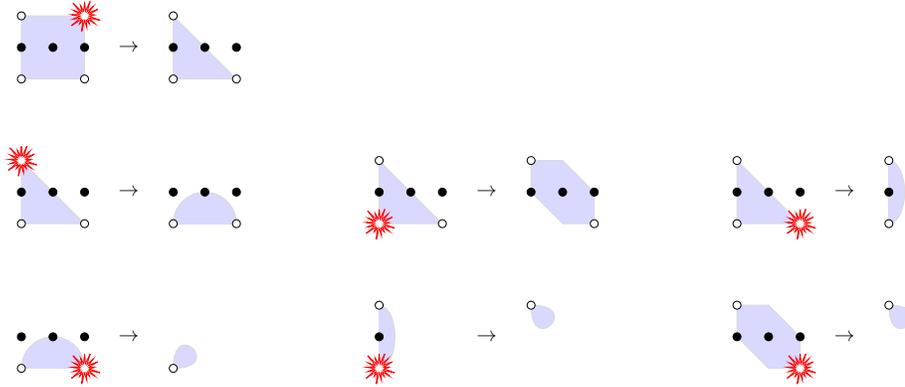}
}
\end{minipage}

\caption{
The possible reductions of $Z$-type $n$-gons for step 1, where dead data qubits (denoted by starbursts) are removed from the code operation.
(Reductions related to these by rotations and reflections are not shown separately.)
The reductions for $X$-type $n$-gons may be obtained from these by $90$ degree rotations.
Even though this step only considers dead data qubits, the auxiliary qubits are displayed to show the collateral removal of live auxiliary qubits that may occur when removing dead data qubits.
Ignoring the auxiliary qubits, the same reduction of $n$-gons can be used for any realization of the surface code.
The full reduction of each $n$-gon in this step can be determined through an iterative process where dead data qubits are removed one at a time until no dead ones remain, i.e working down through this figure.
}
\label{fig:dead_data}
\end{figure}

We now consider the implementation of this strategy for our measurement-based realization of the surface code in detail.
The $n$-gon stabilizer measurement circuits from Sec.~\ref{sec:boundaries} provide all of the circuits that we need in order to implement our dead components strategy.
These $n$-gon stabilizer circuits share some nice features when used for reducing and splitting $n$-gons.
The first is that none of them introduce new physical measurements to the set of measurements needed to operate the code; a given measurement is either removed from the circuit or reduced from a pairwise to a single-qubit measurement of the same type.
(We note that all single-qubit Pauli measurements are required for operation, even though they may not all occur in the $4$-gon circuits.)
Thus, it is a relatively simple matter to change the operation of a plaquette in a manner that reduces or splits the plaquette into smaller $n$-gons.
Additionally, all of these $n$-gon circuits interlock, so we can operate all plaquettes of the same type on the same schedule, without further modification.

\begin{figure}[t!]
\centering

\begin{minipage}{1\linewidth}
{\flushleft
\input{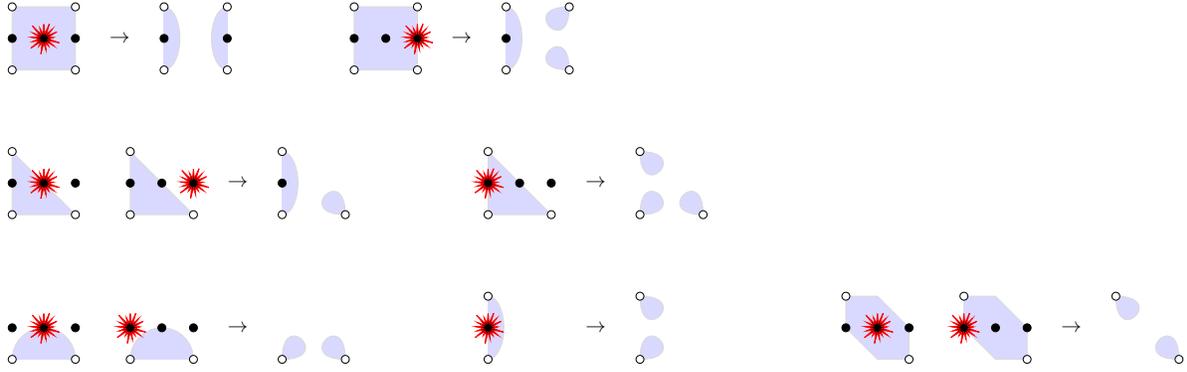}
}
\end{minipage}

\caption{
The possible splittings of $Z$-type $n$-gons for step 2, where dead auxiliary qubits (denoted by starbursts) are removed from the code operation.
(Splittings related to these by rotations and reflections are not shown separately.)
The splittings for $X$-type $n$-gons may be obtained from these by $90$ degree rotations.
In some cases, live auxiliary qubits will be collaterally removed as a result of removing dead auxiliary qubits.
The full splitting of each $n$-gon in this step can be determined through an iterative process where dead auxiliary qubits are removed one at a time (together with any collateral loss), until no dead ones remain, i.e working down through this figure.
}
\label{fig:dead_auxiliary}
\end{figure}

\begin{figure}[t!]
\centering

\begin{minipage}{1\linewidth}
{\flushleft
\input{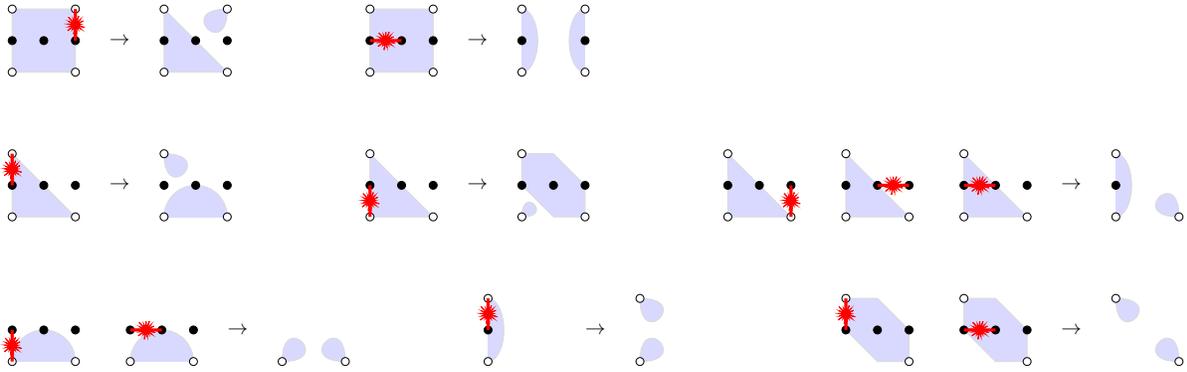}
}
\end{minipage}

\caption{
The possible splittings of $Z$-type $n$-gons for step 3, where dead connections (denoted by starbursts) are removed from the code operation.
(Splittings related to these by rotations and reflections are not shown separately.)
The splittings for $X$-type $n$-gons may be obtained from these by $90$ degree rotations.
In some cases, live auxiliary qubits will be collaterally removed as a result of removing dead connections.
The full splitting of each $n$-gon in this step can be determined through an iterative process where dead connections are removed one at a time (together with any collateral loss), until no dead ones remain, i.e working down through this figure.
}
\label{fig:dead_connection}
\end{figure}

We can now consider each of the three steps for determining the modification of the code in the presence of dead components.
The modifications for step 1, where plaquettes are modified by removing dead data qubits, are shown in Fig.~\ref{fig:dead_data}.
The modifications for step 2, where plaquettes are modified by removing dead auxiliary qubits, are shown in Fig.~\ref{fig:dead_auxiliary}.
The modifications for step 3, where plaquettes are modified by removing dead connections (pairwise measurements), are shown in Fig.~\ref{fig:dead_connection}.
In each of these figures, we display the different possible scenarios (up to rotations and reflections) for removing a single dead component at a time.
In order to remove all dead components from operation, we check components of a given step one at a time and apply the corresponding reduction or splitting shown in the figures, working down through possible scenarios, and repeat for the next component until all dead components are removed.

As a demonstration of the generality of our proposed strategy for surface code realizations, we can apply it to the measurement-based pentagonal tiling realization of Ref.~\onlinecite{Gidney2022a} and the CNOT gate-based realization.
We show the details for these in Appendix~\ref{app:dead_component_realizations}

In order to demonstrate the advantages of our proposed strategy and, in particular, how steps 2 and 3 make our protocol differ from that of Ref.~\onlinecite{Auger2017}, it is useful to consider some example scenarios of dead components in detail.
We show the modifications of plaquettes for several dead component scenarios, as well as the $Z$- and $X$-measurement rounds' corresponding measured stabilizers (damaged and undamaged) and the superplaquette stabilizers that survive from round to round.
(Recall that when there are no dead components, all plaquette stabilizers survive from round to round.)

\begin{figure}[t!]
\centering

\input{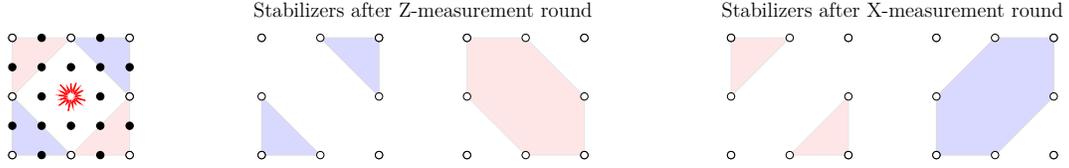}

\caption{Configuration for one dead data qubit.
}
\label{fig:dead_1data}
\end{figure}

\begin{figure}[t!]
\centering

\input{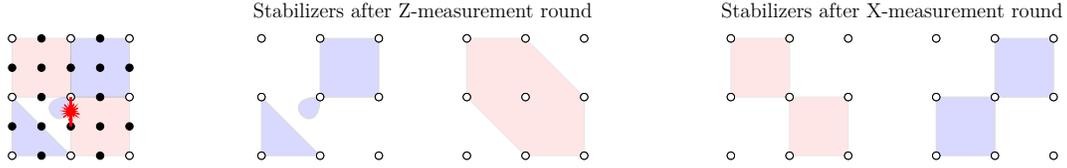}

\caption{Configuration for one dead connection to a data qubit.
}
\label{fig:dead_1connection}
\end{figure}

In Fig.~\ref{fig:dead_1data}, we show the modification of the plaquette stabilizers associated with one dead data qubit.
Following the protocol of Ref.~\onlinecite{Auger2017}, this modification would also be used in the case where that data qubit was live, but any of the four connections involving that qubit were dead.
With this modification, the code distance is reduced by one.
In Fig.~\ref{fig:dead_1connection}, we show our modification of the plaquette stabilizers associated with one dead connection between a data qubit and an auxiliary qubit.
Comparing to the modification in Fig.~\ref{fig:dead_1data}, we see that the stabilizers only differ from the undamaged stabilizers in the $Z$-measurement round and that the code distance is reduced by one for only the $Z$ logical string operators.

\begin{figure}[t!]
\centering
\input{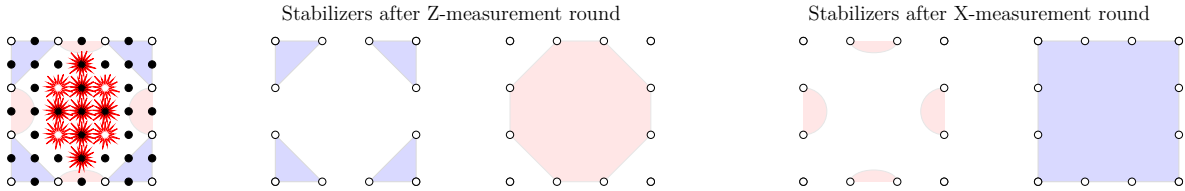}
\caption{Configuration for a dead plaquette.
}
\label{fig:dead_1plaquette}
\end{figure}

\begin{figure}[t!]
\centering

\input{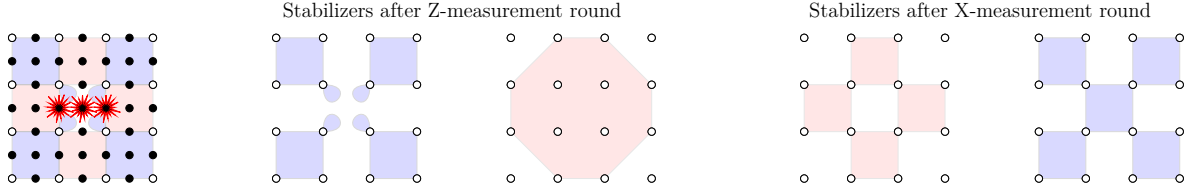}

\caption{Configuration for a plaquette with dead auxiliary qubits $A$ and $C$, or dead connections between data qubits and auxiliary qubits $A$ and $C$.
}
\label{fig:dead_ACauxiliary}
\end{figure}

\begin{figure}[t!]
\centering

\input{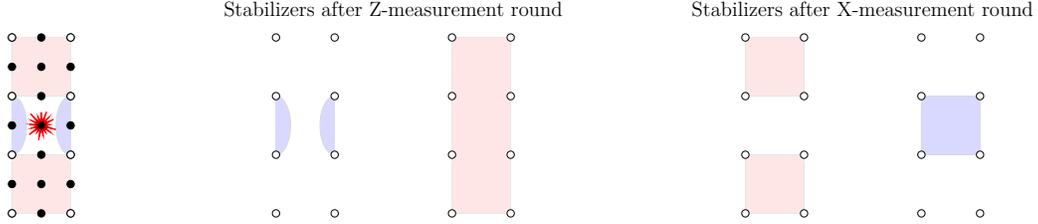}

\caption{Configuration for a plaquette with dead auxiliary qubit $B$ or a dead connection to this qubit.
}
\label{fig:dead_Bauxiliary}
\end{figure}

In Fig.~\ref{fig:dead_1plaquette}, we show the modification of the plaquette stabilizers associated with one dead plaquette, i.e. four dead data qubits.
Following the protocol of Ref.~\onlinecite{Auger2017}, this modification would also be used in the case where the data qubits were live, but any of the auxiliary qubits of the plaquette were dead, or each of those data qubits had a dead connection.
With this modification, the code distance is reduced by two.
In Fig.~\ref{fig:dead_ACauxiliary}, we show our modification of the plaquette stabilizers associated with a plaquette for which the $A$ and $C$ auxiliary qubits are dead, or the four connections between the data qubits and the auxiliary qubits of that plaquette are dead.
We again see that the stabilizers only differ from the undamaged stabilizers in the $Z$-measurement round and that the code distance is reduced by two for only the $Z$ logical string operators.
In Fig.~\ref{fig:dead_Bauxiliary}, we show our modification of the plaquette stabilizers associated with a plaquette for which only the $B$ auxiliary qubit is dead, or for which either of the connections involving the $B$ qubit is dead.
Again, the stabilizers only differ from the undamaged stabilizers in the $Z$-measurement round and, in this case, the distance is reduced by two only for $Z$ logical string operators in the vertical direction.
This may not affect the code distance if the logical qubit is encoded such that the logical $Z$ string operators are aligned in the horizontal direction.

\section{Implementation in Majorana Hardware}
\label{sec:Majorana}

We now consider the implementation of our measurement-based realization of the surface code in Majorana hardware~\cite{Karzig2017}.
As this was the original motivation for generating our new surface code realization, it should not be surprising that the layout and measurements involved are very simple for this hardware.
We consider rectangular arrays of Majorana tetron qubits, which are formed from two Majorana wires connected by a superconducting spine.
Each tetron possesses four Majorana zero modes (MZMs), one at each endpoint of each Majorana wire.
A pair of MZMs combine into a fermionic mode, i.e. the joint fermionic parity of two MZMs corresponds to a two-level system.
However, since each tetron is a floating superconducting island with a charging energy, there is an overall parity constraint of the four MZMs of a tetron.
In this way, the tetrons form a qubit, where the joint parity operators associated with different MZM pairs correspond to the different Pauli operators of the qubit.
In particular, when the $j$th MZM of a tetron has corresponding Majorana operator $\gamma_j$, we use the convention where the joint fermionic parity operator $P_{jk}=i \gamma_j \gamma_k$ of MZMs $j$ and $k$ map to Pauli operators according to
\begin{align}
X &= P_{23} = P_{14} , \label{eq:X} \\
Y &= P_{13} = - P_{24} , \\
Z &= P_{12} = P_{34} \label{eq:Z}
.
\end{align}
See Ref.~\onlinecite{Karzig2017} for more details.

\begin{figure}[t!]
\centering
\includegraphics[width=\linewidth]{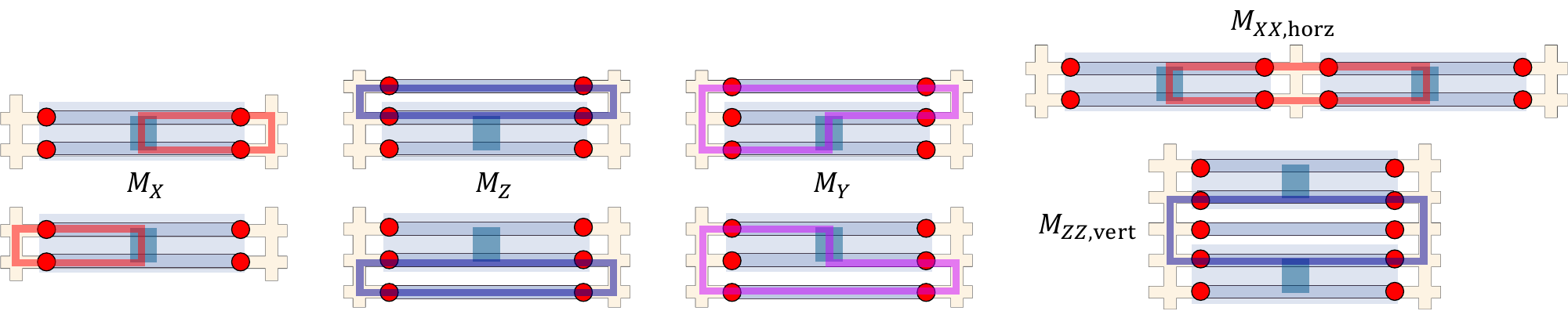}
\caption{Measurement loops in Majorana hardware corresponding to the measurements needed for our code.  MZMs (red circles) are located at the end points of topological wires (gray lines).  Two topological wires connected by a trivial superconducting spine (dark gray) form a tetron.  Tetrons can be connected to each other or coherent links through semiconductor segments (tan).   
}
\label{fig:measurement_loops}
\end{figure}

In order to facilitate measurements of these operators, one needs additional components around the tetrons that can couple to MZMs as desired to form interference loops that enable measurement of the joint parity of all the MZMs in the interference loop.
(Such measurements always involve exactly two MZMs from each tetron qubit involved in the measurement.)
The components utilized for this include semiconductor rails running along the short direction of the tetrons (between columns of tetrons in a rectangular array), which enable (electrostatic) gate-defined quantum dots with gate-controlled couplings to the MZMs.
There are also ``coherent links,'' which are single floating Majorana wires (with a MZM on each of its two endpoints), which are located between tetrons running along the long direction.
These facilitate the creation of interference loops involving MZMs on the opposite sides of tetrons in the long direction. 
We display schematic drawings of this hardware and some of the measurement loops in Fig.~\ref{fig:measurement_loops}.
The measurement loops shown serve to define our basis conventions with respect to the hardware, and the same choice is used for all tetrons.
Once the interference loops are turned on, the joint parity of all the MZMs included in the loop can be measured, for example by probing the quantum capacitance of the coupled tetron-quantum dot system (which will exhibit parity dependence) using microwave resonators.

We expect that the fidelities of such MZM parity measurements will generally decrease with increasing length of semiconductor, number of MZMs, number of coherent links, and number tetrons utilized in the corresponding measurement loops~\cite{Tran2019}, that is
\begin{align}
f_{M_X} > f_{M_Z} > f_{M_Y} > f_{M_{XX:\text{horz}}} > f_{M_{ZZ:\text{vert}}} > f_{M_{YZ:\text{vert}}} > f_{M_{YY:\text{vert}}} >  \cdots
.
\end{align}
Here, we have indicated the horizontal or vertical direction on pairwise measurements, because it makes a significant difference in the difficulty of the measurement, e.g. $M_{ZZ:\text{horz}}$ would require the additional use of two coherent links as compared with $M_{ZZ:\text{vert}}$.
This provides a rough guide for optimizing codes or computations with respect to the measurements utilized.

We now examine the implementation of our measurement-based realization of the surface code in a rectangular array of tetrons.
We notice that our code only uses $M_X$, $M_Z$, $M_{XX:\text{horz}}$, and $M_{ZZ:\text{vert}}$ measurements (for logical memory).
These are the two simplest single-qubit measurements and two simplest two-qubit measurements for this hardware.
Thus, our measurement-based realization of the surface code is highly optimized with respect to the set of measurements in Majorana hardware.

\begin{figure}[t!]
\centering

\scalebox{0.5}{
\input{DualRailSteps}
}

\includegraphics[width=0.49 \linewidth]{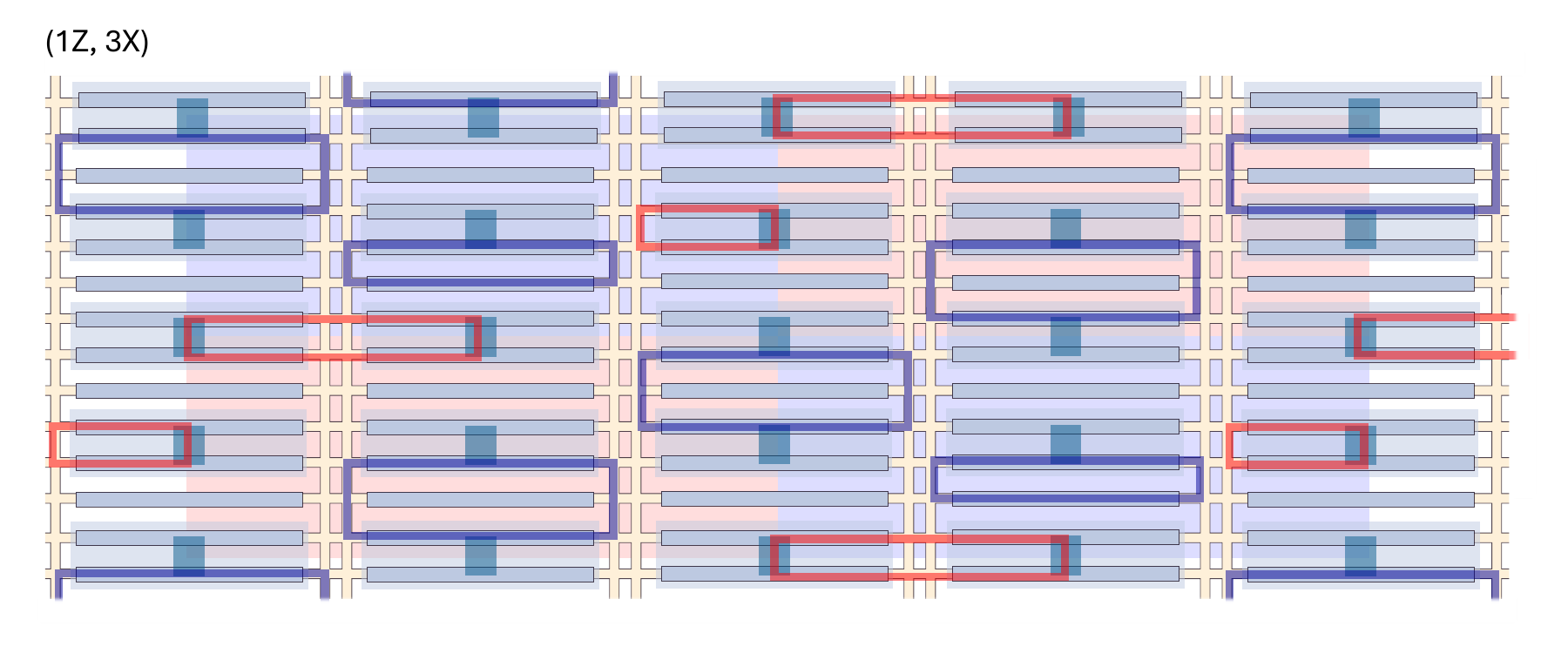}
\includegraphics[width=0.49 \linewidth]
{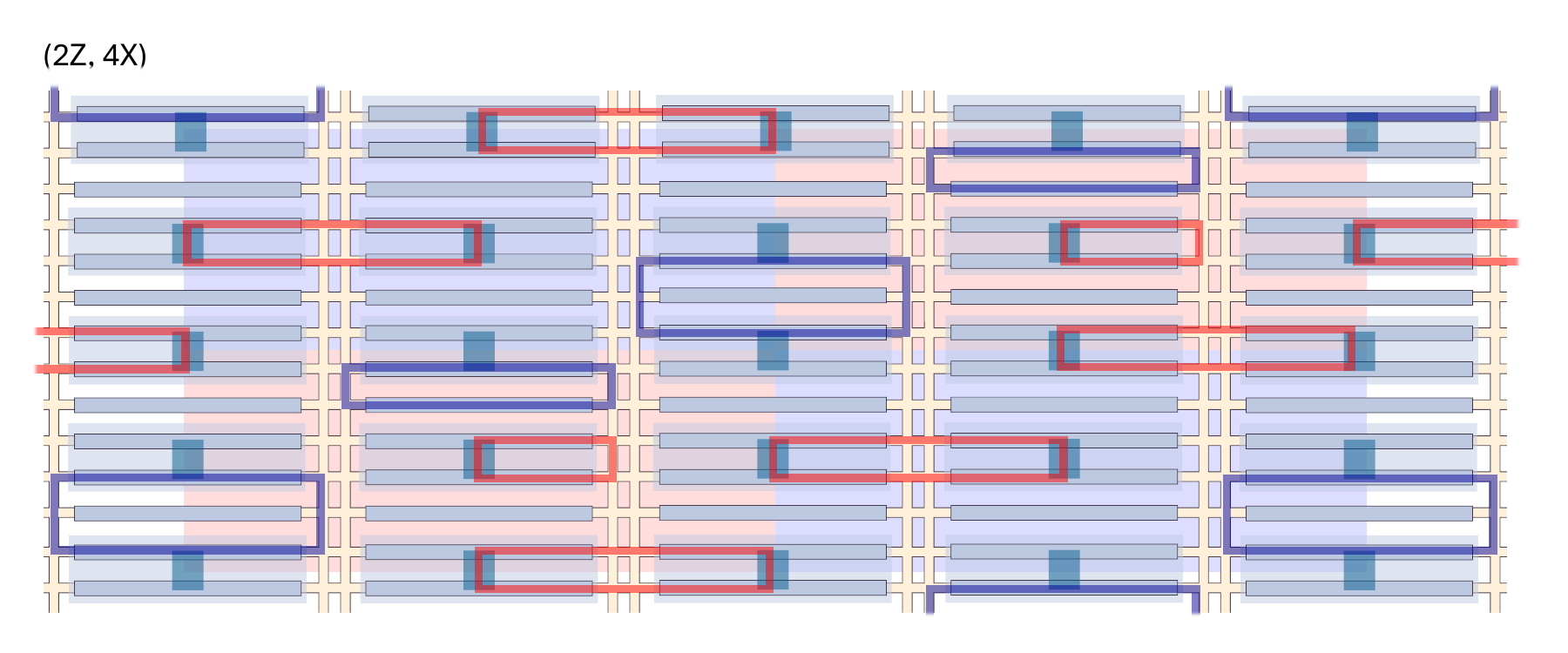}

\includegraphics[width=0.49 \linewidth]{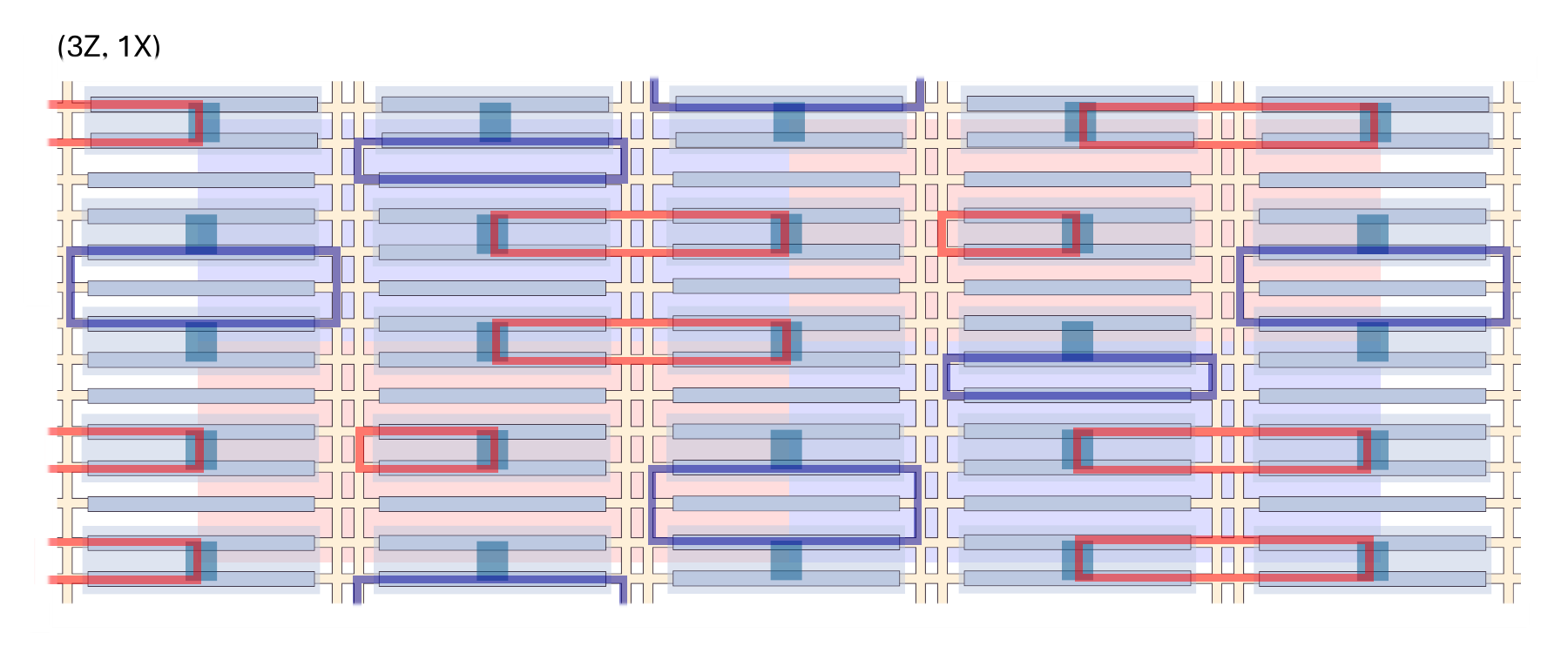}
\includegraphics[width=0.49 \linewidth]{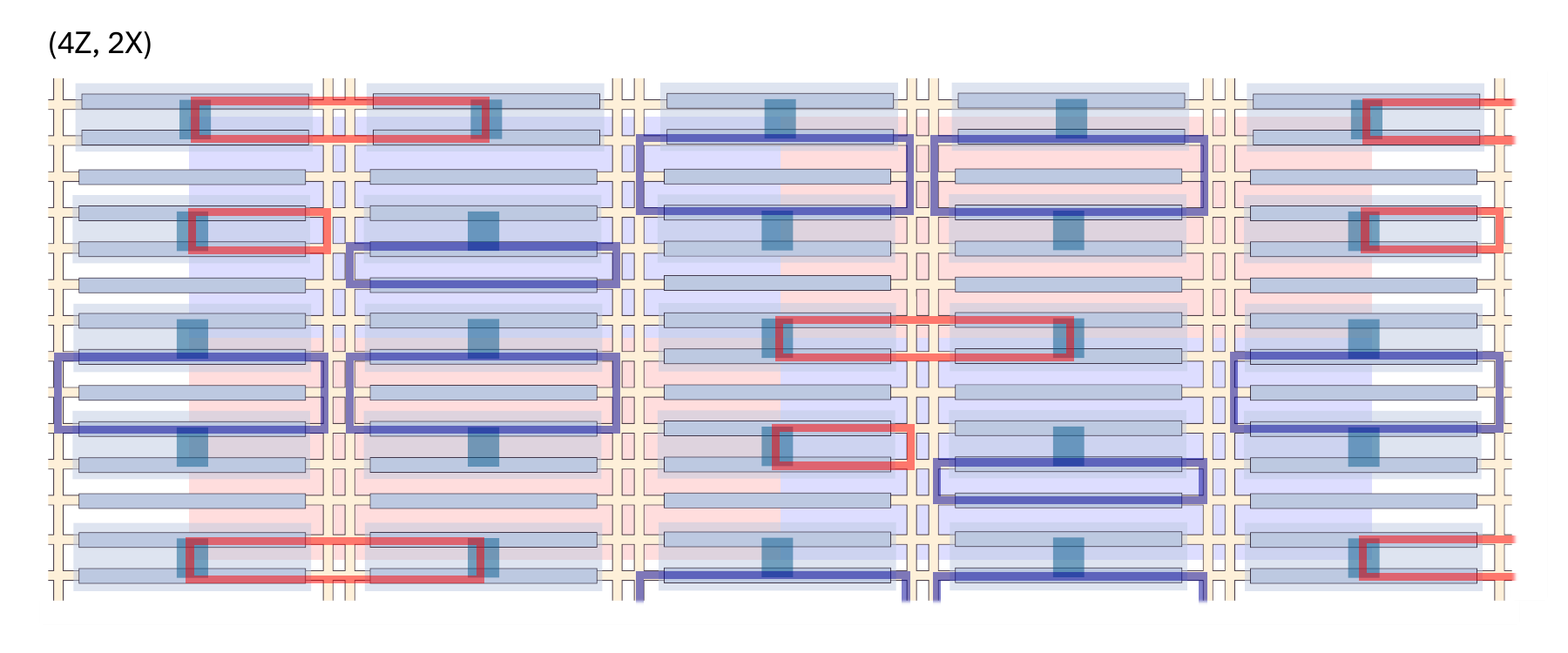}

\caption{
Interference loops of measurements for the implementation of our measurement-based surface code realization in Majorana hardware with double-rail semiconductor layouts.
This layout avoids loop conflicts, allowing for the (minimal) period 4 operation schedule.
}
\label{fig:double_rail}
\end{figure}

One significant aspect of device design for Majorana hardware is whether a single rail or double rail of semiconductor is used between the columns of tetrons.
The semiconductor rails are where most of the operational activity is concentrated in this hardware, e.g. the quantum dots, coupling to MZMs, and measurements.  
As such, utilizing double-rail semiconductors constitutes an increase in fabrication and control difficulties, which may also translate into higher error rates.
However, the positive aspect of the trade-off is that using double-rails allows independent measurements on adjacent tetron columns to be performed without conflict.
For single-rail layouts, the interference loops for certain configurations of measurements would overlap and hence could not be performed at the same time.
(For double-rail layouts, there may, in practice, be unwanted cross-talk between such measurements when performed simultaneously, since their interference loops would necessarily be in close proximity to each other.)
With this in mind, it is useful to consider the implementation of codes in both single-rail and double-rail layouts to appreciate how this difference affects performance in Majorana hardware.

\begin{figure}[t!]
\centering

\scalebox{0.5}{\input{SingleRailSteps}
}

\includegraphics[width=.49 \linewidth]{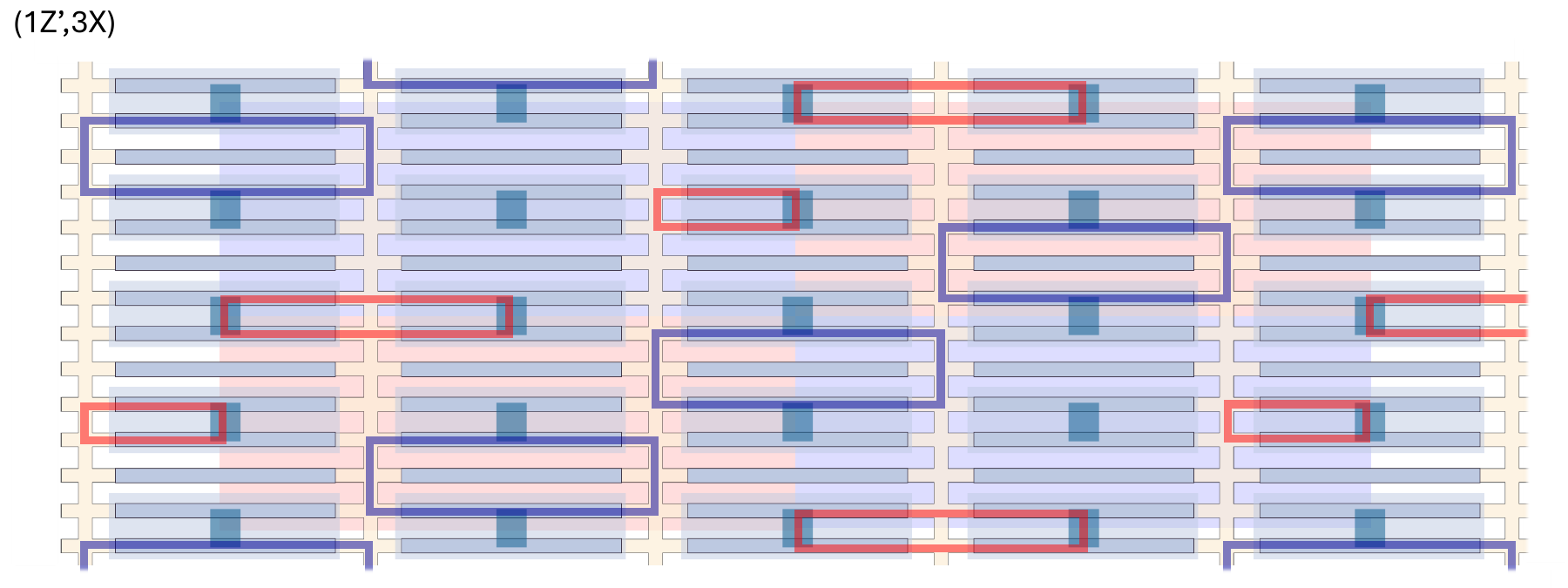}
\includegraphics[width=.49 \linewidth]{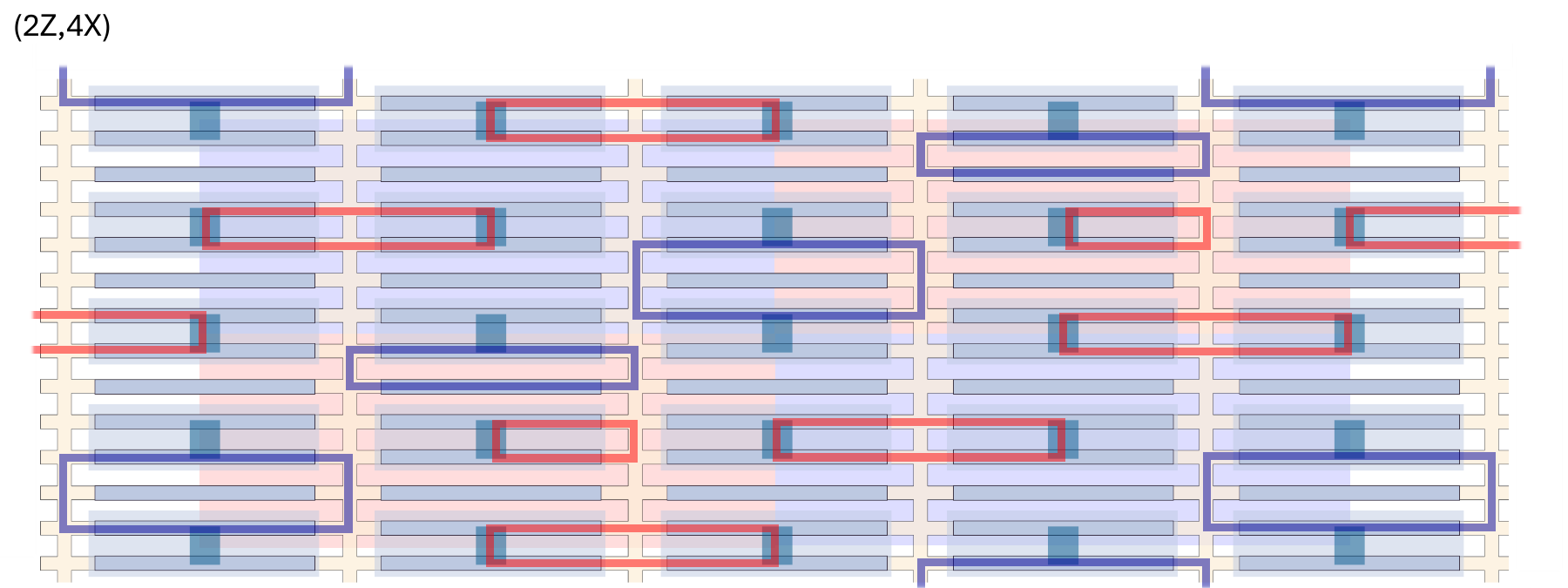}
\includegraphics[width=.49 \linewidth]{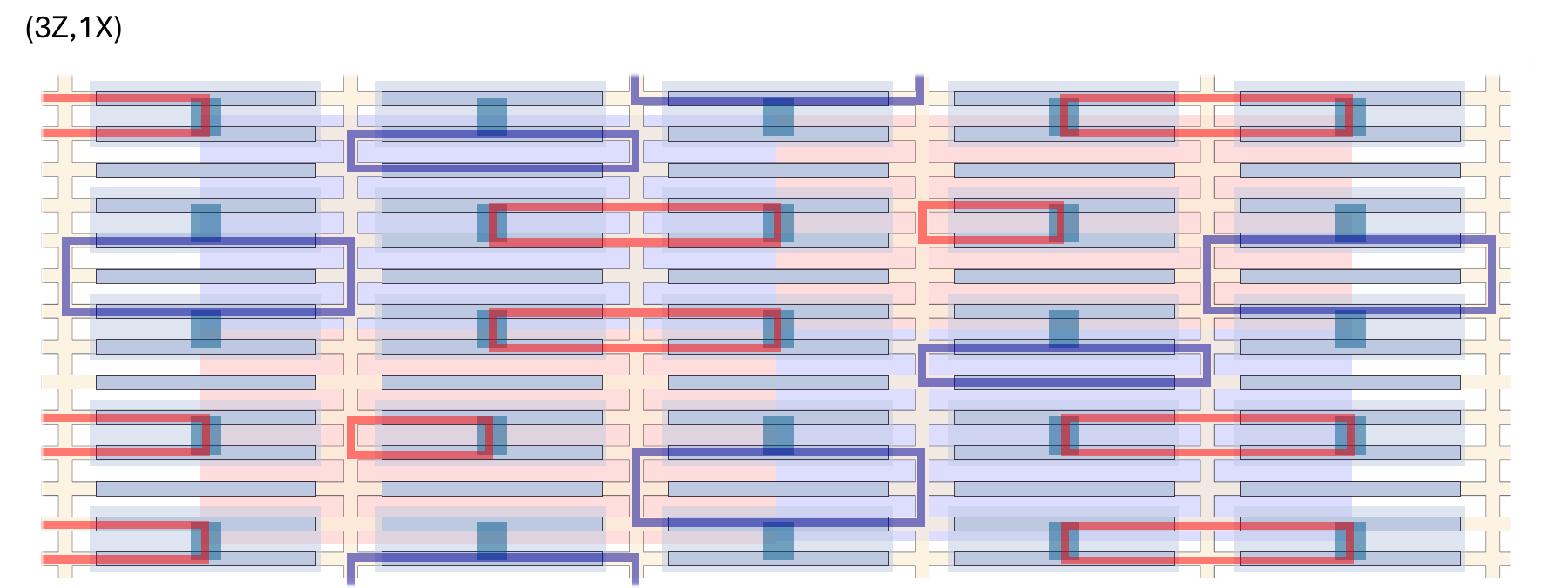}
\includegraphics[width=.49 \linewidth]{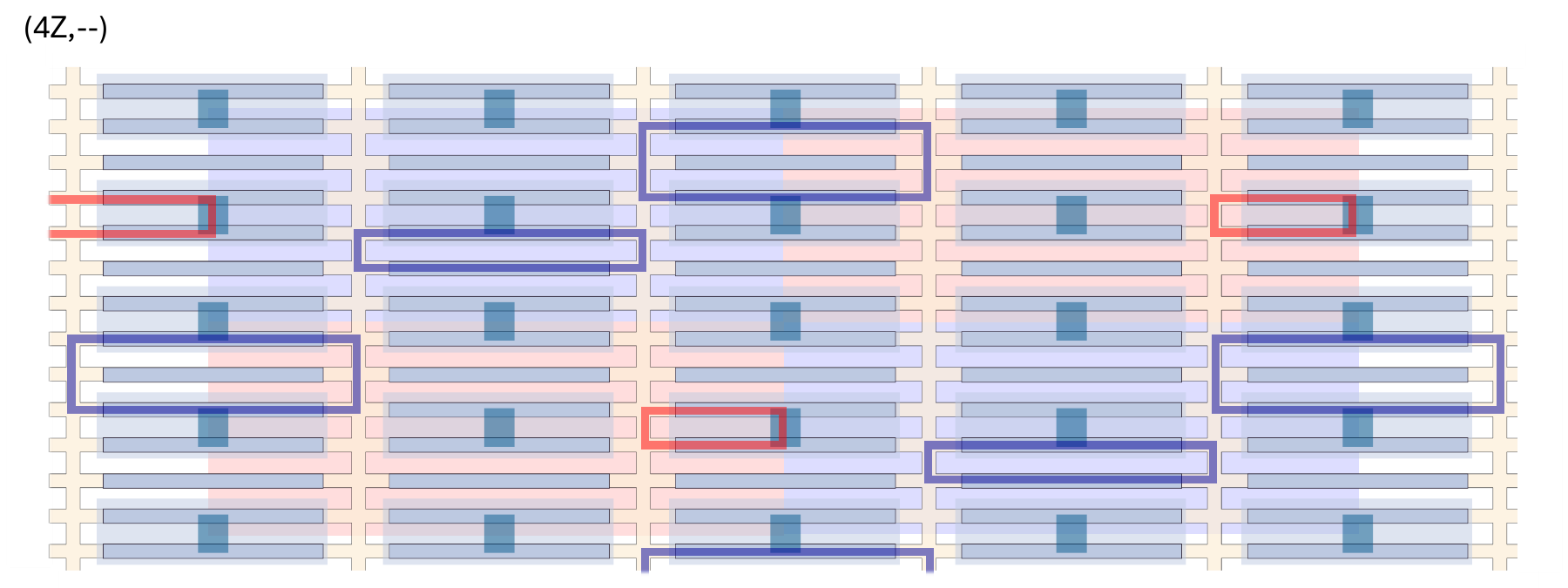}
\includegraphics[width=.49 \linewidth]{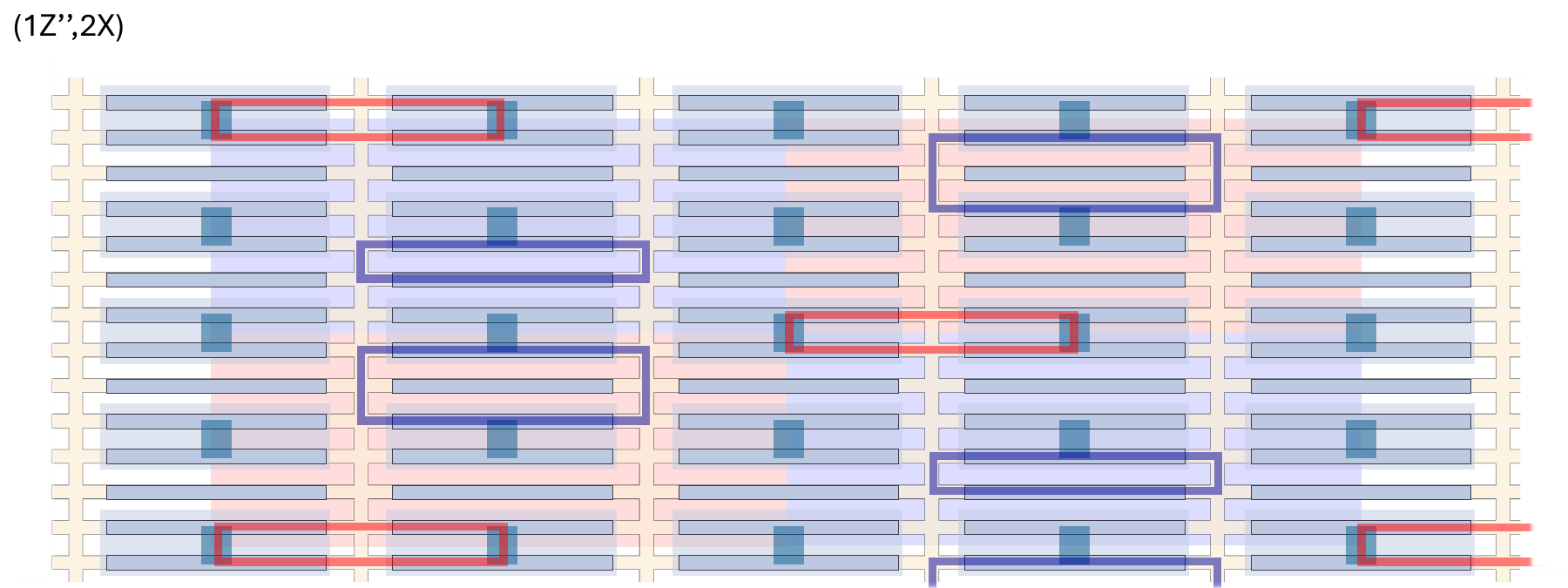} \phantom{\includegraphics[width=.49 \linewidth]{SLSRSC_SR_1Z2X.pdf}}

\caption{The implementation of our measurement-based surface code realization in Majorana hardware with single-rail semiconductor layouts requires resolution of loop conflicts in two steps.
A resolution with period 5 operation schedule is shown here.
}
\label{fig:single_rail}
\end{figure}

We can implement our code in Majorana hardware with the operation schedule of Eq.~\eqref{eq:compressed_pipeline} using rectangular arrays with double-rail semiconductors, as shown in Fig.~\ref{fig:double_rail}.
Examining the interference loops in the steps, we see that attempting to implement this operation schedule on an array with single-rail semiconductors would result in loop conflicts in steps $(1Z,3X)$ and $(4Z,2X)$.
A na\"ive resolution of this would be to split each of the steps with conflicts into two steps, distributing the measurements between them so that no step contains a conflict.
However, a more efficient resolution is possible, which we show in Fig.~\ref{fig:single_rail}.
In particular, we split step $(4Z,2X)$ into two steps, $(4Z,-)$ and $(-,2X)$, and shift one of the conflicting measurements from step $(1Z,3X)$ into the second of these split steps.
Denoting step $1Z$ without the $M_{Z_{B}}$ measurement as $1Z'$ and the $M_{Z_{B}}$ measurement as $1Z''$, the resulting pipelining for single-rail layouts is
\begin{equation}
\label{eq:single_rail_pipeline}
\cdots, (1Z',3X), (2Z,4X), (3Z,1X), (4Z,-),(1Z'',2X), \cdots
.
\end{equation}
This operation schedule for single-rail semiconductor layouts has a five step period, which is a relatively mild slow down.
(We note that there are other possibilities for redistributing the measurements among a five step cycle that will be compatible with single-rail layouts.)

For the hook-preventing circuits described in Sec.~\ref{sec:Hook}, we can implement the code using any of the operation schedules shown in Fig.~\ref{fig:hook_detect_pipeline} with double-rail layouts.
For single-rail semiconductor layouts, we can resolve interference loop conflicts while only increasing the operation period by one step (from seven steps to eight steps), e.g. by modifying the pipelining option 2 given in Eq.~\eqref{eq:hook_preventing_pipelines} in a similar manner as described above.

It is worth comparing the implementation of our surface code realization in Majorana hardware to that of the Floquet code on the $4.8.8$ lattice, as considered in Ref.~\onlinecite{Paetznick2023}.
The Floquet code can also be implemented on a rectangular array of qubits, where it utilizes measurements that we denote as $M_{XX:\text{horz}}$, $M_{ZZ:\text{vert}}$, and $M_{YY:\text{vert}}$ (in equal proportions).
For double-rail semiconductor layouts, the 4.8.8 Floquet code could operate on a six step period, in which no qubits are idle.
However, in order to implement this code on single-rail semiconductor layouts, it appears that interference loop conflicts can only be resolved by splitting each step into two steps, doubling the operation period and introducing the possibility of faults on idle qubits\replace{idle qubit errors} for half the qubits at each step.
As such, the comparative performance of our code vs. 4.8.8 Hastings-Haah is likely to improve when the more realistic hardware conditions are taken into consideration.

\section{Performance}
\label{sec:Performance}

In this section, we assess the performance of the codes presented in Secs.~\ref{sec:Circuits}, \ref{sec:Hook}, and~\ref{sec:Majorana}, using the circuit noise model presented in Ref.~\onlinecite{Chao2020}.
In this noise model, each measurement fails independently with probability $p_\text{physical}$.
When a measurement fails, it acts as an ideal measurement followed by an error drawn uniformly from the set of nontrivial errors supported on the qubits involved in the measurement.
For a single-qubit measurement, the error is drawn from the set 
$\{ (P_1 , F) \} - \{ (I, 0) \}$.
For a two-qubit measurement fault, the error is drawn from the set $\{ (P_1\otimes  P_2 , F) \} - \{ (I\otimes I , 0) \}$.
Here, $P_1,P_2 \in \{ I,X,Y,Z\}$ are Pauli errors acting on the support of the measurement and $F\in \{0,1\}$ corresponds to a readout error, i.e. a bit flip of the measurement outcome.

Some of the code variants we consider include steps in which qubits are idle.
The noise model also includes faults on idle qubits, with errors drawn\replace{idling errors for qubits, which are single-qubit errors drawn} from the set $\{ X, Y, Z \}$.
At each time step that a qubit idles, we assign the same error rate $p_\text{physical}$ for an idling error to occur.
However, since this is likely to overestimate the relative error rate of idling compared to measurements for our hardware of interest, it is useful to compare code performance with and without faults on idle qubits\replace{idle errors included}.
We will specify when our analysis does not include faults on idle qubits\replace{idle errors for idle qubits}.
The distinction will be significant for the hook-preventing circuit and when analyzing implementations in Majorana hardware with single-rail semiconductor layouts.

\subsection{Decoding graph construction}

We first provide an overview of how we construct a decoding graph, on which we use PyMatching~v2 to efficiently perform minimum weight perfect matching via the ``sparse blossom'' algorithm~\cite{higgott2023sparse}. 
We use the spacetime circuit formalism developed in Ref.~\onlinecite{Delfosse2023}, and the splitting decoder of Ref.~\onlinecite{Delfosse2023a}. 
We start with a Clifford circuit consisting of single- and two-qubit measurements. 
In the absence of errors, the measurements within this circuit satisfy nontrivial correlations, called detectors.
More formally, a detector consists of a set of measurements for which the joint parity of their outcomes is fixed in the absence of errors~\cite{Gidney2021b}. 
Let $\{m_\alpha\}$ be the set of measurement outcomes for the circuit, where $\alpha$ runs over all spacetime coordinates. 
The set of detectors form a classical parity check code over the measurement outcomes. 
We define the matrix $D$ to be such that each row forms a check of the corresponding classical parity check code. 
Here, we assume $D_{j\alpha}$ and $m_\alpha$ take values in $\mathbb{F}_2$ with $\cdot$ the usual product and addition mod~2.
Then $\sum_\alpha D_{j\alpha} \cdot m_\alpha$ takes on a fixed value in the absence of errors, and forms the $j$th detector.
We can find all detectors using the fault propagation as described in Ref.~\onlinecite{Delfosse2023} [see Algorithm 1 in this reference, and note that they refer to detectors as ``checks'' (as is also the case in Ref.~\onlinecite{bombinlogical2023})].

A spacetime error chain corresponds to a collection of Pauli errors and readout errors\replace{measurement flips} on the circuit. An error chain is detectable if it triggers one or more detectors. 
In practice, we choose a basis of detectors and use those to form a decoding graph or hypergraph.
We label each detector by an integer $j$, and refer to the set of detectors $\{j,k,...\}$ that are triggered by an error chain (meaning that the sum of the measurement outcomes mod 2 differs from its value in the absence of errors) as the error chain's syndrome.  Our ability to perform error correction depends on the ability to distinguish between equivalence classes of spacetime error chains, based on their syndromes.

Each vertex in the decoding graph can (in our chosen basis) be associated to a $Z$ or $X$ plaquette, as well as a time coordinate. We note that for a repeated plaquette stabilizer measurement, using the measurement circuits defined in  Fig.~\ref{fig:MZ4circuit} (original circuit) or Fig.~\ref{fig:hook_detect_circuit} (hook-preventing circuit), there are several low-weight detectors associated with each plaquette  (meaning detectors with a small number of contributing measurement outcomes), along with the high-weight detector that corresponds to the repeated measurement of the surface code stabilizer associated to the plaquette. The low-weight detectors are formed by repeated single auxiliary qubit measurements---such as the repeated $M_{X_B}$---or (in the hook-preventing circuit) by the alternating repeated pairwise auxiliary qubit measurements.

Given a basis of detectors together with a set of generative fault\replace{error} configurations $\mathcal{F} = \{ f_1, f_2, \cdots \}$, which we take to be degree one faults in the circuit noise model\replace{degree one 
circuit noise errors (meaning errors with probability $\propto p$ in the circuit noise model)}, a decoding hypergraph $H={V,E}$ is defined as follows. Each vertex $v_j\in V$ corresponds to a detector $j$. A set of vertices $\{v_j, v_k,...\}$ are connected by a hyperedge $e\in E$ if there exists a fault\replace{a circuit noise error} $f \in \mathcal{F}$ with the corresponding syndrome $\{j,k,...\}$. The weight of the hyperedge is defined from the probability of the fault\replace{error} occurring.  
This construction generally does not result in a decoding \emph{graph}, as certain faults\replace{circuit noise errors} in $\mathcal{F}$ may trigger three or more detectors and give rise to hyperedges. 
While matching on a graph can be performed in polynomial time, matching on a hypergraph is generally an NP-hard problem~\cite{karp1972}. 
As described in Ref.~\onlinecite{Delfosse2023a}, it is possible for some decoding hypergraphs to split hyperedges into edges in a consistent way that allows for successful decoding. This is done through the construction of a \emph{split noise model}. We define \emph{primitive faults}\replace{errors} as faults\replace{circuit noise errors} that either trigger one detector (1-faults\replace{errors}), or that trigger two detectors without being decomposable into 1-faults\replace{errors} (2-faults\replace{errors}). By construction, a set of generative fault\replace{error} configurations containing only primitive faults\replace{errors} will only give rise to edges. 
To define the split noise model, any non-primitive fault\replace{error} in $\mathcal{F}$ is decomposed into primitive faults\replace{errors}, such that they together trigger the same set of detectors as the original non-primitive fault\replace{error}. Each of these new faults\replace{errors} is assigned the same error rate as the original $n$-fault\replace{error}, and is added to a new set of generative fault\replace{error} configurations, $\tilde{\mathcal{F}}$, together with the primitive faults\replace{errors} in $\mathcal{F}$. $\tilde{\mathcal{F}}$ approximates $\mathcal{F}$ while containing only primitive faults\replace{errors}, and is used to define a decoding graph on which minimum weight matching is performed. 
The matching decoder that uses this graph is referred to as a splitting decoder.

\subsection{Dynamic re-weighting of the decoder graph edges}\label{Dynamic}

While the splitting decoder works straightforwardly for the original circuit defined in Fig.~\ref{fig:MZ4circuit}, the hook-preventing circuit defined in Fig.~\ref{fig:hook_detect_circuit} contains additional detectors, which complicate the use of a splitting decoder. These increase the weight of the syndromes for some circuit noise errors in $\mathcal{F}$, which in turn affects the performance of the splitting decoder such that the fault distance is effectively halved. To avoid this issue, we dynamically change the weights of the edges in the decoding graph in the presence of certain syndromes, as described below.  
This dynamic re-weighting is inspired by Ref.~\onlinecite{Pattison2021}, where soft information is used to dynamically determine edge weights (rather using than syndromes, as in the present context). 
With the dynamic re-weighting added for the hook-preventing circuit, we find that the splitting decoder is successful in all cases considered here.

We illustrate the need for re-weighting with an example. 
We recall the measurement sequence in the hook-preventing circuit for a $Z$-plaquette, shown in Fig.~\ref{fig:hook_detect_circuit}.
The repeated pairwise auxiliary measurements $M_{X_A X_B}$ and $M_{X_B X_C}$ in steps 2-5 give rise to two low-weight detectors. 
Using superscripts to indicate the time steps of measurements, the first low-weight detector is given by the sum of the measurement outcomes from $M_{X_A X_B}^{\text{2Z}}$ and $M_{X_A X_B}^{\text{4Z}}$, and the second by the sum of the measurement outcomes from $M_{X_B X_C}^{\text{3Z}}$ and $M_{X_B X_C}^{\text{5Z}}$. 
In our chosen basis of detectors, a single readout error on the second repetition of the measurements in these pairs, e.g. $M_{X_A X_B}^{4Z}$, will trigger not only the low-weight detector, but also two high-weight detectors associated to neighboring $X$-plaquettes.
As such, a fault that results in such a readout error is a non-primitive fault\replace{As such, it is a non-primitive error}.
It is decomposed onto a 1-fault resulting in a readout error on $M_{X_A X_B}^{2Z}$, which in the chosen basis triggers only the low-weight detector, and two 2-faults that together result in $Z$-errors on two data qubits.\replace{$Z$-errors on two data qubits.}\footnote{These can be reversed by an equally natural basis choice, so that a readout error in the first repetition triggers three detectors, and a readout error in the second repetition only triggers one detector.}
The spatial distribution of the errors is indicated in the following figure, where we denote the faults that result in readout errors\replace{readout errors} on the first and second $M_{X_A X_B}$ by $f_1$ and $f_2$: 
\begin{equation*}
\vcenter{\hbox{
\begin{tikzpicture}
\newcommand{\spacing}{0.5}
\foreach \i in {1}{\foreach \j in {0,2}{
 \draw [fill = red, opacity = 0.1] (\i*\spacing,\j*\spacing) rectangle (\i*\spacing + \spacing,\j*\spacing + \spacing);
 \draw [fill = red, opacity = 0.1] (\j*\spacing,\i*\spacing) rectangle (\j*\spacing + \spacing,\i*\spacing + \spacing);
}}
\foreach \i in {0,2}{\foreach \j in {0,2}{
 \draw [fill = blue, opacity = 0.15] (\i*\spacing,\j*\spacing) rectangle (\i*\spacing + \spacing,\j*\spacing + \spacing);
}}
\foreach \i in {1}{\foreach \j in {1}{
 \draw [fill = blue, opacity = 0.15] (\i*\spacing,\j*\spacing) rectangle (\i*\spacing + \spacing,\j*\spacing + \spacing);
}}
\node at (1.5*\spacing, 1.5*\spacing) {\color{red}\small$f_2$};
\end{tikzpicture}
}}  \quad \rightarrow \quad
\vcenter{\hbox{
\begin{tikzpicture}
\newcommand{\spacing}{0.5}
\foreach \i in {1}{\foreach \j in {0,2}{
 \draw [fill = red, opacity = 0.1] (\i*\spacing,\j*\spacing) rectangle (\i*\spacing + \spacing,\j*\spacing + \spacing);
 \draw [fill = red, opacity = 0.1] (\j*\spacing,\i*\spacing) rectangle (\j*\spacing + \spacing,\i*\spacing + \spacing);
}}
\foreach \i in {0,2}{\foreach \j in {0,2}{
 \draw [fill = blue, opacity = 0.15] (\i*\spacing,\j*\spacing) rectangle (\i*\spacing + \spacing,\j*\spacing + \spacing);
}}
\foreach \i in {1}{\foreach \j in {1}{
 \draw [fill = blue, opacity = 0.15] (\i*\spacing,\j*\spacing) rectangle (\i*\spacing + \spacing,\j*\spacing + \spacing);
}}
\node at (1.5*\spacing, 1.5*\spacing) {\color{red}\small$f_1$};
\node at (2*\spacing, 2*\spacing) {\color{red}\small$Z$};
\node at (2*\spacing, 1*\spacing) {\color{red}\small$Z$};
\end{tikzpicture}
}}
\end{equation*}

The primitive $Z$ 2-faults\replace{errors} above correspond to edges between the decoding graph vertices that represent the high-weight detectors associated to the neighboring $X$-plaquettes. 
The primitive $f_1$ 1-fault\replace{error} corresponds to a ``dangling edge,'' in the graph as it only connects to one single vertex, i.e. the low-weight detector. 
We represent such dangling edges as dots in the pictures below. In terms of edges, the three primitive faults\replace{errors} are represented as
\begin{equation*}
\vcenter{\hbox{
\begin{tikzpicture}
\newcommand{\spacing}{0.5}
\foreach \i in {1}{\foreach \j in {0,2}{
 \draw [fill = red, opacity = 0.1] (\i*\spacing,\j*\spacing) rectangle (\i*\spacing + \spacing,\j*\spacing + \spacing);
 \draw [fill = red, opacity = 0.1] (\j*\spacing,\i*\spacing) rectangle (\j*\spacing + \spacing,\i*\spacing + \spacing);
}}
\foreach \i in {0,2}{\foreach \j in {0,2}{
 \draw [fill = blue, opacity = 0.15] (\i*\spacing,\j*\spacing) rectangle (\i*\spacing + \spacing,\j*\spacing + \spacing);
}}
\foreach \i in {1}{\foreach \j in {1}{
 \draw [fill = blue, opacity = 0.15] (\i*\spacing,\j*\spacing) rectangle (\i*\spacing + \spacing,\j*\spacing + \spacing);
}}
\node at (1.5*\spacing, 1.5*\spacing) {\color{red}\small$f_1$};
\node at (2*\spacing, 2*\spacing) {\color{red}\small$Z$};
\node at (2*\spacing, 1*\spacing) {\color{red}\small$Z$};
\end{tikzpicture}
}}
\quad = \quad 
\vcenter{\hbox{
\begin{tikzpicture}
\newcommand{\spacing}{0.5}
\foreach \i in {1}{\foreach \j in {0,2}{
 \draw [fill = red, opacity = 0.1] (\i*\spacing,\j*\spacing) rectangle (\i*\spacing + \spacing,\j*\spacing + \spacing);
 \draw [fill = red, opacity = 0.1] (\j*\spacing,\i*\spacing) rectangle (\j*\spacing + \spacing,\i*\spacing + \spacing);
}}
\foreach \i in {0,2}{\foreach \j in {0,2}{
 \draw [fill = blue, opacity = 0.15] (\i*\spacing,\j*\spacing) rectangle (\i*\spacing + \spacing,\j*\spacing + \spacing);
}}
\foreach \i in {1}{\foreach \j in {1}{
 \draw [fill = blue, opacity = 0.15] (\i*\spacing,\j*\spacing) rectangle (\i*\spacing + \spacing,\j*\spacing + \spacing);
}}
\node at (1.5*\spacing, 1.5*\spacing) {\color{red}\textbullet};
\draw[thick, red] (1.5*\spacing,0.5*\spacing) -- (2.5*\spacing, 1.5*\spacing); 
\draw[thick, red] (1.5*\spacing,2.5*\spacing) -- (2.5*\spacing, 1.5*\spacing); 
\end{tikzpicture}
}}
\end{equation*}

In the presence of multiple non-primitive faults\replace{errors} of this type, the matching can fail. Consider the following edges representing two $f_2$ faults\replace{errors} on a 6 by 6 torus (left), and another, logically inequivalent edge configuration (right) with the same syndrome:
\begin{equation*}\vcenter{\hbox{
\begin{tikzpicture}
\newcommand{\spacing}{0.3}
\foreach \i in {1,3,5}{\foreach \j in {0,2,4}{
 \draw [fill = red, opacity = 0.1] (\i*\spacing,\j*\spacing) rectangle (\i*\spacing + \spacing,\j*\spacing + \spacing);
 \draw [fill = red, opacity = 0.1] (\j*\spacing,\i*\spacing) rectangle (\j*\spacing + \spacing,\i*\spacing + \spacing);
}}
\foreach \i in {0,2,4}{\foreach \j in {0,2,4}{
 \draw [fill = blue, opacity = 0.15] (\i*\spacing,\j*\spacing) rectangle (\i*\spacing + \spacing,\j*\spacing + \spacing);
}}
\foreach \i in {1,3,5}{\foreach \j in {1,3,5}{
 \draw [fill = blue, opacity = 0.15] (\i*\spacing,\j*\spacing) rectangle (\i*\spacing + \spacing,\j*\spacing + \spacing);
}}
\node at (1.5*\spacing, 1.5*\spacing) {\color{red}\textbullet};
\draw[thick, red] (1.5*\spacing,0.5*\spacing) -- (2.5*\spacing, 1.5*\spacing); 
\draw[thick, red] (1.5*\spacing,2.5*\spacing) -- (2.5*\spacing, 1.5*\spacing); 
\node at (1.5*\spacing, 5.5*\spacing) {\color{red}\textbullet};
\draw[thick, red] (1.5*\spacing,4.5*\spacing) -- (2.5*\spacing, 5.5*\spacing); 
\draw[thick, red] (2*\spacing,6*\spacing) -- (2.5*\spacing, 5.5*\spacing); 
\draw[thick, red, dotted] (2*\spacing,6*\spacing) -- (1.5*\spacing, 6.5*\spacing); 
\draw[thick, red] (2*\spacing,0*\spacing) -- (1.5*\spacing,0.5*\spacing); 
\draw[thick, red, dotted] (2.5*\spacing,-0.5*\spacing) -- (2*\spacing,0*\spacing); 
\end{tikzpicture}}}
\quad  \not\sim   \quad \vcenter{\hbox{
\begin{tikzpicture}
\newcommand{\spacing}{0.3}
\foreach \i in {1,3,5}{\foreach \j in {0,2,4}{
 \draw [fill = red, opacity = 0.1] (\i*\spacing,\j*\spacing) rectangle (\i*\spacing + \spacing,\j*\spacing + \spacing);
 \draw [fill = red, opacity = 0.1] (\j*\spacing,\i*\spacing) rectangle (\j*\spacing + \spacing,\i*\spacing + \spacing);
}}
\foreach \i in {0,2,4}{\foreach \j in {0,2,4}{
 \draw [fill = blue, opacity = 0.15] (\i*\spacing,\j*\spacing) rectangle (\i*\spacing + \spacing,\j*\spacing + \spacing);
}}
\foreach \i in {1,3,5}{\foreach \j in {1,3,5}{
 \draw [fill = blue, opacity = 0.15] (\i*\spacing,\j*\spacing) rectangle (\i*\spacing + \spacing,\j*\spacing + \spacing);
}}
\node at (1.5*\spacing, 1.5*\spacing) {\color{red}\textbullet};
\draw[thick, red] (1.5*\spacing,2.5*\spacing) -- (2.5*\spacing, 3.5*\spacing); 
\node at (1.5*\spacing, 5.5*\spacing) {\color{red}\textbullet};
\draw[thick, red] (1.5*\spacing,4.5*\spacing) -- (2.5*\spacing, 3.5*\spacing); 
\end{tikzpicture}}}
\end{equation*}

As non-primitive $f_2$ faults\replace{errors} appear on \emph{all} plaquettes in the set of generative fault\replace{error} configurations $\mathcal{F}$, all non-dangling edges in the above picture by symmetry come with the same weight. Therefore, the minimum weight matching prioritizes the path with fewer edges, and the decoding induces a logical error. To get around this mismatch, we temporarily assign weight zero to all non-dangling edges that correspond a split $f_2$ fault\replace{error} (meaning that they come for free) whenever the dangling edge is ``lit up'' by a fault configuration\replace{an error chain}, i.e. whenever the corresponding low-weight detector is triggered. With this temporary re-weighting, the matching will not penalize the longer path and the decoding succeeds.

Other faults need to be treated in the same fashion. 
When constructing the decoding graph for the hook-preventing circuits, we first identify the faults whose splitting will require dynamical re-weighting.
These correspond to faults of the form $(I,1)$, i.e. pure readout errors, that light up three detectors.
These faults are split into three primitive faults\replace{errors}.
One of these primitive faults\replace{errors} in the splitting is a 1-fault that triggers only a single detector $v_j$.
The other two primitive faults in the splitting are 2-faults, and during decoding the corresponding two edges in the decoding graph are assigned weight zero whenever $j$ is present in the observed syndrome.

\subsection{Results}
\label{sec:performance_results}

\begin{figure}
    \centering
    \includegraphics[width=0.48\textwidth]{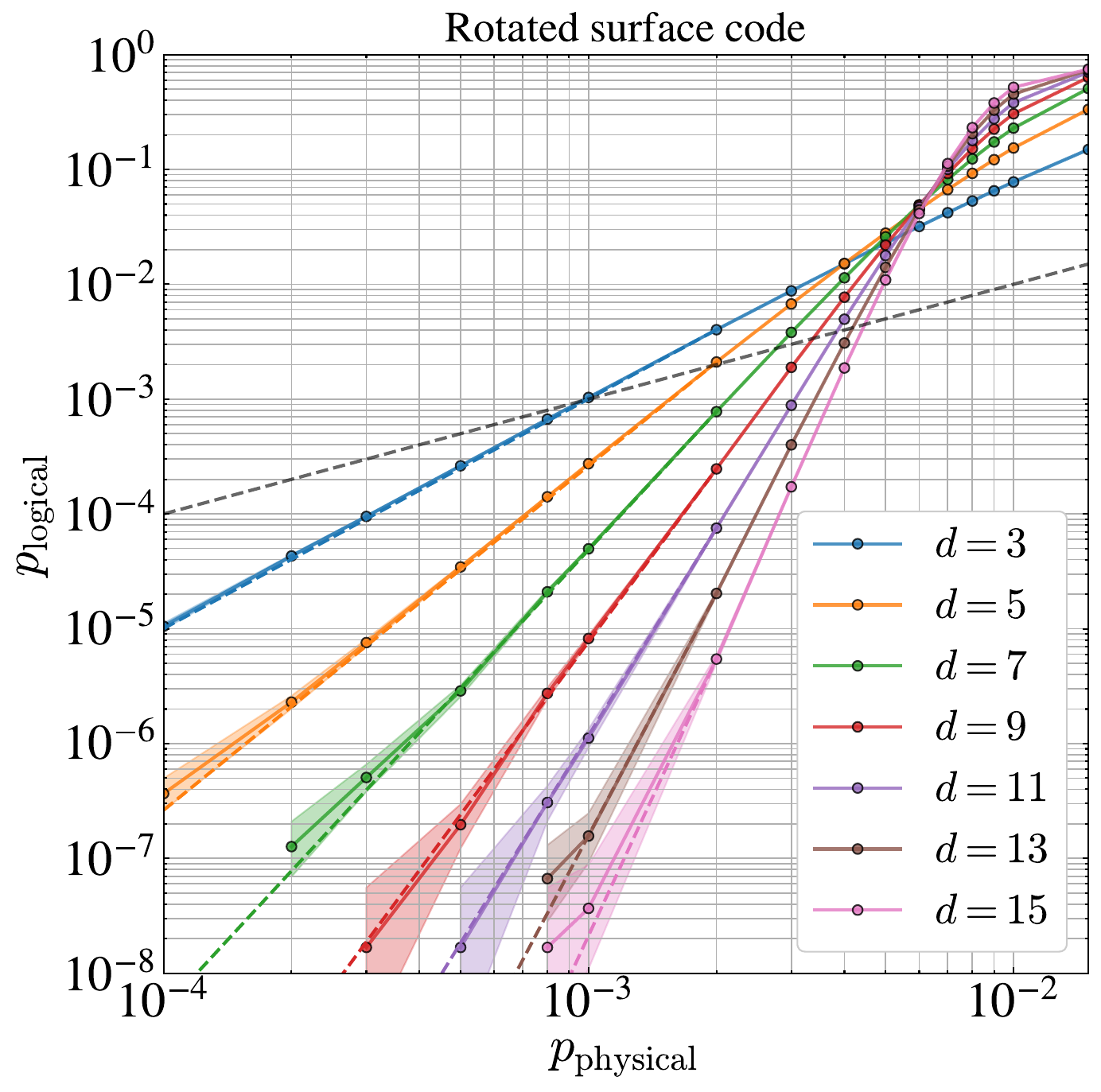}
    \hspace{0.1in}
    \includegraphics[width=0.48\textwidth]{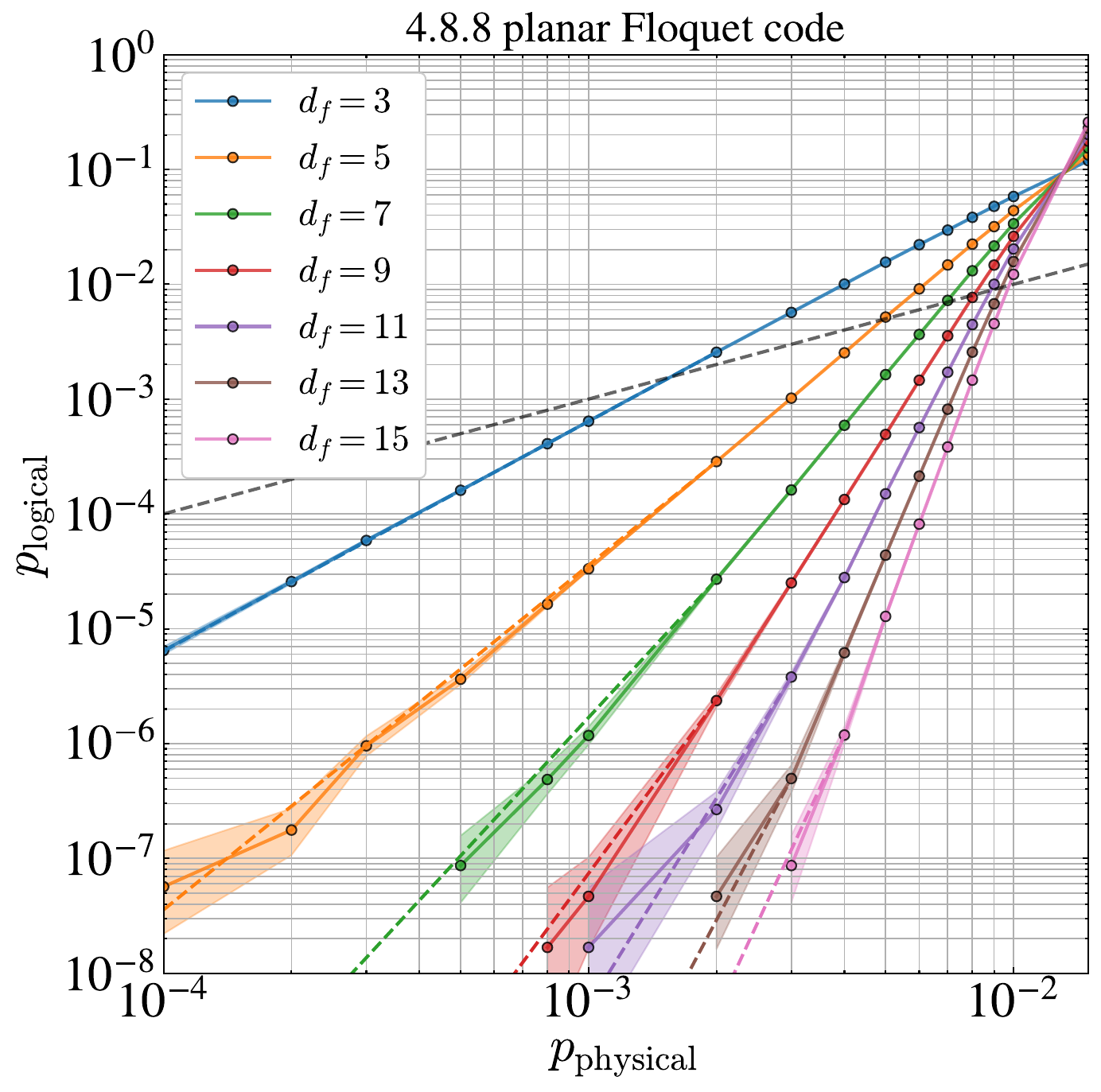}
    \caption{(left) Performance results for our pairwise measurement-based surface code realization described in Sec.~\ref{sec:Circuits} on a rotated surface code patch using boundary conditions that make the hook errors benign (see Sec.~\ref{sec:boundaries}).
    (right) Performance results for the 4.8.8 Floquet code on a planar patch with rectangular boundary conditions (as in Ref.~\protect\onlinecite{Paetznick2023}).
    For implementation in Majorana hardware, these require double-rail semiconductor layouts.
    Fault-tolerance thresholds are found to be $0.66\%$ for our code and  $1.3\%$ for the 4.8.8 Floquet code.
    }
    \label{fig:raw_DR}
\end{figure}

In this section, we analyze the performance of our code and compare it directly with that of the 4.8.8 Floquet code, which represents the state of the art for pairwise measurement-based codes optimized with respect to Majorana hardware.
Files providing computer parsable descriptions of the circuits used in our simulations are provided in the supplementary material.
We estimate the logical failure rates by running Monte Carlo simulations with up to $10^8$ trials for a series of increasing code sizes and a fixed set of physical error rates $10^{-4} \leq p_\text{physical} \leq 1.5\times 10^{-2}$.  
At sufficiently large code size and low $p_\text{physical}$, we observe no failures in the $10^8$ trials and, thus, do not include the corresponding point in the plots.
The shaded regions in performance plots [Figs.~\ref{fig:raw_DR}, \ref{fig:raw_SR}, \ref{fig:raw_hook_malignant}, and \ref{fig:raw_hook_preventing}] represent 95\% credible intervals of a posterior beta distribution given the observed number of logical failures and completed trials, assuming a uniform prior distribution for the logical error rate; the points represent the median of this posterior distribution.
In this paper, we extract threshold values by finding the intersection of the two largest simulated code sizes via a linear interpolation \emph{in log-log space} of the obtained $(p_\mathrm{physical}, p_\mathrm{logical})$ data.
We ignore sampling uncertainties in these threshold estimates, since sampling error is generally negligible (e.g., occurring in the 4th significant figure in the $p_\mathrm{logical}$ estimates) at the relatively high error rates encountered near threshold.
There are, however, systematic uncertainties due to finite code size and our employed interpolation procedure.
We did not attempt to estimate error bars for the threshold values, since the difference of noise model used from a realistic noise model undoubtedly results in more significant deviations.
Note that we expect the finite size effects to cause \emph{underestimation} of the threshold.
For a more rigorous threshold estimation procedure, see, e.g., Ref.~\onlinecite{bombinlogical2023}.
We extract ``pseudo-thresholds'' for a particular system size by locating the error rate where $p_\mathrm{logical} = p_\mathrm{physical}$, again obtained via linear interpolation in log-log space.

We first compare the performance for our original circuits and pipelining (as described in Sec.~\ref{sec:Circuits}) on a rotated surface code patch using boundary conditions that make the hook errors benign (as explained in Sec.~\ref{sec:boundaries}) with the performance of the 4.8.8 Floquet code on a planar patch with rectangular boundary conditions (as in Ref.~\onlinecite{Paetznick2023}).
For implementation in Majorana hardware, these both correspond to realizations that would require double-rail semiconductor layouts.
The performance results for these two cases are shown in Fig.~\ref{fig:raw_DR}.
Notably, we find a fault-tolerance threshold of $0.66\%$ for our code and $1.3\%$ for the 4.8.8 Floquet code.\footnote{The improvement of the threshold value for 4.8.8 Floquet code compared to the previous simulations of Ref.~\onlinecite{Paetznick2023} are likely due to some combination of (a) our use of an improved decoder, (b) larger considered code sizes, and (c) improved sampling statistics.} 
For the smallest code size ($d_{\rm f}=3$), the pseudo-thresholds are comparable for the two codes at approximately $0.096\%$ for the surface code and $0.16\%$ for 4.8.8 Floquet code.

\begin{figure}
    \centering
    \includegraphics[width=\textwidth]{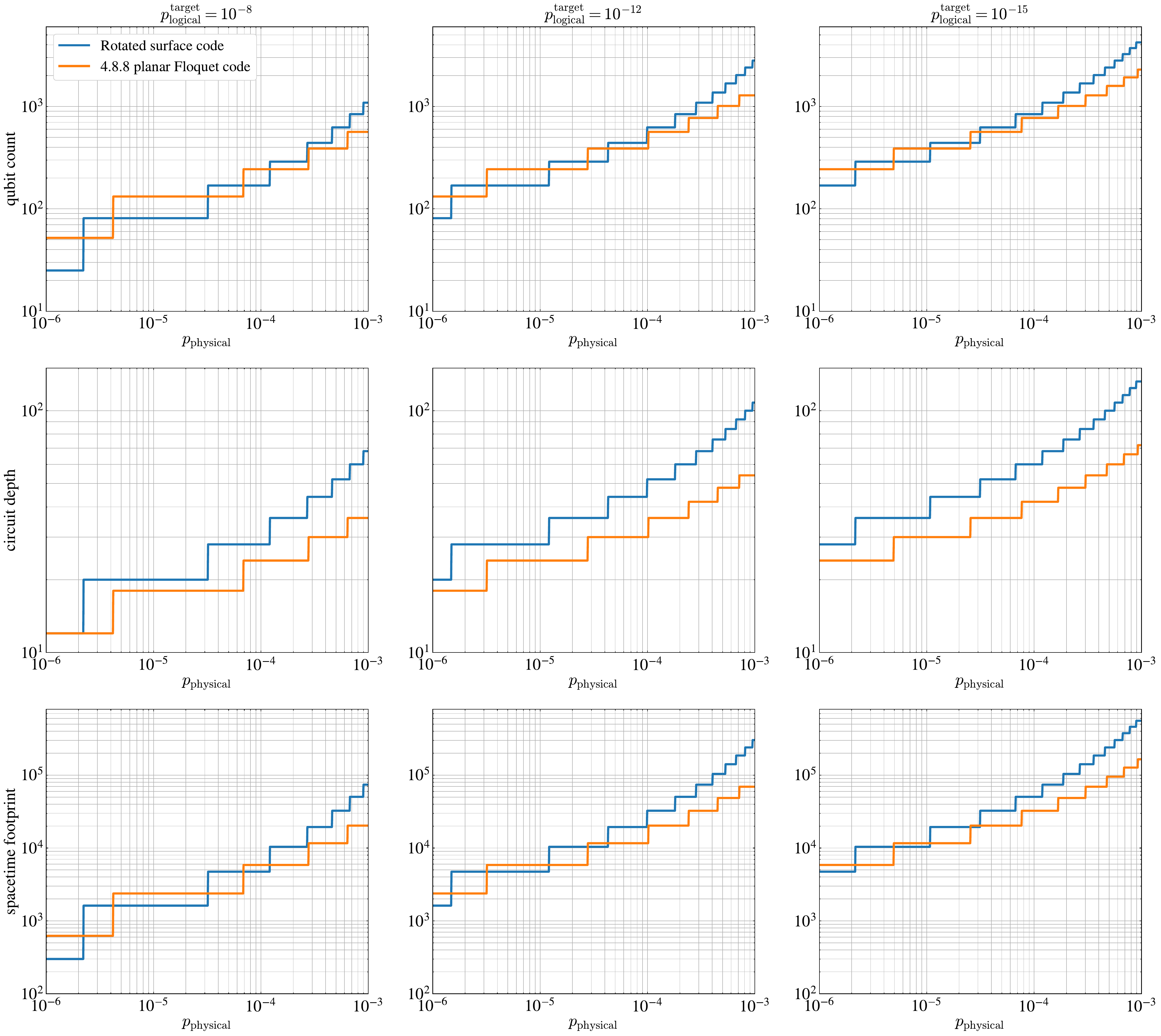}
    \caption{Resource requirements to reach target logical error rates of $p_\mathrm{logical}^\mathrm{target} = 10^{-8}$, $10^{-12}$, and $10^{-15}$ per $d_{\rm f}$ rounds, comparing our realization of the surface code and the 4.8.8 planar Floquet code. The respective target logical error rates correspond to columns from left to right, and the resource quantities being considered correspond to rows, from top to bottom: qubit count, circuit depth, and spacetime footprint (i.e., qubit count times circuit depth). 
    }
    \label{fig:footprint_DR}
\end{figure}

When evaluating code performance, it is helpful to go beyond the threshold and consider the resource requirements for obtaining a given target logical error rate $p_\mathrm{logical}^\mathrm{target}$ for the simulated memory experiment.
In Fig.~\ref{fig:footprint_DR}, we plot the qubit count, circuit depth, and spacetime footprint (i.e., qubit count multiplied by circuit depth) required to achieve logical error rates of $p_\mathrm{logical}^\mathrm{target} = 10^{-8}$, $10^{-12}$, and $10^{-15}$ per $d_{\rm f}$ rounds of syndrome measurement, for physical error rates in the range $10^{-6} \leq p_\text{physical} \leq 10^{-3}$.
In Table~\ref{tab:double_rail_d}, we list the corresponding fault distances $d_{\rm f}$ required to reach these target logical error rates for physical error rates $p_{\rm physical} = 10^{-6}$, $10^{-5}$, $10^{-4}$, and $10^{-3}$. Within the context of this simplified error model, we see that our surface code is competitive in terms of these resource requirements, especially at error rates $p_\mathrm{physical} \lesssim 10^{-4}$, and it has substantially narrowed the gap from the original requirements of the pairwise measurement-based surface codes in Ref.~\onlinecite{Chao2020} at higher error rates $p_\mathrm{physical} \approx 10^{-3}$, where the Floquet code previously enjoyed a nearly two full orders of magnitude reduction in spacetime footprint at $p_\mathrm{logical}^\mathrm{target} = 10^{-12}$ [see Fig.~9 of Ref.~\onlinecite{Paetznick2023}].

To obtain these resource estimates, we have taken the following approach. 
For each code and code size, we expect the low $p_\mathrm{physical}$ behavior to be dominated by circuit noise faults of weight $(d_{\rm f} + 1)/2$~\cite{fowlersurface2012}. 
Rather than assuming that this form persists all the way to $p_\mathrm{physical}$ on the order of the threshold, we select, for each code size, a characteristic reference point $(p_\mathrm{physical}^\mathrm{ref}, p_\mathrm{logical}^\mathrm{ref})$ in the sub-threshold regime of the empirical data and assume that for $p_\mathrm{physical} \leq p_\mathrm{physical}^\mathrm{ref}$, the logical error rate is governed by the form
\begin{equation}
p_\mathrm{logical} = p_\mathrm{logical}^\mathrm{ref}\left(\frac{p_\mathrm{physical}}{p_\mathrm{physical}^\mathrm{ref}}\right)^{(d_{\rm f}+1)/2}
.
\label{eq:p_log_ref}
\end{equation} 
The colored dashed lines in the performance plots represent these chosen sub-threshold estimates of the logical error rate.\footnote{In the plots, a given dashed line terminates at the chosen reference point $(p_\mathrm{physical}^\mathrm{ref}, p_\mathrm{logical}^\mathrm{ref})$. For $p_\mathrm{logical}^\mathrm{ref}$, we use the median of the posterior beta distribution obtained for $p_\mathrm{logical}$.}
To estimate the fault distance required to hit a prescribed $p_\mathrm{logical}^\mathrm{target}$ at a given $p_\mathrm{physical}$, we first fit the values of these obtained scaling forms for the sub-threshold logical error rate for all simulated code sizes $d_{\rm f} > 3$ to the exponential form
\begin{equation}
p_\mathrm{logical}(p_\mathrm{physical}, d_{\rm f}) = \alpha(p_\mathrm{physical}) e^{-\beta(p_\mathrm{physical}) d_{\rm f}}
.
\end{equation}
The fits thereby obtained are of very high quality in the considered window $10^{-6} \leq p_\text{physical} \leq 10^{-3}$, justifying the approach \emph{a posteriori}.
Finally, we determine the smallest (odd) $d_{\rm f} = d_{\rm f}^\mathrm{target}$ necessary for the fitted exponential form to predict $p_\mathrm{logical} \leq p_\mathrm{logical}^\mathrm{target}$. 
The actual footprints can then be read off from the circuit corresponding to the required fault distance $d_{\rm f}$ as follows. The physical qubit count for our code on a rotated surface code patch with hook-benign boundary conditions (i.e. $d_{\rm f}=d$) is $N=4d^2-4d+1$ and the circuit depth is counted as $4d$.
For the 4.8.8 Floquet code on a patch with rectangular boundary conditions~\cite{Paetznick2023}, the qubit count is $N= 4d_{\rm f}^2 + 8(d_{\rm f}-1)$ and the circuit depth is counted as $6 \lceil d_{\rm f}/2 \rceil$.

\begin{table}[]
    \centering
\begin{tabular}{c c c c c c c}
 \multicolumn{1}{c}{\vphantom{\large $ p_{\rm logical}^{\rm target}$ }}   &\multicolumn{2}{c}{$p_{\rm logical}^{\rm target} = 10^{-8}$} & \multicolumn{2}{c}{$p_{\rm logical}^{\rm target} = 10^{-12}$} & \multicolumn{2}{c}{$p_{\rm logical}^{\rm target} = 10^{-15}$} \\ 
\cline{2-7}
 \multicolumn{1}{c}{ $p_{\rm physical}$}   &  $\;\;$SC$\;\;$ & 4.8.8 &  $\;\;$SC$\;\;$  & 4.8.8  & $\;\;$SC$\;\;$ & 4.8.8 \\
    \hline
\multicolumn{1}{c}{$10^{-6}$ } &  3   &   3  &  5   &   5  &  7   &   7   \\
\multicolumn{1}{c}{$10^{-5}$}  &  5   &  5   &  7   &   7  &  9   &   9   \\
\multicolumn{1}{c}{$10^{-4}$}  &  7   &  7   &  13   &   9  &  15   &   13   \\
\multicolumn{1}{c}{$10^{-3}$}  &  17   &  11   &  27   &  17   &  33   &   23   \\
\hline
\end{tabular}
    \caption{Fault distance $d_{\rm f}$ required to reach a target logical error rate $p_{\rm logical}^{\rm target}$ of $10^{-8}$, $10^{-12}$ and $10^{-15}$, respectively, comparing our realization of the surface code and the 4.8.8 planar Floquet code.
    The determination of the required $d_{\rm f}$ is the same as in Fig.~\ref{fig:footprint_DR}.
    (The total number of qubits is $N=O( 4d_{\rm f}^2)$ for both of these codes.)}
    \label{tab:double_rail_d}
\end{table}

As we are interested in implementation of these codes in Majorana hardware, where single-rail semiconductor layouts may be strongly preferable to double-rail layouts, we repeat the above analysis for single-rail variants of the two codes.
For our measurement-based surface code realization, we use a modification of the circuits that is compatible with single-rail layouts, as described in Sec.~\ref{sec:Majorana}.
For the 4.8.8 Floquet code, we must split each step of the original circuits into two steps in such a way to avoid conflicts between measurement loops.
There are different ways to do this, but a convenient choice is to distribute half of the measurements of each type from each step into the resulting two steps.
We will not show further detail of the circuit for the single-rail variant of the 4.8.8 Floquet code, as the main point is simply that there are twice as many steps in the period and half the qubits are idle in each step.
Since the single-rail variants of both of these codes have steps with idle qubits, we now repeat the above analysis for these single-rail variants.
We note that the noise model we use potentially overestimates the relative error rates of idle faults as compared to measurement faults, so one can view the original results and the single-rail results as assessments at the endpoints of a range of possible idle noise rates.
(If we consider the single-rail code variants with the idle noise set to zero, we would obtain the original performance results.)

\begin{figure}
    \centering
    \includegraphics[width=0.48\textwidth]{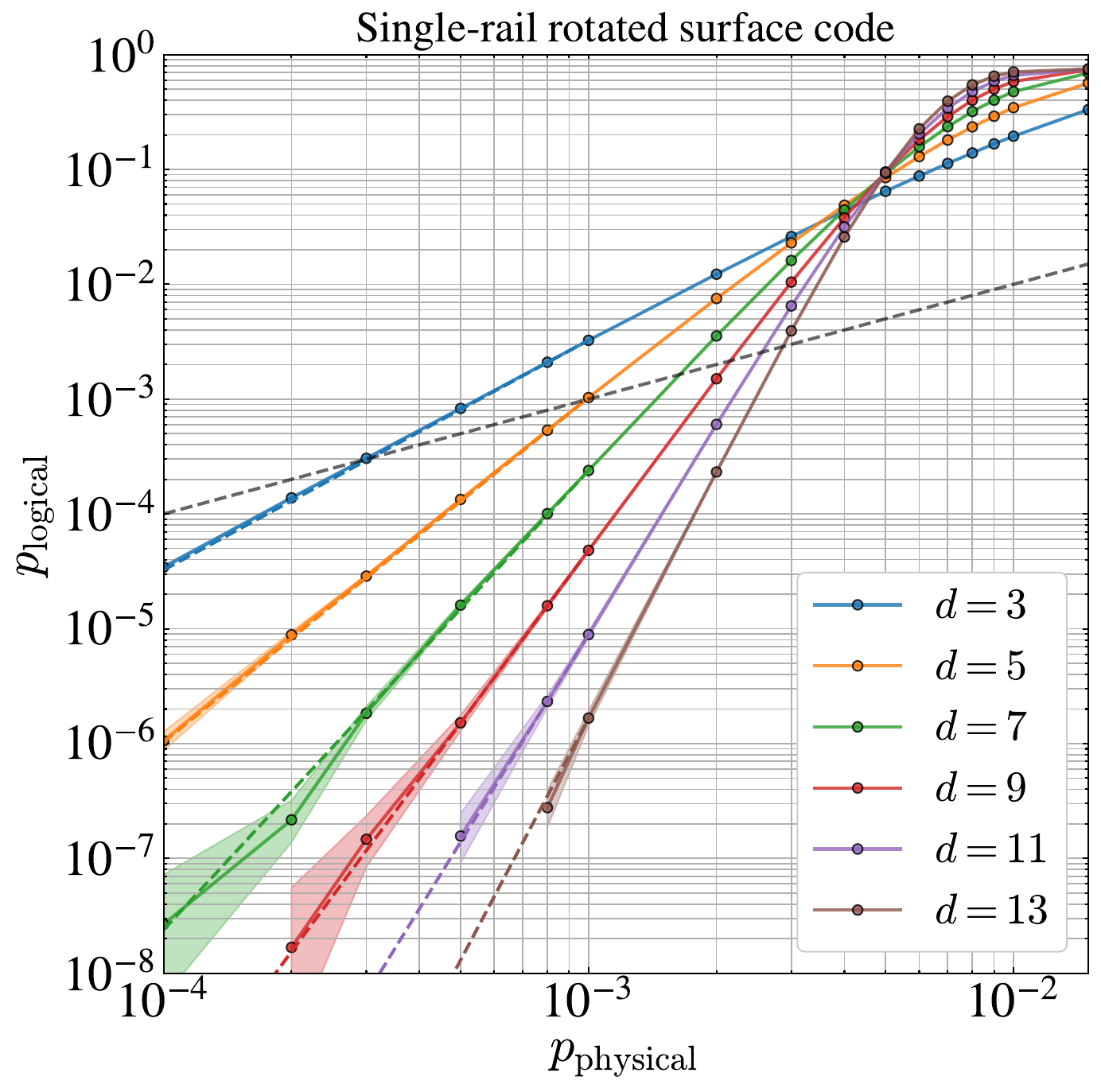}
    \hspace{0.1in}
    \includegraphics[width=0.48\textwidth]{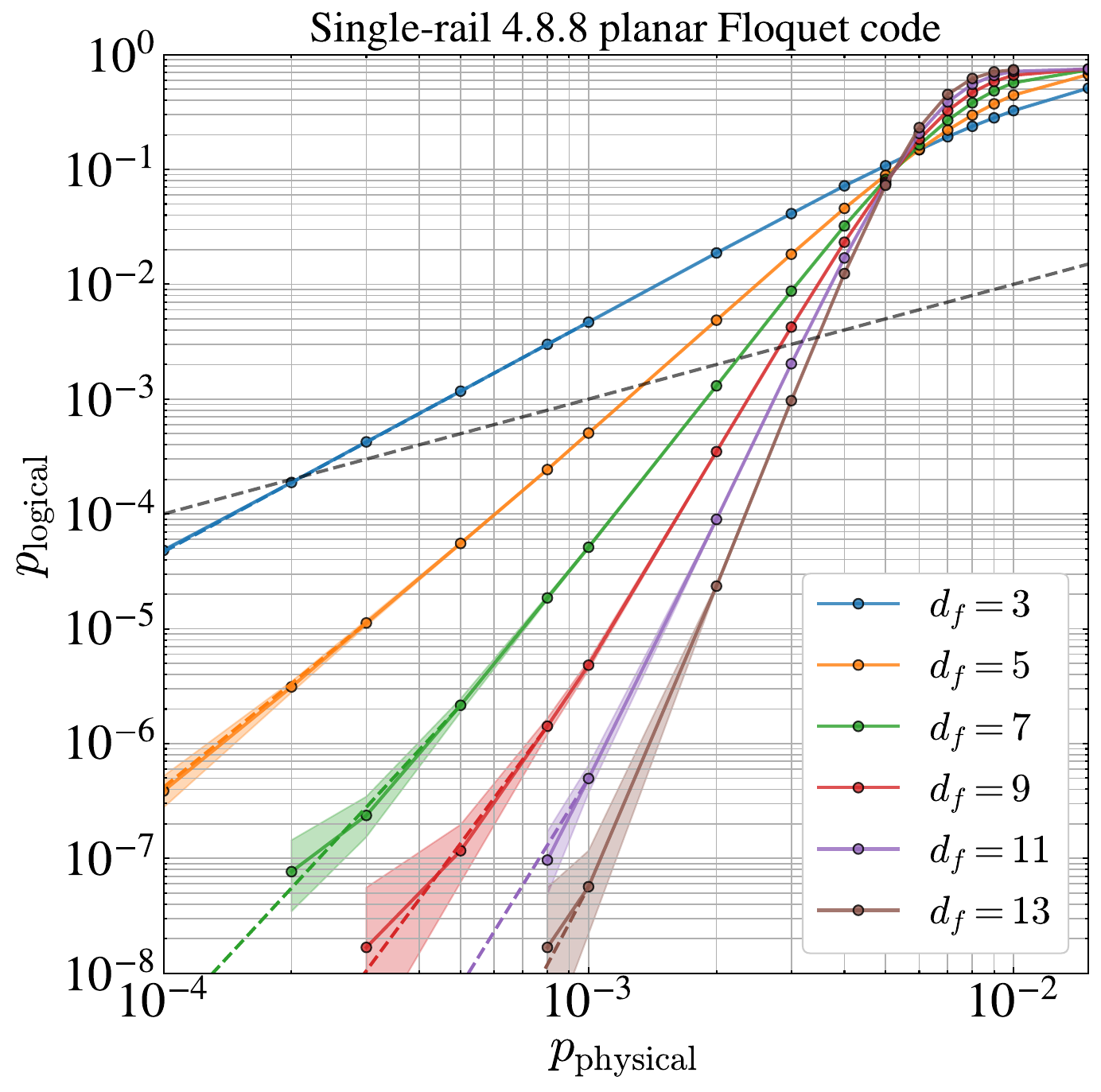}
    \caption{(left) Performance results for the single-rail variant of our pairwise measurement-based surface code realization described in Sec.~\ref{sec:Majorana} on a rotated surface code patch using boundary conditions that make the hook errors benign.
    (right) Performance results for the single-rail variant of 4.8.8 Floquet code on a planar patch with rectangular boundary conditions (see description in text).
    These variants can be implemented in Majorana hardware with single-rail semiconductor layouts.
    Fault-tolerance thresholds for these single-rail variants are found to be $0.51\%$ for our code and  $0.52\%$ for the 4.8.8 Floquet code.
    }
    \label{fig:raw_SR}
\end{figure}

\begin{figure}
    \centering
    \includegraphics[width=\textwidth]{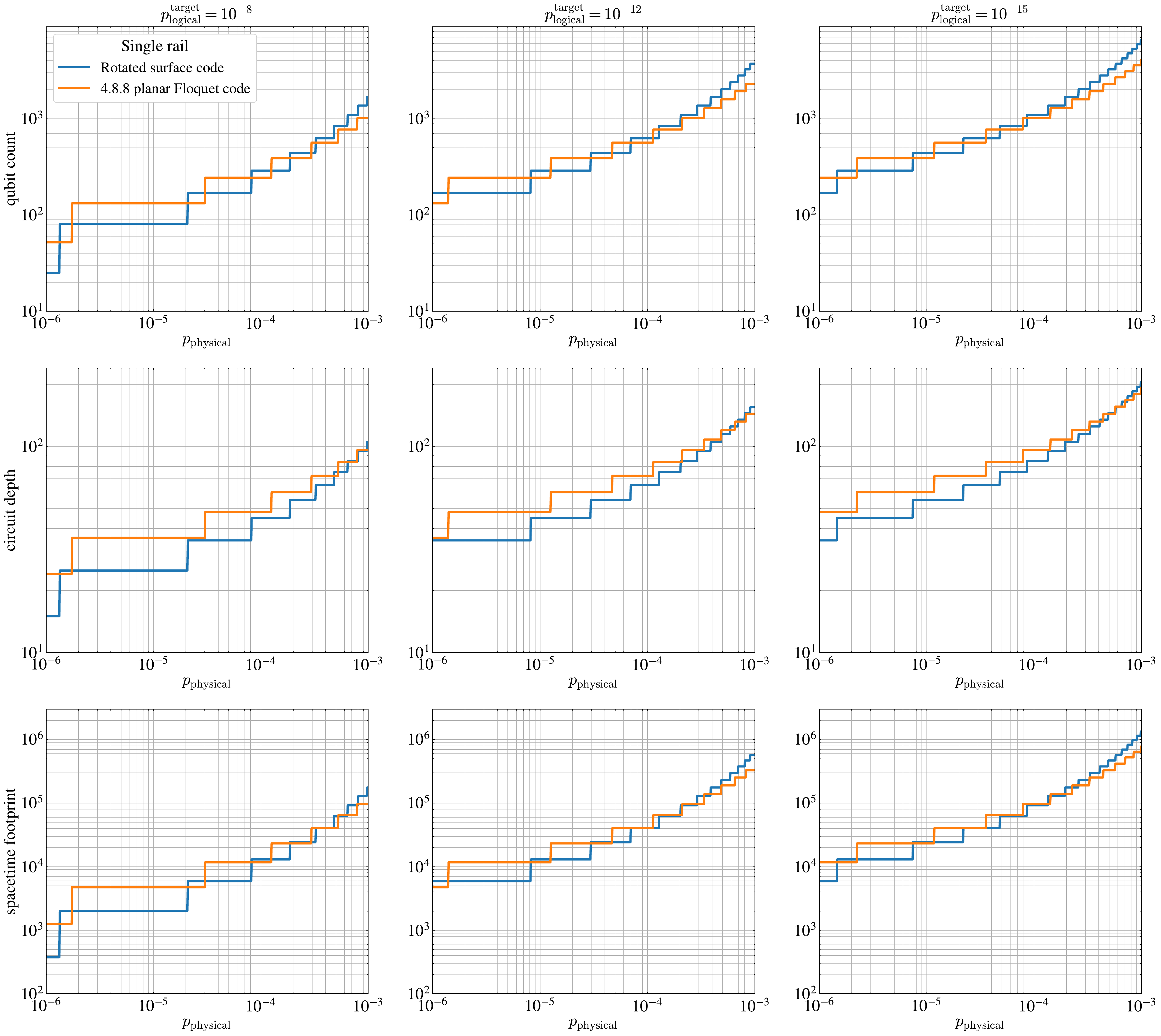}
    \caption{Resource requirements to reach target logical error rates of $p_\mathrm{logical}^\mathrm{target} = 10^{-8}$, $10^{-12}$, and $10^{-15}$, comparing single-rail variants of our realization of the surface code and the 4.8.8 planar Floquet code.
    }
    \label{fig:footprint_SR}
\end{figure}

\begin{figure}[]
    \centering
    \includegraphics[width=0.48\textwidth]{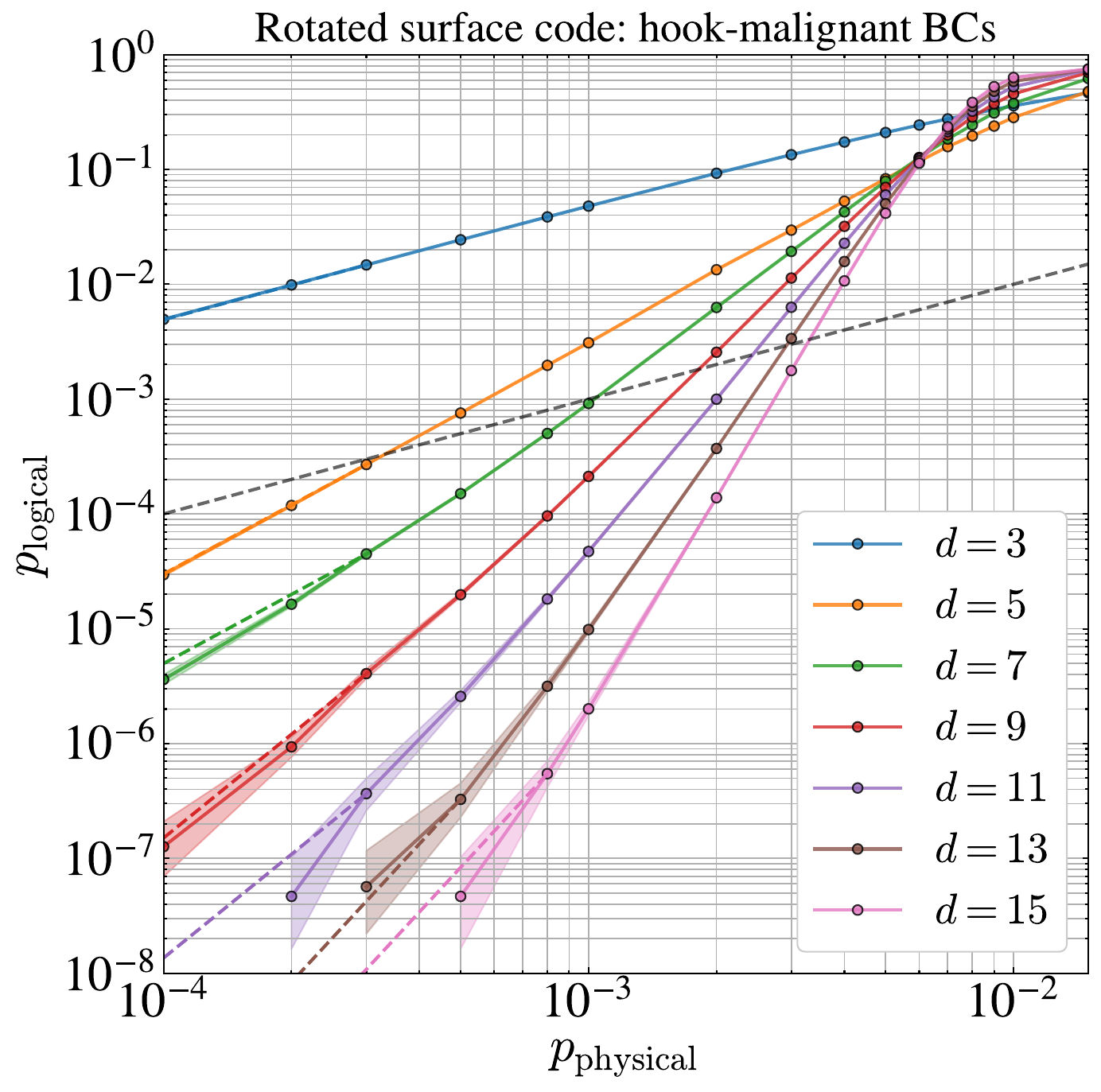}
    \hspace{0.1in}
    \includegraphics[width=0.48\textwidth]{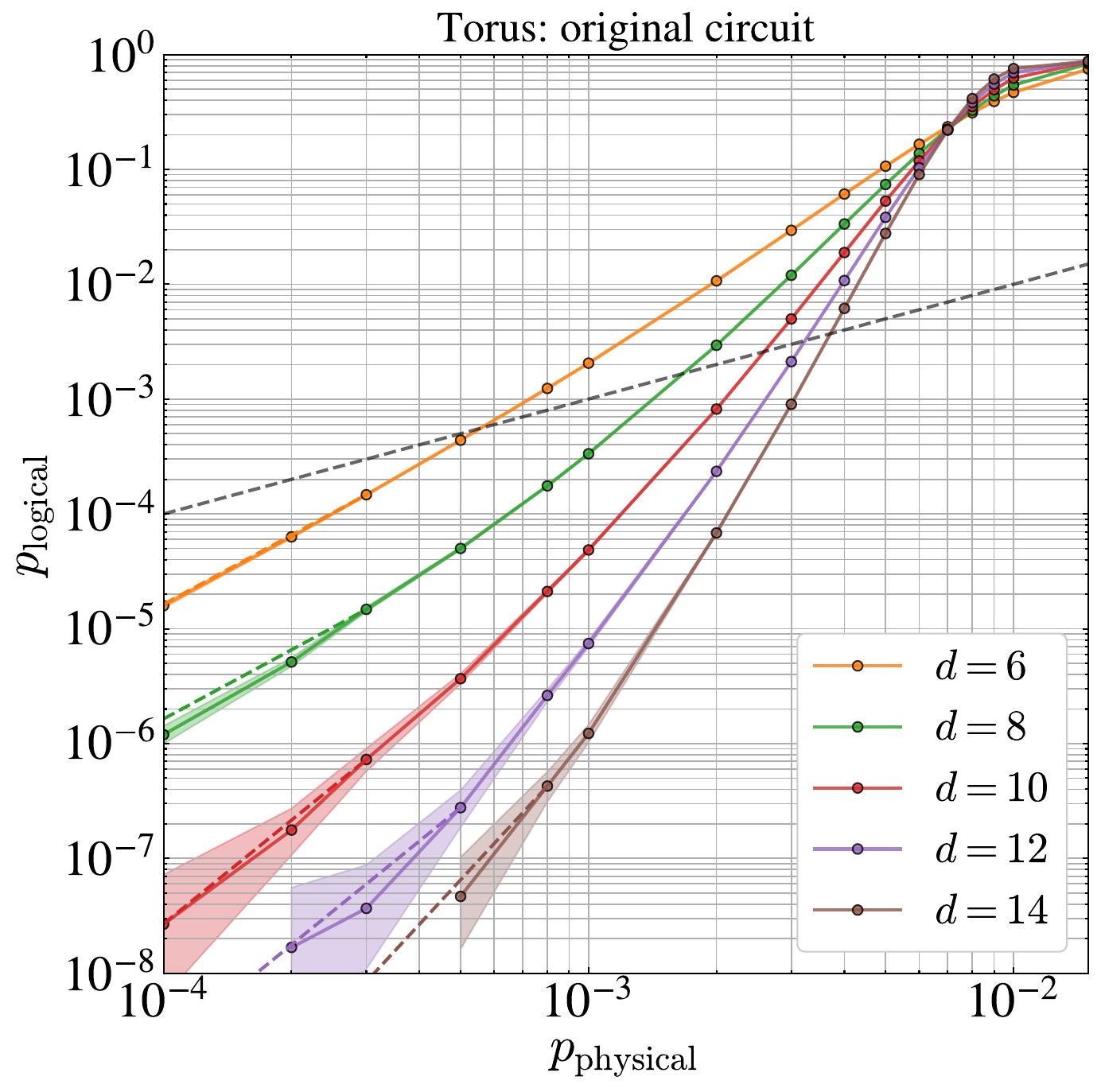}    
    \caption{(left) Performance results for our pairwise measurement-based surface code realization on a rotated surface code patch using boundary conditions that make the hook errors malignant (see Sec.~\ref{sec:boundaries}).
    We note that the $d=3$ curve is not expected to intersect with the other curves at (or near) threshold, as the code is not error-correcting at $d=3$, because the the fault distance is $d_{\rm f} = 2$.
    (right) Performance results for our code on a torus.
    Fault-tolerance thresholds are found to be $0.65\%$ for the planar patch and $0.70\%$ for the torus.
    }
    \label{fig:raw_hook_malignant}
\end{figure}

\begin{figure}[]
    \centering
    \includegraphics[width=0.48\textwidth]{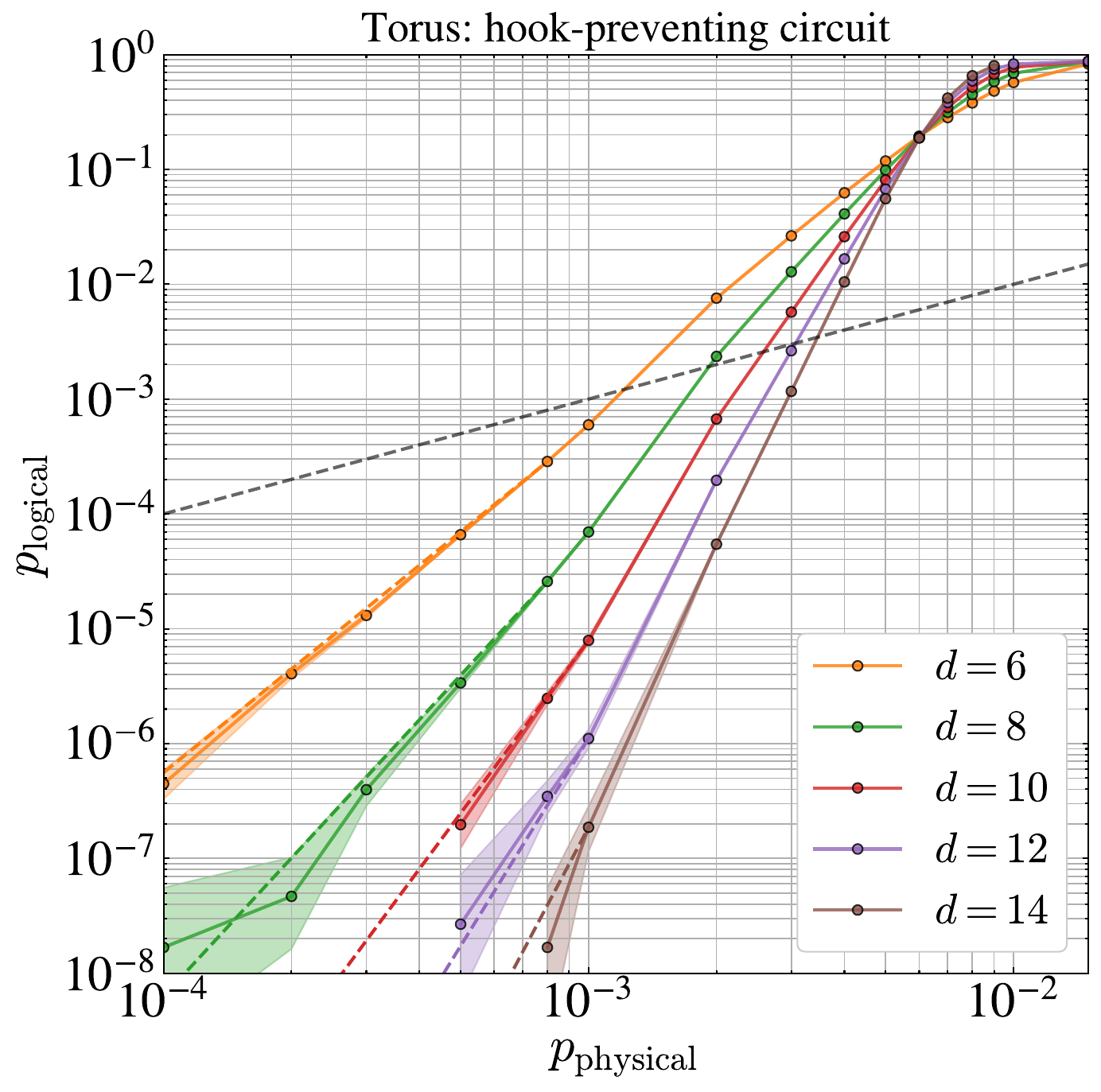}
    \hspace{0.1in}
    \includegraphics[width=0.48\textwidth]{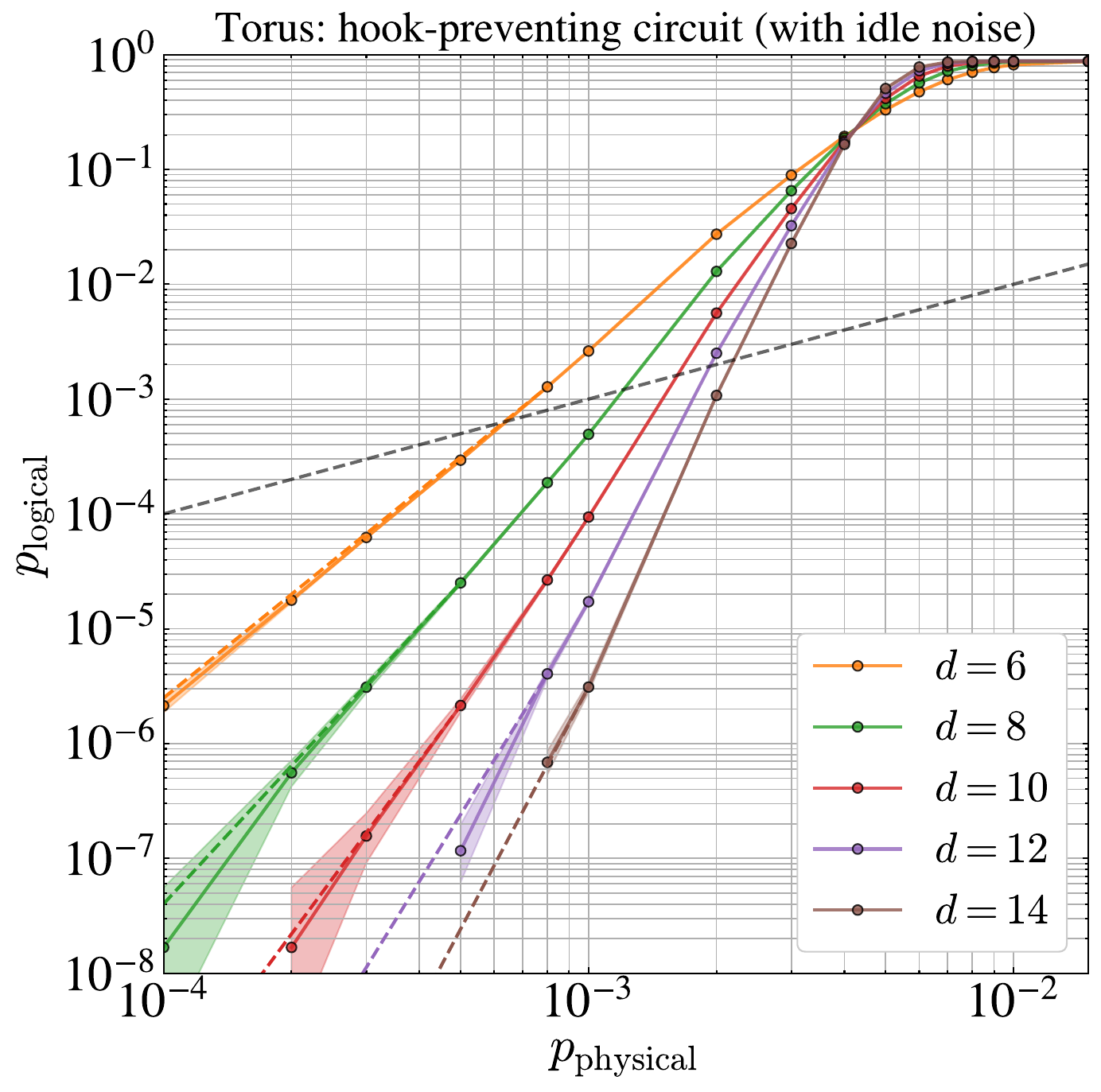}    
    \caption{Performance results for the hook-preventing variant of our pairwise measurement-based surface code realization described in Sec.~\ref{sec:Hook} on a torus for noise model without (left) and with (right) idle noise.
    The full code distance is recovered and performance is improved with respect to the code using the original circuits when hook errors are malignant (see Fig.~\ref{fig:raw_hook_malignant}). 
    Fault-tolerance thresholds are found to be $0.61\%$ where there are no idle errors and $0.43\%$ when idle errors are included.
    }
    \label{fig:raw_hook_preventing}
\end{figure}

Comparing the single-rail variants of these two codes, we find that our code becomes even more competitive.
From the performance results in Fig.~\ref{fig:raw_SR}, we find much closer fault-tolerance thresholds of $0.51\%$ and $0.52\%$ for our surface code and the 4.8.8 planar Floquet code, respectively.
We note that these decreases from the original thresholds are, respectively, in rough agreement with a $4/5$ and $1/2$ decrease that one might na\"ively anticipate from the syndrome extraction period increases of four to five steps for the surface code and three to six for the 4.8.8 Floquet code, with extra idle noise on introduced for each additional step.  
For the smallest code size ($d_{\rm f}=3$), the pseudo-thresholds are again comparable, though now favoring the surface code at approximately $0.03\%$ for the surface code and $0.02\%$ for 4.8.8 Floquet code.
In Fig.~\ref{fig:footprint_SR}, we plot the qubit count, circuit depth, and spacetime footprint resource estimates required to achieve logical error rates of $p_\mathrm{logical}^\mathrm{target} = 10^{-8}$, $10^{-12}$, and $10^{-15}$ for physical error rates in the range $10^{-6} \leq p_\text{physical} \leq 10^{-3}$.
In Table~\ref{tab:single_rail_d}, we list the corresponding fault distances $d_{\rm f}$ required to reach these target logical error rates for physical error rates $p_{\rm physical} = 10^{-6}$, $10^{-5}$, $10^{-4}$, and $10^{-3}$.

\begin{table}[]
    \centering
\begin{tabular}{c c c c c c c}
 \multicolumn{1}{c}{\vphantom{\large $ p_{\rm logical}^{\rm target}$ }}   &\multicolumn{2}{c}{$p_{\rm logical}^{\rm target} = 10^{-8}$} & \multicolumn{2}{c}{$p_{\rm logical}^{\rm target} = 10^{-12}$} & \multicolumn{2}{c}{$p_{\rm logical}^{\rm target} = 10^{-15}$} \\ 
\cline{2-7}
 \multicolumn{1}{c}{ $p_{\rm physical}$}   &  $\;\;$SC$\;\;$ & 4.8.8 &  $\;\;$SC$\;\;$  & 4.8.8  &  $\;\;$SC$\;\;$  & 4.8.8 \\
    \hline
\multicolumn{1}{c}{$10^{-6}$ } &  3   &   3    &  7   &   5   &  7   &   7   \\
\multicolumn{1}{c}{$10^{-5}$}  &  5   &  5     &  9   &   7   &  11   &   9   \\
\multicolumn{1}{c}{$10^{-4}$}  &  9   &  7     &  13   &   11  &  17   &   15   \\
\multicolumn{1}{c}{$10^{-3}$}  &  21   &  15   &  31   &  23   &  41   &   29   \\
\hline
\end{tabular}
    \caption{Fault distance $d_{\rm f}$ required to reach a target logical error rate $p_{\rm logical}^{\rm target}$ of $10^{-8}$, $10^{-12}$ and $10^{-15}$, respectively, comparing single-rail variants of our realization of the surface code and the 4.8.8 planar Floquet code.
    (The total number of qubits is $N=O( 4d_{\rm f}^2)$ for both of these codes.)
    }
    \label{tab:single_rail_d}
\end{table}

It is worth noting that we anticipate the comparison of our code's performance to that of the 4.8.8 Floquet code to improve when we use a noise model that better reflects the physical errors affecting Majorana hardware. 
This is due to to natural assumptions, such as two-qubit measurements having higher fault rates than single-qubit measurements, and measurements having higher fault rates than idling error rates; moreover, our code does not utilize $M_{YY:\text{vert}}$ measurements, which are used in the 4.8.8 Floquet code and can be expected to have higher fault rates than the $M_{XX:\text{horz}}$ and $M_{ZZ:\text{vert}}$ measurements.

As our code provides an interesting test bed for exploring the effect of hook errors, we now investigate this matter in simulation.
In  Fig.~\ref{fig:raw_hook_malignant}, we present the performance for our code (using the original circuits and pipelining of Sec.~\ref{sec:Circuits}) on a rotated surface code patch with boundary conditions intentionally chosen to align the hook errors with the corresponding logical operators (left) and on a torus (right).\footnote{We use $p_\mathrm{logical}$ to denote the error rate for incorrect recovery of \emph{any} of the logical membranes~\cite{bombinlogical2023}, i.e.~which applies for a logical error on either of the two qubits encoded on a torus.}
The hook errors are malignant for both of these systems.
These can be compared to the code performance on a patch with boundary conditions chosen to make hook errors benign, as shown in Fig.~\ref{fig:raw_DR}(left).
The choice of boundary conditions should not affect the fault-tolerance threshold, as the circuits implement the same bulk operations.
Indeed, the thresholds for the hook-malignant systems in Fig.~\ref{fig:raw_hook_malignant} are estimated to be approximately $0.65\%$ for the planar patch and $0.70\%$ for the torus.
The discrepancy between planar and torus is likely due to finite-size effects, and is perhaps not extremely surprising at these sizes.

On the other hand, when hook errors are malignant in the code, it should impact the scaling of the logical failure rate curves in the deep sub-threshold (low $p_\mathrm{physical}$) regime, due to the fault distance $d_{\rm f}$ being \emph{halved} as compared to the code distance $d$.
One interesting consequence of this distance-halving is that, in the low-error regime, we expect the curves to ``pair up'' in terms of slope (when plotted on a log-log scale): the code can correct up to $\lfloor \frac{d_{\rm f} -1}{2} \rfloor$ faults, with $d_{\rm f} = \lceil \frac{d}{2} \rceil$, meaning that $d$ must increase by four for the slope to increase by one.
Indeed, we observe all of this expected behavior in Fig.~\ref{fig:raw_hook_malignant}, where the colored dashed lines represent $p_\mathrm{logical} \sim p_\mathrm{physical}^{\lfloor (d_{\rm f} + 1)/2 \rfloor}$ scaling, chosen to intercept a reference empirical data point, as previously discussed in the context of Figs.~\ref{fig:raw_DR} and \ref{fig:raw_SR} [see Eq.~\eqref{eq:p_log_ref}---although here, the dashed lines are only drawn for visual reference and not used for any subsequent calculations].
More rigorously investigating the deep sub-threshold scaling of these hook-malignant codes with extensive, large-scale \emph{stratified sampling}~\cite{Paetznick2023} targeting only the dominant subpopulations expected to contribute to $p_\mathrm{logical}$ at $p_\mathrm{physical} \ll 1$ is an interesting topic for future work. 
Initial such studies on the hook-malignant rotated surface code patch in Fig.~\ref{fig:raw_hook_malignant}(left) indicate that we can indeed at least find fault configurations contributing to $p_\mathrm{logical}$ at the expected powers in $p_\mathrm{physical}$.
For example, at $d=11$, we can see logical failures for subpopulations contributing to $p_\mathrm{logical}$ at $O(p_\mathrm{physical}^3)$, where $3 = \lfloor \frac{\lceil 11/2 \rceil + 1}{2} \rfloor$ is the slope of the dashed curve for $d=11$ in Fig~\ref{fig:raw_hook_malignant}(left).
Finally, we remark that hook errors of course have the most dramatic consequence at the smallest $d=3$, where the code is now no longer even error-correcting in the circuit noise model, and thus we expect $p_\mathrm{logical} \propto p_\mathrm{physical}$, as observed in the data in Fig.~\ref{fig:raw_hook_malignant}(left).

Finally, we evaluate the performance of our code variant utilizing the hook-preventing circuits and pipelining described in Sec.~\ref{sec:Hook}. 
For this, we have performed simulations for the code on a torus for the noise model without and with idle noise.
The performance results in Fig.~\ref{fig:raw_hook_preventing} demonstrate the expected recovery of the full distance, i.e. $d_{\rm f}=d$, in the deep sub-threshold regime.
Moreover, the performance is overall better than that of the original circuits when hook errors are malignant.
Since the hook-preventing measurement circuits are different from the original circuits, we no longer expect the thresholds to be the same as before.
For these hook-preventing variants of our surface code realization on the torus, we find fault-tolerance thresholds of approximately $0.61\%$ when there is no idle noise and $0.43\%$ when there is idle noise.
This represents a modest decrease from the threshold value of the code using the original measurement circuits.

\begin{acknowledgments}

We are very grateful to A.~Paetznick for many useful discussions and help with the decoder implementation and performance assessment.   
We also thank N.~Delfosse for helpful discussions about the decoder construction, M.~Beverland for suggesting dynamic weight assignment of decoding graphs for the hook-preventing circuits, J.~Weston for assistance setting up the large-scale simulation pipeline on Azure, and A.~Paz for providing the parsable circuit format we use to share the simulated circuits.
We thank J.~Haah, M.~Hastings, C.~Nayak, and K.~Svore for helpful feedback.

\end{acknowledgments}

\clearpage
\appendix

\section{Interleaved Stabilizer Measurement Circuits}
\label{app:interleave}

\begin{figure}[t!]
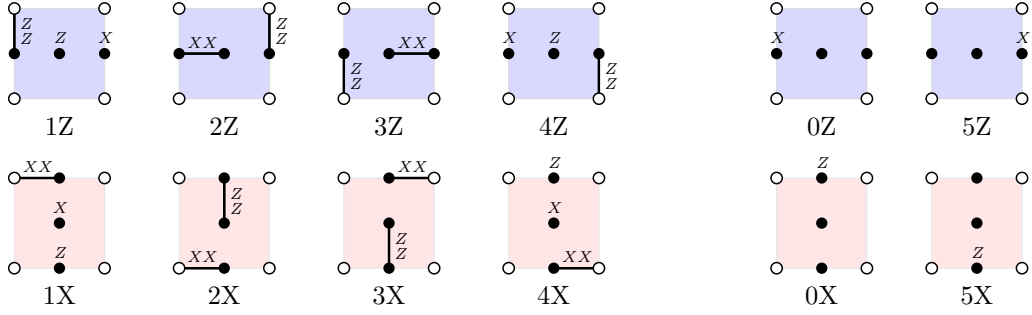

\centering

\input{ZZZZ_inter}\hspace{1.5cm}\input{ZZZZ_0}\input{ZZZZ_5}

\input{XXXX_inter}\hspace{1.5cm}\input{XXXX_0}\input{XXXX_5}

\caption{
The stabilizer measurement circuits can be interleaved and run on a synchronous schedule, i.e. $(0Z,0X),(1Z,1X),\ldots$, by addressing the data qubits in a different order than the pipelined measurement schedule presented in Sec.~\ref{sec:Circuits}.
This interleaved measurement schedule has significant disadvantages compared to the pipelined schedule.
}
\label{fig:interleaved_circuit}
\end{figure}

We can interleave $M_{ZZZZ}$ and $M_{XXXX}$ circuits while running them on the same schedule, i.e. $(0Z,0X),(1Z,1X),\ldots$, by choosing the order in which data qubits are addressed as shown in Fig.~\ref{fig:interleaved_circuit}. 
This has the slight benefit of requiring $4r+2$ steps for $r$ rounds of stabilizer measurement, but it turns out to have significant disadvantages when compared to the pipelined measurement schedule presented in Sec.~\ref{sec:Circuits}.
One disadvantage is that pipelined circuits are expected to perform better than interleaved circuits when using dead component protocols, as discussed in Sec.~\ref{sec:DeadComponents}.
This is because using interleaved circuits with the dead component protocols will yield fewer total measurements of the superplaquette operators than using pipelined circuits.
Another disadvantage arises when using single-rail semiconductor layouts in Majorana hardware, as discussed in Sec.~\ref{sec:Majorana}.
For the interleaved measurement schedule, we find that single-rail layouts would require each circuit step to be split into two steps in order to avoid physically conflicting measurements (overlapping measurement loops).
This would double the measurement period to 8 steps, in contrast with the pipelined circuit which could be implemented in single-rail layouts with a 5 step period, introducing greater opportunity for idling errors to damage performance.
In terms of code performance, for the best case scenario, i.e. ignoring these dead component and single-rail issues, we find that the interleaved and pipelined scheduling yield very similar performance data, so there is essentially no upside to using the interleaved measurement-based circuits.

\section{Hook Preventing Modifications for the Pentagonal Tiling Realization of the Surface Code}
\label{app:pentagonal_hook_prevention}

\begin{figure}[t!]
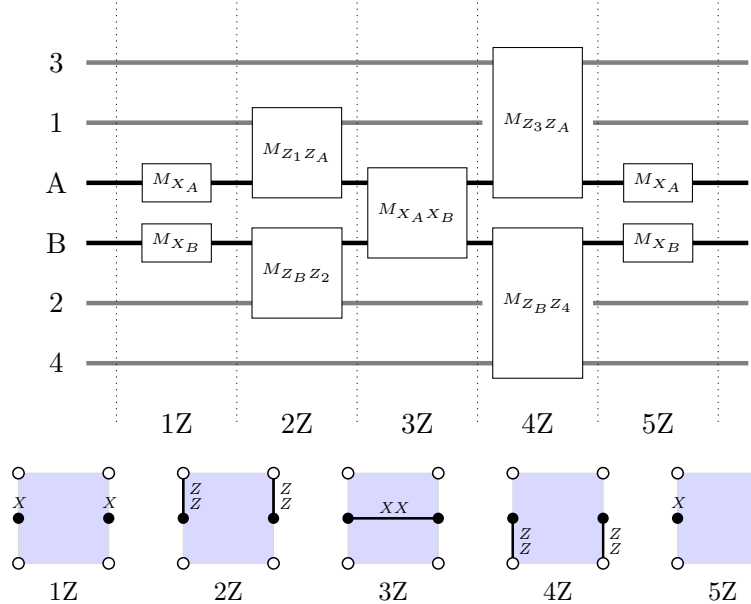

\centering
\input{ZZZZ_diagram_Gidney}

\input{ZZZZ_Gidney}
\caption{
The $M_{ZZZZ}$ circuit from Ref.~\protect\onlinecite{Gidney2022a} for the pentagonal tiling realization of the surface code.
}
\label{fig:pentagonal}
\end{figure}

\begin{figure}[t!]
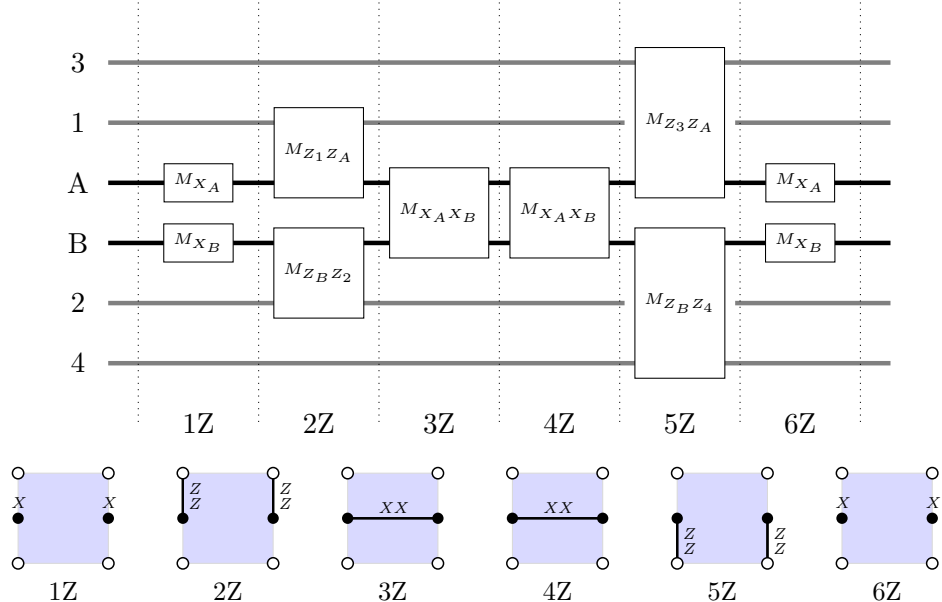

\centering
\input{ZZZZ_diagram_Gidney_hook}

\input{ZZZZ_Gidney_hook}
\caption{
A modification of the $M_{ZZZZ}$ circuit from Ref.~\protect\onlinecite{Gidney2022a} that prevents the problematic hook errors associated with readout error at the $M_{X_A X_B}$ measurement.
Incorporating this and a similar modification of the $M_{XXXX}$ circuit reduces the problem of bidirectional hook errors to unidirectional hook errors in the pentagonal tiling realization of the surface code.
}
\label{fig:pentagonal_hook_detect}
\end{figure}

The pentagonal tiling surface code realization of Ref.~\onlinecite{Gidney2022a} utilizes two auxiliary qubits for each 4-gon stabilizer measurement, the circuit of which is shown in Fig.~\ref{fig:pentagonal}.
In contrast to our realization, circuit noise for the pentagonal tiling circuits results in bidirectional hook errors, as discussed in Ref.~\onlinecite{Gidney2022a}.
In particular, for the $M_{ZZZZ}$ circuit, a readout error at the $M_{X_A X_B}$ measurement is equivalent to a $Z_1 Z_3$ or $Z_2 Z_4$ error on the data qubits, while a $Z_A Z_B$ error at the same measurement is equivalent to a $Z_1 Z_2$ or $Z_3 Z_4$ error on the data qubits.
This bidirectionality makes hook errors more problematic for the pentagonal tiling realization.
For example, one cannot align these hook errors to be perpendicular to the direction of the corresponding logical operators.
Applying our hook-preventing idea to the pentagonal tiling circuit by repeating the pairwise auxiliary qubit measurement, as shown in Fig.~\ref{fig:pentagonal_hook_detect}, will eliminate the hook errors corresponding to the readout errors, though not the two-qubit Pauli errors.
The remaining hook errors of our hook-preventing pentagonal tiling circuits are unidirectional with the direction correlated with the error type (though oppositely correlated with our non-hook-preventing circuits): the $Z$-plaquettes' hook errors correspond to $ZZ$ data qubit errors in the horizontal direction and the $X$-plaquettes' hook errors correspond to $XX$ data qubit errors in the vertical direction.
We note that this is the opposite directionality of hook errors that we found for our stabilizer measurement circuits in Sec.~\ref{sec:Hook}.
Again, one way of addressing these remaining unidirectional hook errors is to choose logical operators to be aligned perpendicular to the corresponding type of hook errors.

There is another, somewhat more drastic modification one can make to the stabilizer measurement circuits of Ref.~\onlinecite{Gidney2022a} that prevents their hook errors in the other direction.
Comparing our $M_{ZZZZ}$ circuit in Fig.~\ref{fig:MZ4circuit} with the $M_{ZZZZ}$ circuit in Fig.~\ref{fig:pentagonal}, we can retrospectively view our circuit as a modification of the pentagonal tiling $M_{ZZZZ}$ circuit by introducing an additional auxiliary qubit and appropriate measurements to obtain an equivalent circuit.
This modification has the effect of trading the horizontal hook error due to a $Z_A Z_B$ error at the pairwise auxiliary qubit measurement in the pentagonal tiling circuit for a vertical hook error due to a $Z_B$ error between the two pairwise auxiliary qubit measurements in our circuit, again leaving only unidirectional hook errors.

\section{Dead Components in Other Surface Code Realizations}
\label{app:dead_component_realizations}

\begin{figure}[t!]
\centering

\begin{minipage}{1\linewidth}
{\flushleft
\input{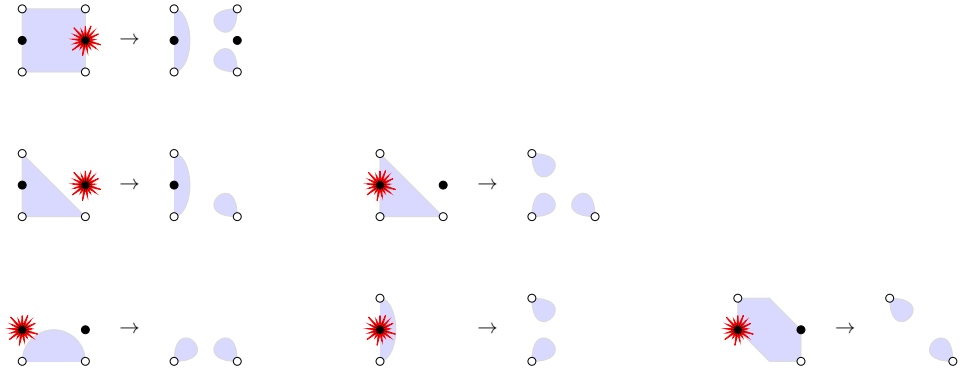}
}
\end{minipage}

\caption{The possible splittings of $Z$-type $n$-gons for step 2, where dead auxiliary qubits are removed from the code operation for the measurement-based pentagonal tiling realization of the surface code. 
(Splittings related to these by rotations and reflections are not shown separately.) 
The splittings for $X$-type $n$-gons may be obtained from these by 90 degree rotations.
}
\label{fig:dead_pent}
\end{figure}

In this appendix, we demonstrate our dead components strategy for the measurement-based pentagonal tiling surface code realization of Ref.~\onlinecite{Gidney2022a} and the CNOT gate-based realization of the surface code.
Step 1 is the same for all realizations, so we can use the modifications shown in Fig.~\ref{fig:dead_data} for removing dead data qubits, with the understanding that the array of auxiliary qubits and their collateral removals should be replaced with that of the given realization.
For the measurement-based pentagonal tiling realization, the splitting of plaquettes for steps 2 and 3 are shown in Figs.~\ref{fig:dead_pent} and \ref{fig:dead_pent_2}, respectively.
For the CNOT gate-based realization, the splitting of plaquettes for steps 2 and 3 are shown in Fig.~\ref{fig:dead_CNOT}.
We note that our strategy may not constitute a desirable trade-off for the CNOT gate-based realization in hardware where measurements could be a prohibitively costly resource, as it would increase in the number of measurements performed for each splitting.

\begin{figure}[t!]
\centering

\begin{minipage}{1\linewidth}
{\flushleft
\input{dead_connection_pent}
}
\end{minipage}

\caption{The possible splittings of $Z$-type $n$-gons for step 3, where dead connections are removed from the code operation for the measurement-based pentagonal tiling realization of the surface code.
(Splittings related to these by rotations and reflections are not shown separately.) 
The splittings for $X$-type $n$-gons may be obtained from these by 90 degree rotations.
}
\label{fig:dead_pent_2}
\end{figure}

\begin{figure}[t!]
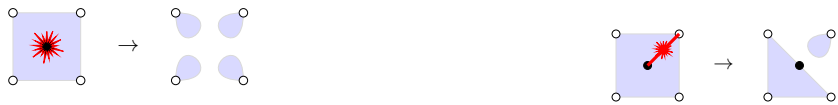

\centering

\begin{minipage}{0.35\linewidth}
\input{dead_auxiliary_CNOT}
\end{minipage}
\hspace{2cm}
\begin{minipage}{0.35\linewidth}
{\flushleft
\input{dead_connection_CNOT}}
\end{minipage}

\caption{
The splittings of $Z$-type or $X$-type plaquettes for steps 2 and 3, where dead auxiliary qubits and connections are removed from the code operation for the CNOT gate-based realization of the surface code.
All possible $n$-gons splittings are not shown because they all follow the same pattern: for step 2, a dead auxiliary qubit splits the $n$-gon into $n$ 1-gons; for step 3, a dead connection splits the $n$-gon into a $(n-1)$-gon and a 1-gon, according to which connection is dead.
}
\label{fig:dead_CNOT}
\end{figure}

We note that the pentagonal tiling realization of Ref.~\onlinecite{Gidney2022a} is pipelined in a manner that exhibits a natural alternation between $Z$-type and $X$-type plaquette measurements, similar to our surface code realization.
As such, it also has the advantage of measuring superplaquette operators at the maximum rate when using dead component protocols.
It is worth mentioning that a similar advantage could potentially be obtained for the CNOT gate-based realization of the surface code by using an appropriate pipelining of the $Z$-type and $X$-type circuits.
In particular, by offsetting the $Z$-type and $X$-type $n$-gon measurement circuits by three steps (and carefully choosing the order that data qubits are addressed in a circuit), the $n$-gon measurements effectively alternate between $Z$-type and $X$-type.
The advantage of this may be undone for hardware in which the measurements time is substantially longer than the CNOT gate time, as measurements will occur during four steps, rather than two steps per cycle with such pipelining.



\bibliographystyle{quantum}
\bibliography{references}

\end{document}

%% file: figurecommands.tex
\newcommand{\deadancilla}[1]{
\node at #1 [starburst, fill=black, draw=red, line width=0.1pt,scale=0.5] {};}

\newcommand{\deaddata}[1]{
\node at #1 [starburst, fill=white, draw=red, line width=0.1pt,scale=0.5] {};}

\newcommand{\noqubit}[1]{
\draw #1 node[circle, draw, fill=white, white, minimum size = 2*\nodesize  mm, inner sep = 0]{};
}

\newcommand{\deadconnection}[2]{
\draw[red, line width = 0.6mm] #1 -- #2 node () [midway,starburst, fill=red, draw=red, line width=0.1pt,scale=0.3] {};
}

\newcommand{\standardscales}{
\providecommand{\scaling}{1.2}
\providecommand{\halfscaling}{0.6}
\providecommand{\nodesize}{1.5}
}

\newcommand{\standardspacing}{
\draw [fill=white,opacity=0] (-0.5*\scaling,0.5*\scaling) rectangle (1.5*\scaling,-1.5*\scaling);
}

\newcommand{\vspacing}{\vspace{0.2cm}}

\newcommand{\hspacing}{\hspace{-0.3cm}}

\newcommand{\arrowspacing}{\hspace{-1.1cm}}

\newcommand{\inbetweenspacing}{\hspace{1cm}}

\newcommand{\diagramscale}{0.7}

\newcommand{\ZSWsmall}{
\draw[scale=1.25,fill = blue, opacity = 0.15] (0,-0.8*\scaling)  to[in=90,out=0,loop] (0,-0.8*\scaling);
}

\newcommand{\ZNW}[0]{
\draw[scale=2,fill = blue, opacity = 0.15] (0,0)  to[in=0,out=-90,loop] (0,0);
}

\newcommand{\ZNE}{
\draw[scale=2,fill = blue, opacity = 0.15] (0.5*\scaling,0)  to[in=-90,out=180,loop] (0.5*\scaling,0);
}

\newcommand{\ZSW}{
\draw[scale=2,fill = blue, opacity = 0.15] (0,-0.5*\scaling)  to[in=90,out=0,loop] (0,-0.5*\scaling);
}

\newcommand{\ZSE}{
\draw[scale=2,fill = blue, opacity = 0.15] (0.5*\scaling,-0.5*\scaling)  to[in=90,out=180,loop] (0.5*\scaling,-0.5*\scaling);
}

\newcommand{\ZZright}{
\begin{scope}
    \clip (0,0) rectangle (1*\scaling,-1*\scaling);
    \draw[fill = blue, opacity = 0.15] (0,-0.5*\scaling) ellipse (0.25*\scaling cm and 0.5*\scaling cm);
\end{scope}
}

\newcommand{\ZZleft}{
\begin{scope}
    \clip (0,0) rectangle (1*\scaling,-1*\scaling);
    \draw[fill = blue, opacity = 0.15] (1*\scaling,-0.5*\scaling) ellipse (0.25*\scaling cm and 0.5*\scaling cm);
\end{scope}
}

\newcommand{\ZZtop}{
\begin{scope}
    \clip (0,0) rectangle (1*\scaling,-1*\scaling);
    \draw[fill = blue, opacity = 0.15] (0.5*\scaling,-1*\scaling) circle(0.5*\scaling);
\end{scope}
}

\newcommand{\ZZslash}{
\draw [fill = blue, opacity = 0.15]  
(0.5*\scaling,0) -- (1*\scaling,0) -- (1*\scaling, -0.5*\scaling) 
  -- (0.5*\scaling,-1*\scaling) -- (0,-1*\scaling) -- (0, -0.5*\scaling) 
  -- cycle;
}

\newcommand{\ZZbackslash}{
\draw [fill = blue, opacity = 0.15]  
(0.5*\scaling,0) -- (0*\scaling,0) -- (0*\scaling, -0.5*\scaling) 
  -- (0.5*\scaling,-1*\scaling) -- (1*\scaling,-1*\scaling) -- (1*\scaling, -0.5*\scaling) 
  -- cycle;
}

\newcommand{\ZZZSW}{
\draw [fill = blue, opacity = 0.15]  (0,0) -- (0, -1*\scaling) 
  -- (1*\scaling,-1*\scaling) 
  -- cycle;
}

\newcommand{\ZZZNW}{
\draw [fill = blue, opacity = 0.15]  (0,0*\scaling) -- (1*\scaling, 0*\scaling) 
  -- (0*\scaling,-1*\scaling) 
  -- cycle;
}

\newcommand{\ZZZSE}{
\draw [fill = blue, opacity = 0.15]  (1*\scaling,0) -- (1*\scaling, -1*\scaling) 
  -- (0*\scaling,-1*\scaling) 
  -- cycle;
}

\newcommand{\ZZZNE}{
\draw [fill = blue, opacity = 0.15]  (0,0) -- (1*\scaling,0) 
  -- (1*\scaling,-1*\scaling) 
  -- cycle;
}

\newcommand{\ZZZZ}{
\draw [fill=blue,opacity=0.15] (0,0) rectangle (1*\scaling,-1*\scaling);
}

\newcommand{\XXright}{
\begin{scope}
    \clip (0,0) rectangle (1*\scaling,-1*\scaling);
    \draw[fill = red, opacity = 0.1] (0,-0.5*\scaling) circle(0.5*\scaling);
\end{scope}
}

\newcommand{\XXleft}{
\begin{scope}
    \clip (0,0) rectangle (1*\scaling,-1*\scaling);
    \draw[fill = red, opacity = 0.1] (1*\scaling,-0.5*\scaling) circle(0.5*\scaling);
\end{scope}
}

\newcommand{\XXtop}{
\begin{scope}
    \clip (0,0) rectangle (1*\scaling,-1*\scaling);
    \draw[fill = red, opacity = 0.1]  (0.5*\scaling, -1*\scaling)  ellipse (0.5*\scaling cm and 0.25*\scaling cm);
\end{scope}
}

\newcommand{\XXbottom}{
\begin{scope}
    \clip (0,0) rectangle (1*\scaling,-1*\scaling);
    \draw[fill = red, opacity = 0.1] (0.5*\scaling,0) ellipse (0.5*\scaling cm and 0.25*\scaling cm);
\end{scope}
}

\newcommand{\XXbackslash}{
\draw [fill = red, opacity = 0.1]  
(0.5*\scaling,0) -- (0*\scaling,0) -- (0*\scaling, -0.5*\scaling) 
  -- (0.5*\scaling,-1*\scaling) -- (1*\scaling,-1*\scaling) -- (1*\scaling, -0.5*\scaling) 
  -- cycle;
}

\newcommand{\XXXNW}{
\draw [fill = red, opacity = 0.1]  (0,0*\scaling) -- (1*\scaling, 0*\scaling) 
  -- (0*\scaling,-1*\scaling) 
  -- cycle;
}

\newcommand{\XXXSE}{
\draw [fill = red, opacity = 0.1]  (1*\scaling,0) -- (1*\scaling, -1*\scaling) 
  -- (0*\scaling,-1*\scaling) 
  -- cycle;
}

\newcommand{\XXXX}{
\draw [fill=red,opacity=0.1] (0,0) rectangle (1*\scaling,-1*\scaling);
}